\documentclass[11pt]{article}
\usepackage[margin=1in]{geometry}
\usepackage{amsmath,amssymb,mathtools}
\usepackage{booktabs}
\usepackage{float}
\usepackage[table]{xcolor}
\usepackage{tikz}
\usetikzlibrary{arrows.meta}
\usepackage{pgfplots}
\pgfplotsset{compat=1.18}
\definecolor{cD}{HTML}{2A78D6}   
\definecolor{cR}{HTML}{EB6834}   
\definecolor{cS}{HTML}{1BAF7A}   
\definecolor{cW}{HTML}{B0306A}   
\definecolor{cG}{HTML}{52514E}   
\definecolor{cM}{HTML}{6250D6}   
\pgfplotsset{report/.style={
  axis line style={black!45}, tick style={black!45}, grid=major, grid style={black!8},
  label style={font=\small}, tick label style={font=\footnotesize},
  legend style={font=\footnotesize,draw=black!25,fill=white,cells={anchor=west}}}}
\newcounter{rfile}
\newcommand{\rfitem}[2]{\refstepcounter{rfile}\label{rf:#1}\item[F\therfile] \path{#2}}
\newcommand{\rf}[1]{F\ref{rf:#1}}
\newcommand{\gridfill}{black!5}
\usepackage{array}
\usepackage{url}
\usepackage{xspace}
\usepackage{needspace}
\usepackage[colorlinks=true,linkcolor=blue,citecolor=blue,urlcolor=blue]{hyperref}
\hypersetup{pdftitle={Fast matrix multiplication: mixing fast, accurate schemes over a vendor GEMM, chosen by a calibrated planner},pdfauthor={Greg Sidebottom}}
\newcommand{\R}{\mathbb{R}}
\newcommand{\mm}[1]{\langle #1\rangle}
\newcommand{\ginf}{\gamma_{\infty,2}}
\newcommand{\amp}{\operatorname{amp}}
\newcommand{\vect}{\operatorname{vec}}
\newcommand{\lv}[1]{\texttt{#1}}
\newcommand{\blockgrid}[6]{%
  \fill[\gridfill] (#1,#2) rectangle ++(#3,#4);
  \foreach \i in {1,...,#5} {\draw[black!55,line width=0.3pt] (#1,{#2+#4*\i/#5}) -- ++(#3,0);}
  \foreach \j in {1,...,#6} {\draw[black!55,line width=0.3pt] ({#1+#3*\j/#6},#2) -- ++(0,#4);}
  \draw[black,line width=0.6pt] (#1,#2) rectangle ++(#3,#4);
}
\newcommand{\matbox}[8]{%
  \pgfmathsetmacro{\mbx}{#1-#3/2}%
  \pgfmathsetmacro{\mby}{#2-0.8-#4/2}%
  \blockgrid{\mbx}{\mby}{#3}{#4}{#5}{#6}%
  \node[anchor=south] at ({#1},{#2+0.05}) {#7};
  \node[anchor=north,align=center] at ({#1},{#2-1.68}) {#8};
}
\newcommand{\flatbox}[6]{%
  \pgfmathsetmacro{\mbx}{#1-#3/2}%
  \pgfmathsetmacro{\mby}{#2-0.8-#4/2}%
  \fill[cG!22] (\mbx,\mby) rectangle ++(#3,#4);
  \draw[black,line width=0.6pt] (\mbx,\mby) rectangle ++(#3,#4);
  \node[anchor=south] at ({#1},{#2+0.05}) {#5};
  \node[anchor=north,align=center] at ({#1},{#2-1.68}) {#6};
}
\title{Fast matrix multiplication:\\
mixing fast, accurate schemes over a vendor GEMM,\\
chosen by a calibrated planner}
\author{Greg Sidebottom}
\date{November 24, 2026}

\begin{document}
\maketitle

\begin{abstract}
Fast matrix multiplication is in practice a few levels of a Strassen-like
scheme over a vendor's General Matrix Multiplication (GEMM) routine.  We
treat each scheme as an operator that takes a problem one level down, and
let every level of a plan use a different one.  On Apple's M4~Pro, over
Accelerate's GEMM, a calibrated planner chooses among accurate forms of the
$\mm{2\times2\times2:7}$, $\mm{4\times4\times4:48}$ and
$\mm{3\times3\times6:40}$ schemes.  A level runs as three fused passes over
memory, so its cost is the blocks it moves, not its arithmetic.  At
$n=13824$ the plans run $1.71\times$ (f32) and $1.88\times$ (f64) faster than
Accelerate's GEMM with $0.38$ and $0.50$ of its error on well-scaled inputs.  At $6912$ and
$13824$ they are 5--16\% faster than the best plans built from the
$\mm{4\times4\times4:48}$ scheme alone, with $0.57$--$0.75$ of their error.
The planner's error model predicts a plan's error from where its levels sit
(a level's own rounding is multiplied by the levels above it), and the
planner uses it to choose between plans of equal speed.
A conventional implementation of Strassen--Winograd (one block operation at a time), the algorithm of IBM ESSL's \texttt{\_GEMMS}, over the same GEMM is $1.21$--$1.54\times$ slower than the plans and has $1.9$--$9.7$ times their error; fused $2\times2$ levels, which the library also has, are level with the $4\times4$ scheme at half the depth.  On badly scaled inputs the plans need exact power-of-two prescaling,
at up to 4\% of the time, and inputs whose magnitudes cross between blocks
defeat them.  In f32, sharing a level's products with the GPU brings the
speedup over Accelerate to $3.1\times$.
\end{abstract}

\section{Introduction}\label{sec:intro}

Throughout, the problem is $C=AB$ with $A$ an $m\times k$ and $B$ a
$k\times n$ matrix; a problem is \emph{square} when $m=k=n$, and a square
problem is named by that size $n$.  A scheme $\mm{g_1\times g_2\times g_3:r}$
multiplies a $g_1\times g_2$ grid of blocks by a $g_2\times g_3$ grid of
blocks with $r<g_1g_2g_3$ block products.  Applied to the problem, it cuts $A$
and $B$ into those grids and hands down $r$ products of
$(m/g_1)\times(k/g_2)$ by $(k/g_2)\times(n/g_3)$ blocks, and applied again to
those products it saves products at every level of recursion.  On a real
machine a level is not free: its linear combinations of blocks are passes
over memory, and the products it hands down are General Matrix
Multiplications (GEMMs, the matrix product routine of the BLAS, the Basic
Linear Algebra Subprograms) whose rate depends on their size
and shape.  Libraries that use fast schemes therefore apply one to three
levels on top of a tuned GEMM and stop \cite{HSvdG16,DPSST26}.  Which scheme, how
many levels, in what order, and at what cost in accuracy, are the questions
this report answers by measurement for one machine, \emph{the machine}
from here on: Apple's M4~Pro with Accelerate's GEMM (details in
Section~\ref{sec:setup} and in the glossary, Appendix~\ref{sec:glossary}).

A scheme is one member of a family.  For invertible matrices $U$, $V$ and
$W$, the product of $UAV^{-1}$ and $VBW^{-1}$ is $U(AB)W^{-1}$, so running a
scheme on those two transformed grids and transforming its result back is
again a scheme for $AB$, with the same $r$ products and other
coefficients.  The schemes obtained this way from one scheme are its
\emph{orbit}, or \emph{class}, under the group $GL_{g_1}\times
GL_{g_2}\times GL_{g_3}$ of such triples \cite{deGroote78}, and each of them
is a \emph{point} of the orbit.  Every point multiplies exactly and costs
the same products; the points differ in their coefficients, and so in how
much they amplify rounding errors.  The accurate schemes of Dumas, Pernet
and Sedoglavic (DPS) are points chosen to amplify them least
\cite{DPS24,DPS26a}.

\begin{figure}[H]
\centering\hyphenpenalty=10000 \exhyphenpenalty=10000
\begin{tikzpicture}[font=\small,
  tok/.style={rounded corners=2pt,inner xsep=5pt,inner ysep=4pt,line width=0.8pt,font=\ttfamily\large},
  note/.style={align=left,text width=3.45cm,font=\footnotesize,anchor=north},
  lead/.style={line width=0.6pt}]
\node[tok,draw=cS,fill=cS!15] (a) at (0,0) {SW};
\node[font=\ttfamily\large] at (1.9,0) {/};
\node[tok,draw=cR,fill=cR!15] (b) at (3.8,0) {R363+3};
\node[font=\ttfamily\large] at (5.7,0) {/};
\node[tok,draw=cR,fill=cR!15] (c) at (7.6,0) {R633};
\node[font=\ttfamily\large] at (9.5,0) {/};
\node[tok,draw=cG,fill=cG!12] (d) at (11.4,0) {G6};
\draw[-{Stealth[length=5pt]},black!55,line width=0.6pt] (-0.6,0.75) -- (12.0,0.75)
  node[midway,above,font=\footnotesize,text=black] {from the whole problem (left) down to the leaf products (right)};
\draw[lead,cS] (a.south) -- ++(0,-0.45);
\node[note] at (0,-0.85) {\textbf{a $2\times2$ level}\\ $A$, $B$, $C$ cut into $2\times2$ grids;\\ $7$ block products instead of~$8$};
\draw[lead,cR] (b.south) -- ++(0,-0.45);
\node[note] at (3.8,-0.85) {\textbf{a $3\times6\times3$ level}\\ on each of those $7$ products;\\ $40$ products instead of $54$.\\ \lv{+3}: its passes run on $3$ threads while the products below are computed};
\draw[lead,cR] (c.south) -- ++(0,-0.45);
\node[note] at (7.6,-0.85) {\textbf{a $6\times3\times3$ level}\\ on each of those $280$ products;\\ $40$ instead of $54$ again};
\draw[lead,cG] (d.south) -- ++(0,-0.45);
\node[note] at (11.4,-0.85) {\textbf{the leaf}\\ the $7\cdot40\cdot40=11{,}200$ products that remain go to Accelerate's GEMM, six at a time};
\end{tikzpicture}
\caption{How to read a plan.  The colours mark the scheme families in every
figure of the report: the $2\times2$ level, the $3\times3\times6$ levels,
the $4\times4$ level (\lv{D48}, not in this plan) and the GEMM.  The counts
of products do not depend on the size of the problem, only the size of each
product does: at $n=13824$ (Figure~\ref{fig:plan}) a leaf product is
$384\times384$ by $384\times768$, and the classical algorithm on the same
grids would need $8\cdot54\cdot54=23{,}328$ of them.}
\label{fig:read}
\end{figure}
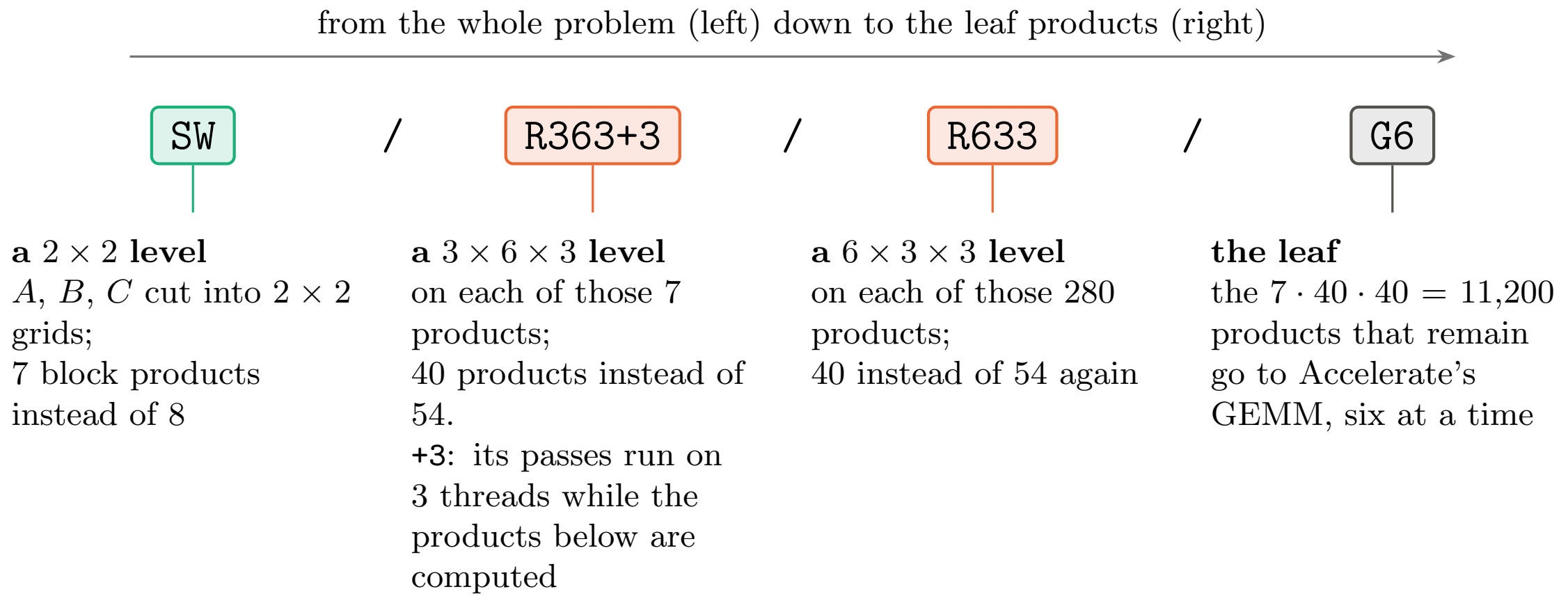

The approach is to mix.  Each scheme is an operator for one step down: a
\emph{level} cuts the current problem's operands into a grid of blocks and
replaces the product of the two grids by the scheme's block products, each a
smaller problem for the next level.  A \emph{plan} is a sequence of levels
from the top of the problem down to a GEMM leaf, and every level may use a
different operator from the menu: Strassen's $\mm{2\times2\times2:7}$
scheme at the accurate point of DPS in $\mathbb{Q}(\sqrt3)$ \cite{DPS24} (Section~\ref{sec:sw}); the
accurate $\mm{4\times4\times4:48}$ scheme of DPS in its $262$-operation form
\cite{DPS26a,PLinOpt}; or their accurate $\mm{3\times3\times6:40}$ scheme
\cite{DPS26b}, which is Tichavsk\'y's Decomposition~A \cite{Tichavsky21}, a
point in the class of Smirnov's \cite{Smirnov13}, in any of
its three orientations, which differ in which of the problem's dimensions
$m$, $k$, $n$ the level cuts into six parts and which into three.  The menu is these schemes because they measured
useful: a screen of the published schemes (Section~\ref{sec:menu}) leaves
nothing else that could make a plan faster at the same accuracy, and the
planner (Section~\ref{sec:planner}) decides per problem which of them to use,
in what order and how deep.  A plan is written from the top of the problem
down to the leaves, the levels separated by slashes
(Figure~\ref{fig:read}; Table~\ref{tab:notation} in Section~\ref{sec:setup}
lists every code).  Figure~\ref{fig:plan} draws the same plan at $n=13824$,
from the problem at the top to its $11{,}200$ leaf GEMMs at the bottom.

\begin{figure}[!tbp]
\centering
\begin{tikzpicture}
\begin{axis}[report, width=14.2cm, height=9.2cm,
  xmin=0.92, xmax=2.08, ymode=log, ymin=0.25, ymax=13,
  log ticks with fixed point, ytick={0.5,1,2,5,10}, xtick={1,1.2,1.4,1.6,1.8,2},
  xlabel={speed, as a multiple of Accelerate's GEMM (to the right is faster)},
  ylabel={error, as a multiple of the GEMM's (lower is better)},
  legend pos=north east, clip=false]
\fill[cR!7] (axis cs:1,0.25) rectangle (axis cs:2.08,1);
\draw[black!45,line width=0.5pt] (axis cs:0.92,1) -- (axis cs:2.08,1);
\draw[black!45,line width=0.5pt] (axis cs:1,0.25) -- (axis cs:1,13);
\node[anchor=south west,font=\footnotesize,text=black!70] at (axis cs:1.02,0.255) {faster \emph{and} more accurate than the GEMM};
\addplot[cW,line width=0.9pt,mark=diamond*,mark size=3.2pt,mark options={fill=cW,draw=white,line width=0.4pt}]
  coordinates {(1.107,1.715) (1.217,3.05) (1.289,4.89) (1.163,9.35)};
\addlegendentry{Strassen--Winograd, our implementation}
\addplot[only marks,mark=square*,mark size=2.9pt,mark options={fill=cD,draw=white,line width=0.4pt}]
  coordinates {(1.22,0.634) (1.62,0.673)};
\addlegendentry{the $4\times4$ scheme alone}
\addplot[only marks,mark=*,mark size=2.9pt,mark options={fill=cR,draw=white,line width=0.4pt}]
  coordinates {(1.23,0.586) (1.20,0.404) (1.69,0.546) (1.75,0.472) (1.87,0.502) (1.90,0.502) (1.87,0.546)};
\addlegendentry{plans with $3\times3\times6$ levels}
\addplot[only marks,mark=triangle*,mark size=3.4pt,mark options={fill=cS,draw=white,line width=0.4pt}]
  coordinates {(1.80,0.902) (1.88,0.653)};
\addlegendentry{a $2\times2$ level on top}
\addplot[only marks,mark=*,mark size=3.4pt,mark options={fill=cG,draw=white,line width=0.4pt},forget plot] coordinates {(1,1)};
\addplot[only marks,mark=*,mark size=3.6pt,mark options={fill=cR,draw=black,line width=1pt},forget plot] coordinates {(1.88,0.502)};
\node[anchor=south west,font=\footnotesize] at (axis cs:1.01,1.03) {Accelerate's GEMM};
\node[anchor=north,font=\footnotesize,align=center] at (axis cs:1.88,0.41) {the planner's choice:\\ $1.88\times$ faster, half the error};
\node[anchor=south,font=\footnotesize] at (axis cs:1.60,0.71) {two $4\times4$ levels};
\node[anchor=south,font=\footnotesize] at (axis cs:1.22,0.67) {one $4\times4$ level};
\node[anchor=north,font=\footnotesize] at (axis cs:1.20,0.385) {one \lv{R363} level};
\node[anchor=west,font=\footnotesize] at (axis cs:1.125,1.64) {1 level};
\node[anchor=west,font=\footnotesize] at (axis cs:1.232,3.05) {2};
\node[anchor=west,font=\footnotesize] at (axis cs:1.304,4.89) {3};
\node[anchor=east,font=\footnotesize] at (axis cs:1.148,9.35) {4 levels};
\end{axis}
\end{tikzpicture}
\caption{Speed against error at $n=13824$ in f64 on uniform inputs, both as
multiples of Accelerate's GEMM (the dot at $1$, $1$).  Every plan with the
accurate levels is in the shaded corner, faster and more accurate than the
GEMM; Strassen--Winograd, the algorithm of ESSL's \texttt{\_GEMMS}, implemented conventionally over the same GEMM, at one to four levels, gains
less speed and multiplies the error.  The points are the plans of
Table~\ref{tab:results} and the depths of Table~\ref{tab:essl}'s run.}
\label{fig:headline}
\end{figure}
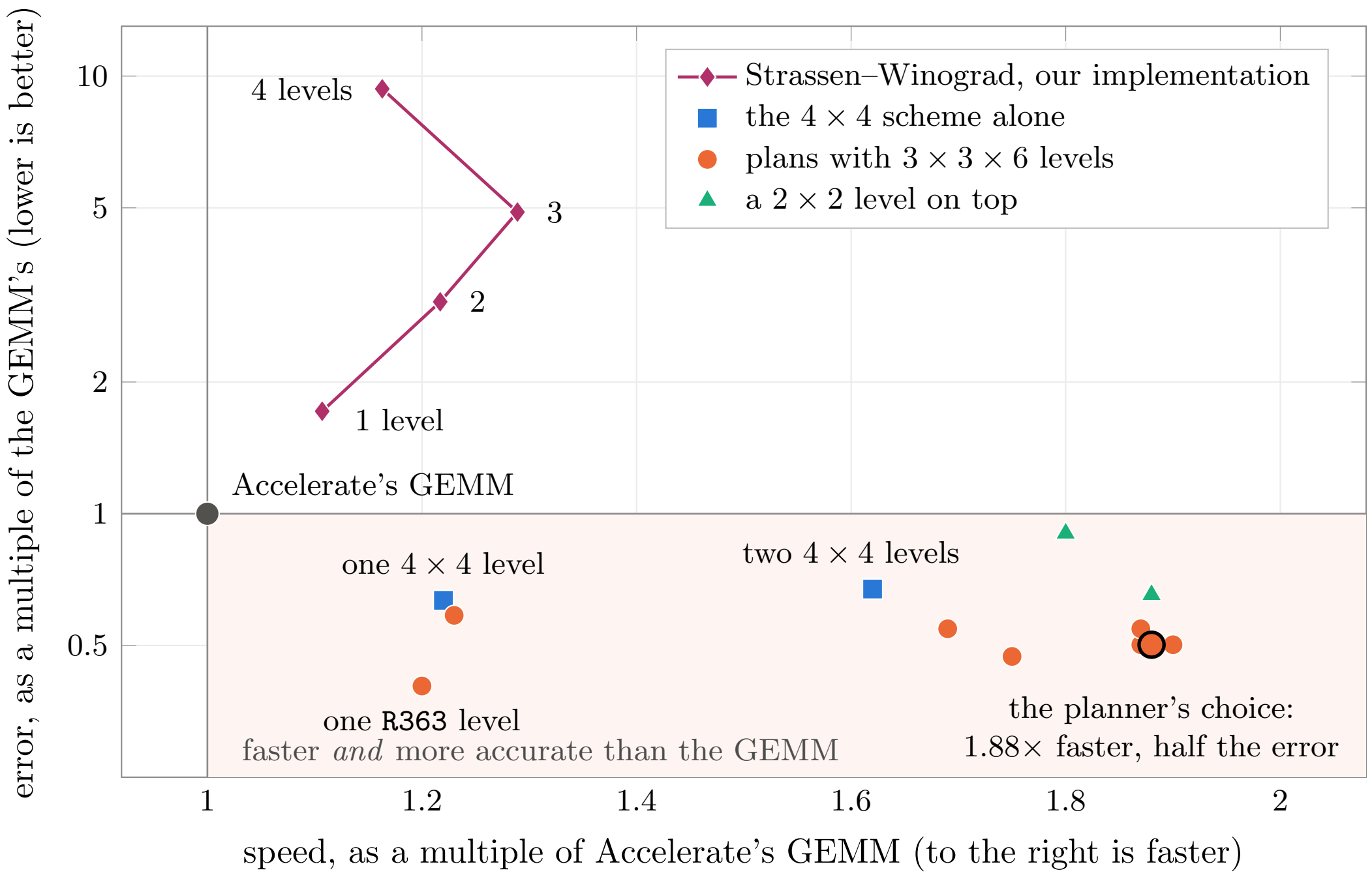

\paragraph{The result.}
Figure~\ref{fig:headline} and Table~\ref{tab:intro-acc} are the headline: the
planner's own choice against Accelerate's GEMM alone on the same problem, at
every size and precision tried, is $1.20$--$1.88\times$ faster and has
$0.38$--$0.61$ of the GEMM's error at once.  The \emph{error} of a product
is its largest deviation from the exact product, relative to the exact
product's largest entry, on uniform random inputs
(Section~\ref{sec:setup}).  Such inputs are the favourable case for a fast
scheme: Section~\ref{sec:scaled} measures graded and structured inputs,
where the plans need prescaling and one structure defeats them.
Table~\ref{tab:intro-matched} is the claim against the alternative: at
$6912$ and $13824$, in each precision, the planner's plan is $5$--$16\%$
faster than the best plan built from the $\mm{4\times4\times4:48}$ scheme
alone and has $0.57$--$0.75$ of its error.  Table~\ref{tab:gpu}
(Section~\ref{sec:gpu}) brings in the other engine on the machine, Apple's
GPU, in f32, the only precision it has: alone, as a second engine for the
classical product, and as a second engine for a level's products.

\begin{table}[!tbp]
\centering\footnotesize
\setlength{\tabcolsep}{5pt}
\begin{tabular}{llrrrrrr}
\toprule
$n$, prec. & mixed plan & time (s) & Accelerate (s) & faster & error & Accelerate's & of the GEMM's\\
\midrule
3456 f32 & \lv{R363/G6} & 0.0204 & 0.0244 & $1.20\times$ & $1.28\cdot10^{-6}$ & $3.04\cdot10^{-6}$ & $0.42$\\
6912 f32 & \lv{R363+4/R633/G6} & 0.137 & 0.194 & $1.41\times$ & $2.20\cdot10^{-6}$ & $3.85\cdot10^{-6}$ & $0.57$\\
13824 f32 & \lv{R363+4/R363/G6} & 0.907 & 1.551 & $1.71\times$ & $2.46\cdot10^{-6}$ & $6.40\cdot10^{-6}$ & $0.38$\\
3456 f64 & \lv{R363/G4} & 0.0715 & 0.0974 & $1.36\times$ & $2.76\cdot10^{-15}$ & $4.75\cdot10^{-15}$ & $0.58$\\
6912 f64 & \lv{R633+3/R363/G6}$^\dagger$ & 0.445 & 0.765 & $1.72\times$ & $4.07\cdot10^{-15}$ & $6.69\cdot10^{-15}$ & $0.61$\\
13824 f64 & \lv{R363+4/R336/G3} & 3.265 & 6.138 & $1.88\times$ & $6.18\cdot10^{-15}$ & $1.23\cdot10^{-14}$ & $0.50$\\
\bottomrule
\end{tabular}
\caption{The planner's choice under its default budget against Accelerate's
GEMM alone, from the campaign of Table~\ref{tab:results}; ``Accelerate's''
is the GEMM's own error and the last column the plan's error as a multiple
of it.  The choices are those of the library as it stands, whose tie rule
(Section~\ref{sec:planner}) changed after the campaign: five of the six were
among the campaign's fixed plans and have its times; $^\dagger$the sixth was measured
afterwards (Table~\ref{tab:results}).  Without a budget the
planner chooses otherwise at three sizes: \lv{SW+4/R363/G6} at $3456$ f64
($0.0706$~s, $3.29\cdot10^{-15}$), \lv{R363+2/R336/G6} at $6912$ f64
($0.449$~s, $4.32\cdot10^{-15}$) and \lv{R363+3/R336/G6} at $13824$ f32
($0.901$~s, $3.53\cdot10^{-6}$).}
\label{tab:intro-acc}
\end{table}

\begin{table}[!tbp]
\centering\footnotesize
\begin{tabular}{llrrlrrr}
\toprule
$n$, prec. & mixed plan & time (s) & error & best $4\times4$-only & time (s) & error & error ratio\\
\midrule
6912 f32 & \lv{R363+4/R633/G6} & 0.137 & $2.20\cdot10^{-6}$ & \lv{D48+4/D48/G6} & 0.146 & $2.94\cdot10^{-6}$ & $0.75$\\
13824 f32 & \lv{R363+4/R363/G6} & 0.907 & $2.46\cdot10^{-6}$ & \lv{D48+4/D48/G6} & 0.952 & $4.35\cdot10^{-6}$ & $0.57$\\
6912 f64 & \lv{R633+3/R363/G6}$^\dagger$ & 0.445 & $4.07\cdot10^{-15}$ & \lv{D48+4/D48/G6} & 0.471 & $5.60\cdot10^{-15}$ & $0.73$\\
13824 f64 & \lv{R363+4/R336/G3} & 3.265 & $6.18\cdot10^{-15}$ & \lv{D48+4/D48/G6} & 3.784 & $8.28\cdot10^{-15}$ & $0.75$\\
\bottomrule
\end{tabular}
\caption{Matched time: the planner's choice under its default budget against
the best plan built from the $\mm{4\times4\times4:48}$ scheme alone, from the
same campaign (Accelerate's times and errors in Table~\ref{tab:intro-acc});
the last column is the mixed plan's error as a multiple of the
$4\times4$-only plan's.  $^\dagger$Both times of this row are from the later run of Table~\ref{tab:results}'s note.}
\label{tab:intro-matched}
\end{table}

\begin{figure}[!tbp]
\centering
\begin{tikzpicture}[font=\footnotesize,
  blk/.style={draw=black!60,fill=black!8,minimum width=0.62cm,minimum height=0.42cm,inner sep=0pt},
  outb/.style={draw=cD!80!black,fill=cD!18,minimum width=0.62cm,minimum height=0.42cm,inner sep=0pt},
  ar/.style={-{Stealth[length=4pt]},black!60,line width=0.5pt}]
\node[font=\small\bfseries,anchor=west] at (-0.3,3.4) {One operation at a time};
\node[anchor=west,text=black!70] at (-0.3,2.95) {every addition is its own sweep over whole blocks};
\foreach \y/\t in {2.2/{sweep 1}, 1.4/{sweep 2}, 0.6/{sweep 3}} {
  \node[anchor=east] at (0.75,\y) {\t};
  \node[blk] at (1.3,\y) {}; \node at (1.85,\y) {$+$}; \node[blk] at (2.4,\y) {};
  \draw[ar] (2.85,\y) -- (3.55,\y);
  \node[outb] at (4.0,\y) {};
}
\node[anchor=west,align=left,text width=6.3cm] at (-0.3,-0.55)
  {\dots\ and so on: a program of $N$ operations is $N$ sweeps, each reading two blocks from main memory and writing one back};
\begin{scope}[xshift=7.3cm]
\node[font=\small\bfseries,anchor=west] at (-0.3,3.4) {Fused: one pass};
\node[anchor=west,text=black!70] at (-0.3,2.95) {one loop over the element positions of a block};
\foreach \i in {0,...,3} \foreach \j in {0,...,3}
  \node[blk,minimum width=0.3cm,minimum height=0.3cm] at ({0.2+0.34*\j},{2.1-0.34*\i}) {};
\node[anchor=north] at (0.71,0.92) {$16$ blocks of $A$};
\node[draw=black!70,rounded corners=3pt,align=center,inner sep=5pt,fill=white] (reg) at (3.25,1.6)
  {registers:\\ the whole\\ $L$ program};
\draw[ar,line width=1.4pt] (1.5,1.6) -- (reg.west);
\foreach \i in {0,...,5} \foreach \j in {0,...,7}
  \node[outb,minimum width=0.3cm,minimum height=0.3cm] at ({5.05+0.34*\j},{2.45-0.34*\i}) {};
\draw[ar,line width=1.4pt] (reg.east) -- (4.8,1.6);
\node[anchor=north] at (6.24,0.58) {$48$ left factors};
\node[anchor=west,align=left,text width=7.4cm] at (-0.3,-0.55)
  {every input block is read once and every output block written once, whatever $N$ is:
   the cost is the $16+48=64$ blocks moved, not the arithmetic};
\end{scope}
\end{tikzpicture}
\caption{What ``fused'' means, for the $L$ program of the $4\times4$ level:
$16$ blocks of $A$ in, $48$ left factors out.  One operation at a time, the
program's operations are as many sweeps through main memory; fused, they
are one pass.  (Figure~\ref{fig:machine} draws the machine: the registers
are a core's, and main memory is two caches away.)}
\label{fig:fused}
\end{figure}
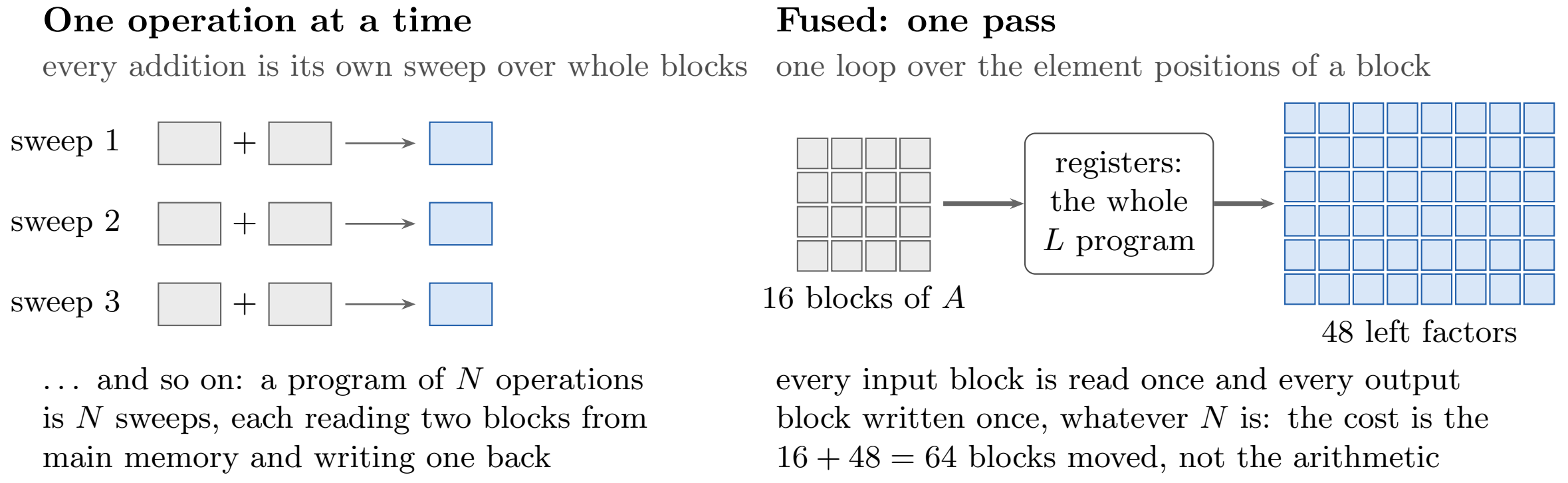

\begin{figure}[!tbp]
\centering
\begin{tikzpicture}[x=1cm,y=1cm,scale=0.88,
  every node/.style={font=\footnotesize,inner sep=1.5pt},
  down/.style={-{Stealth[length=5pt]},black!45,line width=0.6pt},
  up/.style={-{Stealth[length=6pt]},black,line width=1pt}]
\renewcommand{\gridfill}{cS!16}
\node[anchor=east,align=right] at (2.65,-0.8)
  {Level 1: \lv{SW}\\ $\mm{2\times2\times2:7}$\\ $7$ products of $8$};
\matbox{4.6}{0}{1.6}{1.6}{2}{2}{$A$}{$2\times2$ grid\\ $6912^2$ blocks}
\node at (6.25,-0.8) {$\times$};
\matbox{7.9}{0}{1.6}{1.6}{2}{2}{$B$}{$2\times2$ grid\\ $6912^2$ blocks}
\node at (9.55,-0.8) {$=$};
\matbox{11.2}{0}{1.6}{1.6}{2}{2}{$C=AB$}{$2\times2$ grid\\ $6912^2$ blocks}
\node[anchor=west,align=left] at (13.0,-0.8) {$1$ problem\\ $n=13824$, f64};
\draw[down] (2.95,-1.75) -- (2.95,-3.1);
\node[anchor=east,align=right] at (2.8,-2.42) {$L$, $R$ passes:\\ $7$ factors $T_i$, $S_i$};
\draw[up] (12.85,-3.1) -- (12.85,-1.75);
\node[anchor=west,align=left] at (13.0,-2.42) {$P$ pass: $C$'s $4$ blocks\\ from the $7$ products $M_i$};
\renewcommand{\gridfill}{cR!16}
\node[anchor=east,align=right] at (2.65,-4.1)
  {Level 2: \lv{R363+3}\\ $\mm{3\times6\times3:40}$\\ $40$ products of $54$};
\matbox{4.6}{-3.3}{1.6}{1.6}{3}{6}{$T_i$}{$3\times6$ grid\\ $2304\times1152$ blocks}
\node at (6.25,-4.1) {$\times$};
\matbox{7.9}{-3.3}{1.6}{1.6}{6}{3}{$S_i$}{$6\times3$ grid\\ $1152\times2304$ blocks}
\node at (9.55,-4.1) {$=$};
\matbox{11.2}{-3.3}{1.6}{1.6}{3}{3}{$M_i=T_iS_i$}{$3\times3$ grid\\ $2304^2$ blocks}
\node[anchor=west,align=left] at (13.0,-4.1) {$7$ sub-problems\\ $6912\times6912$};
\draw[down] (2.95,-5.05) -- (2.95,-6.4);
\node[anchor=east,align=right] at (2.8,-5.72) {$L$, $R$ passes:\\ $40$ factors $T_{ij}$, $S_{ij}$};
\draw[up] (12.85,-6.4) -- (12.85,-5.05);
\node[anchor=west,align=left] at (13.0,-5.72) {$P$ pass: $M_i$'s $9$ blocks\\ from the $40$ products $M_{ij}$};
\node[anchor=east,align=right] at (2.65,-7.4)
  {Level 3: \lv{R633}\\ $\mm{6\times3\times3:40}$\\ $40$ products of $54$};
\matbox{4.6}{-6.6}{0.8}{1.6}{6}{3}{$T_{ij}$}{$6\times3$ grid\\ $384^2$ blocks}
\node at (6.25,-7.4) {$\times$};
\matbox{7.9}{-6.6}{1.6}{0.8}{3}{3}{$S_{ij}$}{$3\times3$ grid\\ $384\times768$ blocks}
\node at (9.55,-7.4) {$=$};
\matbox{11.2}{-6.6}{1.6}{1.6}{6}{3}{$M_{ij}=T_{ij}S_{ij}$}{$6\times3$ grid\\ $384\times768$ blocks}
\node[anchor=west,align=left] at (13.0,-7.4) {$7\cdot40=280$ sub-problems\\ $(2304\times1152)(1152\times2304)$};
\draw[down] (2.95,-8.35) -- (2.95,-9.7);
\node[anchor=east,align=right] at (2.8,-9.02) {$L$, $R$ passes:\\ $40$ factors $T_{ijl}$, $S_{ijl}$};
\draw[up] (12.85,-9.7) -- (12.85,-8.35);
\node[anchor=west,align=left] at (13.0,-9.02) {$P$ pass: $M_{ij}$'s $18$ blocks\\ from the $40$ products $M_{ijl}$};
\node[anchor=east,align=right] at (2.65,-10.7)
  {Leaf: \lv{G6}\\ Accelerate's GEMM\\ six streams};
\flatbox{4.6}{-9.9}{0.8}{0.8}{$T_{ijl}$}{$384\times384$\\ flat}
\node at (6.25,-10.7) {$\times$};
\flatbox{7.9}{-9.9}{1.6}{0.8}{$S_{ijl}$}{$384\times768$\\ flat}
\node at (9.55,-10.7) {$=$};
\flatbox{11.2}{-9.9}{1.6}{0.8}{$M_{ijl}=T_{ijl}S_{ijl}$}{$384\times768$\\ flat}
\node[anchor=west,align=left] at (13.0,-10.7)
  {$280\cdot40=11{,}200$ GEMMs\\ ($8\cdot54\cdot54=23{,}328$ classical)\\ 85--95\% of the time};
\end{tikzpicture}
\caption{The plan \lv{SW/R363+3/R633/G6} at $n=13824$ in f64.  ``$7$
products of $8$'' and ``$40$ products of $54$'' count a level's block
products against the classical algorithm's on the same grids.  Going down,
each level's $L$ and $R$ passes write the factors of the level below
(linear combinations of its blocks); going up, its $P$ pass composes their
products into the blocks of its own product (Figure~\ref{fig:lrp}).  Colours as in Figure~\ref{fig:read}.}
\label{fig:plan}
\end{figure}
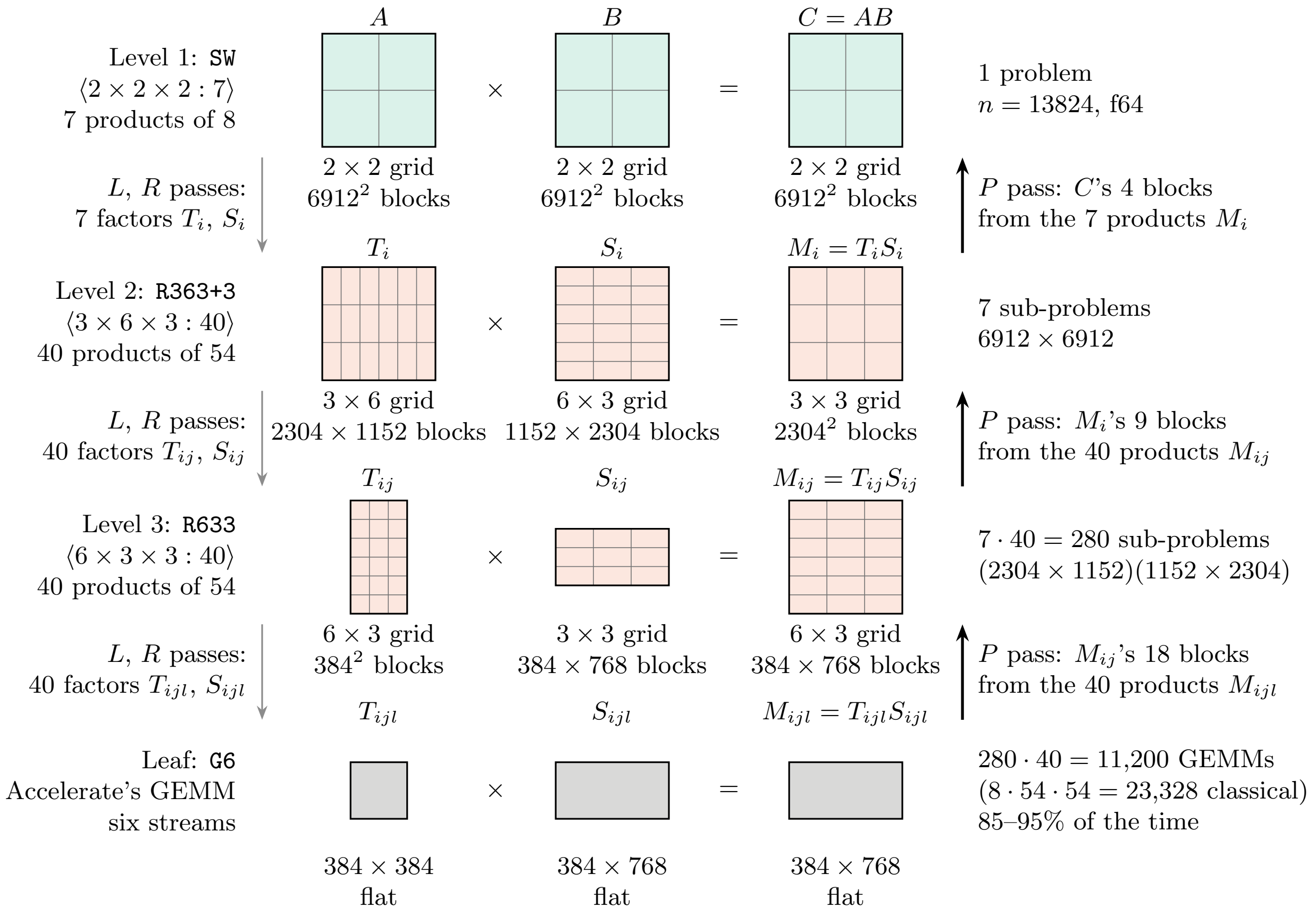

\paragraph{Fused passes.}
A level runs three linear programs (Figure~\ref{fig:lrp}): $L$ forms the left
factors from the blocks of $A$, $R$ the right factors from the blocks of $B$,
and $P$ the blocks of $C$ from the products.  Each program is a sequence of
block additions and scalings.  Done one operation at a time, as in the conventional schedules of Douglas et al.\ \cite{DHSS94} and Boyer et al.\ \cite{BDPZ09} (which our implementation of ESSL's Strassen--Winograd algorithm follows, Section~\ref{sec:essl}; ESSL's own code is closed), every addition is its own sweep over whole blocks,
reading its operands from main memory and writing its result back.  We
\emph{fuse} the operations of a program into a single \emph{pass}
(Figure~\ref{fig:fused}): one loop
over the element positions of a block that, at each position, loads that
element of every input block, evaluates the whole program on those scalars
in registers, and stores that element of every output block.
(Huang et al.\ \cite{HSvdG16} avoid the separate sweeps another way, by folding a level's additions into the operand packing of an open GEMM; Section~\ref{sec:related}.)

Two facts about the costs organize everything.  First, a fused pass is bound
by memory traffic -- the words it moves between main memory and the
processor -- not by arithmetic.  A pass reads every input block once and
writes every output block once, however many operations its program has;
the blocks stream from main memory, being far
larger than the caches (at $n=13824$ one block of a $4\times4$ level in f64
is $95$~MB), and the arithmetic in registers is nearly free.  So programs of
$201$, $284$ and $315$ operations for the $\mm{4\times4\times4:48}$ scheme
all run at the same speed as fused passes (Section~\ref{sec:passspeed}): the
operation counts that scheme searches minimize have little value here, while
the error growth of the program matters \cite{S26b}.  Arithmetic shows only when a program is wasteful enough to make a pass compute-bound.  Fusing matters most for the large schemes: against the $15$ additions of Strassen--Winograd it gains only a few percent (Section~\ref{sec:essl}), but one at a time the $262$ operations of a $4\times4$ level would move four times the blocks of its three fused passes.  Second, the leaf GEMM dominates
the time (85--95\% of a plan), so a level pays only when its products are
large enough for the GEMM to run near its peak, and a scheme that saves
products beats one that saves additions.  The planner trades the two costs:
the products a level saves against the words its passes move, both priced
from measurements, choosing the levels, their order and depth, and how the
passes overlap the products below them.  (The GEMM's own movement of data
into the matrix units is Accelerate's.)  The ternary $\mm{3\times3\times6:42}$ scheme of Section~\ref{sec:ternary},
with coefficients $-1$, $0$ and $1$ only, is the controlled experiment: a third of
the pass arithmetic, two products more, and it loses.

\paragraph{Why mixing pays.}
There are three reasons, which the measurements separate:

\begin{enumerate}
\item \emph{Order.}  The pass of a $g\times g$ level over an input matrix
  reads its $n^2$ elements and writes one $(n/g)^2$-element block for each of
  the scheme's $r$ products, $r/g^2$ times as much: $48/16=3$ for the
  $\mm{4\times4\times4:48}$ scheme, $7/4=1.75$ for a $\mm{2\times2\times2:7}$
  scheme.  The larger expansion belongs at the
  lower level, nearer the leaf.  Per input matrix, \lv{SW/D48/G} moves
  $(1+1.75)+(1.75+5.25)=9.75\,n^2$ words (matrix elements) and
  \lv{D48/SW/G} moves $(1+3)+(3+5.25)=12.25\,n^2$, with the same $336$
  products.
\item \emph{Shape.}  A $\mm{3\times3\times6:40}$ level hands down rectangular
  products.  It keeps $40/54=0.741$ of the $54$ classical products against
  $\mm{4\times4\times4:48}$'s $48/64=0.75$ of $64$, and it has three further
  advantages.
  \begin{itemize}
  \item \emph{Its passes move fewer words.}  On a square problem the pass
    over $A$ reads $n^2$ and writes $40$ blocks of $n^2/9$, the pass over
    $B$ reads $n^2$ and writes $40$ of $n^2/18$, and the pass that assembles
    $C$ reads those $40$ products and writes $n^2$:
    $(1+40/9)+(1+40/18)+(40/18+1)=11.89\,n^2$ against
    $\mm{4\times4\times4:48}$'s $3\times(1+3)=12\,n^2$.
  \item \emph{Its passes keep fewer blocks in flight at once}
    (\emph{streams}, Section~\ref{sec:levelcost}): at most $18+40=58$, in the
    passes over $B$ and $C$, against $16+48=64$.
  \item \emph{Its products are larger.}  After two levels at $n=13824$ they
    are $768\times768$ by $768\times1536$ against $864\times864$ by
    $864\times864$ after two $4\times4$ levels, $1.4\times$ the work each,
    which Accelerate runs at a 13\% higher rate.
  \end{itemize}
\item \emph{Accuracy.}  Each level multiplies the leaf error by a factor set by
  the scheme and the point chosen in its orbit, the family of equivalent schemes defined above.  The measured \emph{level
  factor} (Section~\ref{sec:accuracy}) of the $\mm{3\times3\times6:40}$
  scheme is $0.72$--$1.00$ of the $\mm{4\times4\times4:48}$ scheme's in f32
  and $0.71$--$0.90$ in f64, depending on
  the orientation (\lv{R363} the lowest), and
  both are below one, because their products are shorter dot
  products than the GEMM's.  Under an error budget an accurate level admits
  deeper, faster plans.
\end{enumerate}

\needspace{9\baselineskip}
\paragraph{Contributions.}
The report contributes the following.
\begin{enumerate}
\item A cost model of fused levels in blocks moved and GEMMs priced inside
  the level that sits on them, calibrated per machine and precision, with
  predictions within 0.3--3\% at $n\ge 8192$ (Section~\ref{sec:levels}).
\item A screen of 65 published schemes and six Kronecker products of them, by saving per block moved and by leaf
  error amplification per $4\times4$-equivalent saving, which leaves the
  accurate $\mm{3\times3\times6:40}$ scheme, in its three orientations, as
  the one addition to the $4\times4$ menu and rules out every larger shape
  and every Kronecker composite (Section~\ref{sec:menu}).
\item An error model that knows the order of the levels: the leaf's error
  amplified by every level, plus each level's own rounding amplified by the
  levels above it.  Calibrated at one size, it predicts the errors of plans
  one to three levels deep at $3456$--$13824$ within $0.8$--$1.25$ and the
  ratio between two plans to about $10\%$.  Also the orbit floors of
  the amplification for the two accurate classes, and a study of badly
  scaled inputs with exact prescaling (Section~\ref{sec:accuracy}).
\item A planner over $(m,k,n)$ shapes with GEMM streams, pipelines, padding
  to multiples of $2^i3^j$, an error budget, and shape estimates that price
  uncalibrated rectangular problems within 3.5\% (Section~\ref{sec:planner}).
\item Measurements on the M4~Pro (Section~\ref{sec:results}): the headline
  results above; against a conventional implementation of ESSL's Strassen--Winograd algorithm at its fastest depth, plans $1.21$--$1.54\times$ faster, where the algorithm has
  $1.9$--$9.7$ times the plans' error; rectangular problems at
  $1.09$--$1.53\times$ over the GEMM; and the GPU as a second engine for the
  f32 products, $3.1\times$ over Accelerate at $13824$ with the error the
  same plan has on the CPU.
\item The ternary trial: a scheme with a third of the pass arithmetic and
  two more products, which is slower and less accurate
  (Section~\ref{sec:ternary}).
\item Two implementations, C and Rust, that agree in plans, predicted times
  and results bit for bit, with the GPU as without it
  (Appendix~\ref{sec:impl}).
\item An assessment of the commercial value, in software and as a GPU or
  TPU feature (Appendix~\ref{sec:value}).
\end{enumerate}
Code, programs, calibrations and raw results are in the repository
\cite{logic}; Appendix~\ref{sec:data} lists the result files, cited in the
text as F1, F2 and so on.

\section{Related work}\label{sec:related}

Strassen-type levels over a tuned GEMM are old practice.  IBM's ESSL has long
shipped \texttt{\_GEMMS}, Winograd's variant of Strassen's algorithm in up to
four levels over its GEMM, as a routine beside \texttt{\_GEMM} with a warning
that it can be inaccurate for inputs of divergent scale \cite{ESSL}
(Section~\ref{sec:essl} measures its algorithm against the plans), and
Huss-Lederman et al.\ gave a portable \texttt{DGEFMM} over any BLAS
\cite{HLJJTT96}.  Benson and Ballard \cite{BB15} generate code for many
schemes, rectangular ones included, over Intel's MKL, with several levels of
recursion, and find that the best scheme depends on the shape of the problem
as well as its size; Ballard et al.\ \cite{BBDLS16} analyse the accuracy of
the same algorithms and improve it by diagonal scaling and by the choice of
algorithm.  Huang et al.\ fold a level's additions into the operand packing
of BLIS \cite{HSvdG16} and generate families of one- and two-level
algorithms, the latter as Kronecker products, with a performance model
\cite{HRMvdG17}, and carry the design to GPUs \cite{HYvdG20}.  What this
report adds: DPS's accurate points as the menu, so that the plans are more
accurate than the GEMM rather than less; fused passes priced by the blocks
they move, with each output cache line zeroed before it is written by Arm's
\texttt{DC~ZVA} instruction (Data Cache Zero by Virtual Address), so that it
is not first read from memory; a planner
calibrated per machine that mixes schemes per level under an error budget;
and the GPU as a second engine inside a level.

\section{Setting, measures and notation}\label{sec:setup}

\begin{figure}[!tbp]
\centering
\resizebox{\textwidth}{!}{%
\begin{tikzpicture}[x=1cm,y=1cm,font=\scriptsize,
  core/.style={draw=black!70,fill=white,line width=0.5pt,inner sep=0pt,minimum height=0.6cm,font=\footnotesize},
  lone/.style={draw=black!55,fill=black!7,line width=0.4pt,inner sep=0pt,minimum height=0.4cm},
  ltwo/.style={draw=black!60,fill=black!13,line width=0.5pt,inner sep=0pt,minimum height=0.6cm,font=\footnotesize},
  sme/.style={draw=cG,fill=cG!18,line width=0.6pt,inner sep=1pt,align=center},
  frame/.style={draw=black!45,fill=black!2,rounded corners=3pt,line width=0.5pt},
  ln/.style={black!65,line width=0.5pt},
  bus/.style={black!70,line width=1.6pt},
  lab/.style={anchor=east,align=right,inner sep=1pt},
  lead/.style={black!45,line width=0.3pt,densely dotted}]
\newcommand{\pcluster}[2]{%
  \draw[frame] (#1,5.85) rectangle ++(5.1,4.9);
  \node[font=\footnotesize\bfseries] at ({#1+2.55},10.47) {Performance cluster #2};
  \node[align=center] at ({#1+2.55},9.68) {$5$ cores; NEON on each:\\ $0.10$--$0.11$ (f32), $0.05$--$0.06$ (f64)};
  \foreach \i in {0,...,4} {
    \node[core,minimum width=0.5cm] at ({#1+0.4+0.57*\i},8.7) {P};
    \node[lone,minimum width=0.5cm] at ({#1+0.4+0.57*\i},7.5) {L1};
    \draw[ln] ({#1+0.4+0.57*\i},8.4) -- ++(0,-0.7);
    \draw[ln] ({#1+0.4+0.57*\i},7.3) -- ++(0,-0.7);
  }
  \node[sme,text width=1.82cm,inner sep=0pt,minimum height=1.7cm] at ({#1+4.04},8.15) {\textbf{SME unit}\\ $1.6$--$2.1$ f32\\ $0.38$--$0.47$ f64};
  \draw[ln] ({#1+2.93},8.7) -- ({#1+3.13},8.7);
  \draw[ln] ({#1+4.04},7.3) -- ++(0,-0.7);
  \node[ltwo,minimum width=4.8cm] at ({#1+2.55},6.3) {L2 cache, $16$~MB};
  \draw[bus] ({#1+2.55},6.0) -- ++(0,-1.0);
}
\pcluster{2.6}{1}
\pcluster{7.9}{2}
\draw[frame] (13.2,5.85) rectangle ++(3.8,4.9);
\node[font=\footnotesize\bfseries] at (15.1,10.47) {Efficiency cluster};
\node[align=center] at (15.1,9.68) {$4$ cores; NEON on each:\\ about $0.03$ (f32), $0.02$ (f64)};
\foreach \i in {0,...,3} {
  \node[core,minimum width=0.36cm] at ({13.53+0.42*\i},8.7) {E};
  \node[lone,minimum width=0.36cm,font=\tiny] at ({13.53+0.42*\i},7.5) {L1};
  \draw[ln] ({13.53+0.42*\i},8.4) -- ++(0,-0.7);
  \draw[ln] ({13.53+0.42*\i},7.3) -- ++(0,-0.7);
}
\node[sme,fill=cG!8,text width=1.75cm,inner sep=0pt,minimum height=1.7cm] at (15.975,8.15) {\textbf{SME unit}\\ $0.13$--$0.14$ f32\\ $0.03$--$0.04$ f64};
\draw[ln] (14.97,8.7) -- (15.1,8.7);
\draw[ln] (15.975,7.3) -- ++(0,-0.7);
\node[ltwo,minimum width=3.5cm] at (15.1,6.3) {L2 cache, $4$~MB};
\draw[bus] (15.1,6.0) -- ++(0,-1.0);
\node[draw=black!70,fill=black!20,line width=0.6pt,minimum width=14.4cm,minimum height=1.0cm,align=center,font=\footnotesize]
  (mem) at (9.8,4.5) {\textbf{Unified memory}, $64$~GB: one pool for the processor's clusters and the GPU\\
  {\scriptsize the library's fused passes move $120$--$190$~GB/s through it; Apple's figure for the chip is $273$}};
\node[draw=cM,fill=cM!18,line width=0.6pt,minimum width=8.6cm,minimum height=1.0cm,align=center,font=\footnotesize]
  (gpu) at (9.8,2.75) {\textbf{GPU}, $20$ cores, f32 only: $4.0$--$6.6$ alone,\\ $7.3$--$9.1$ sharing products with the SME units};
\draw[bus] (9.8,4.0) -- (9.8,3.25);
\node[anchor=west,inner sep=3pt] at (9.85,3.63) {shared buffers, no copy};
\node[lab] at (2.35,9.83) {\textbf{rates}: TFLOP/s\\ of a product,\\ $2mkn$ counted};
\node[lab,text=black!60] at (2.35,8.7) {\emph{cores}};
\node[lab] (l1) at (2.35,8.05) {$180$--$240$~GB/s\\ per core};
\node[lab,text=black!60] at (2.35,7.5) {\emph{L1 caches}};
\node[lab] (l2) at (2.35,6.95) {$100$--$175$~GB/s\\ per core};
\node[lab] (l3) at (2.35,5.5) {$120$~GB/s (one\\ thread), $210$--$230$\\ (all the cores)};
\draw[lead] (l1.east) -- (3.0,8.05);
\draw[lead] (l2.east) -- (3.0,6.95);
\draw[lead] (l3.east) -- (5.0,5.5);
\end{tikzpicture}}
\caption{The machine, schematically: what the operating system reports
(cores, cache sizes) with what we measured on it.  Each core has its own
level-1 cache ($128$~KB of data on a performance core, $64$~KB on an
efficiency core); the cores of a cluster share a level-2 cache and one SME
matrix unit; everything reaches the one memory through the level-2 caches,
and the GPU reads and writes that same memory.  Processing rates are in
TFLOP/s of a matrix product, the same two kinds in every cluster: per core
for our NEON GEMM on the cores' own vector units, and per SME unit for
Accelerate's GEMM (on a performance cluster from one thread up to three
streams a unit; the two performance clusters' units together give Accelerate's
$3.2$--$3.3$ in f32 and $0.81$--$0.85$ in f64, and $3.3$--$4.2$ and
$0.86$--$0.94$ under the library's streams); the GPU's rate is that of Apple's GEMM on it.
The efficiency cluster is the slower one on both counts.  Its numbers are
indirect: the NEON rate is what four more threads add to ten, and the SME
rate is that of one GEMM stream held on those cores at background
priority, four times what one of their NEON units gives, which is how we know the
cluster has a matrix unit of its own.  The
bandwidths on the left are those of a plain pass $d=a+b$ on the performance
cores, by where its three arrays live (in a core's level-1 cache, in the
level-2 cache, in memory), counting the bytes read and written; Apple's
figure is from \cite{AppleM4Pro}
(files~\rf{cacheprobe}, \rf{probes}, \rf{machprobes}, \rf{gpushare}).}
\label{fig:machine}
\end{figure}
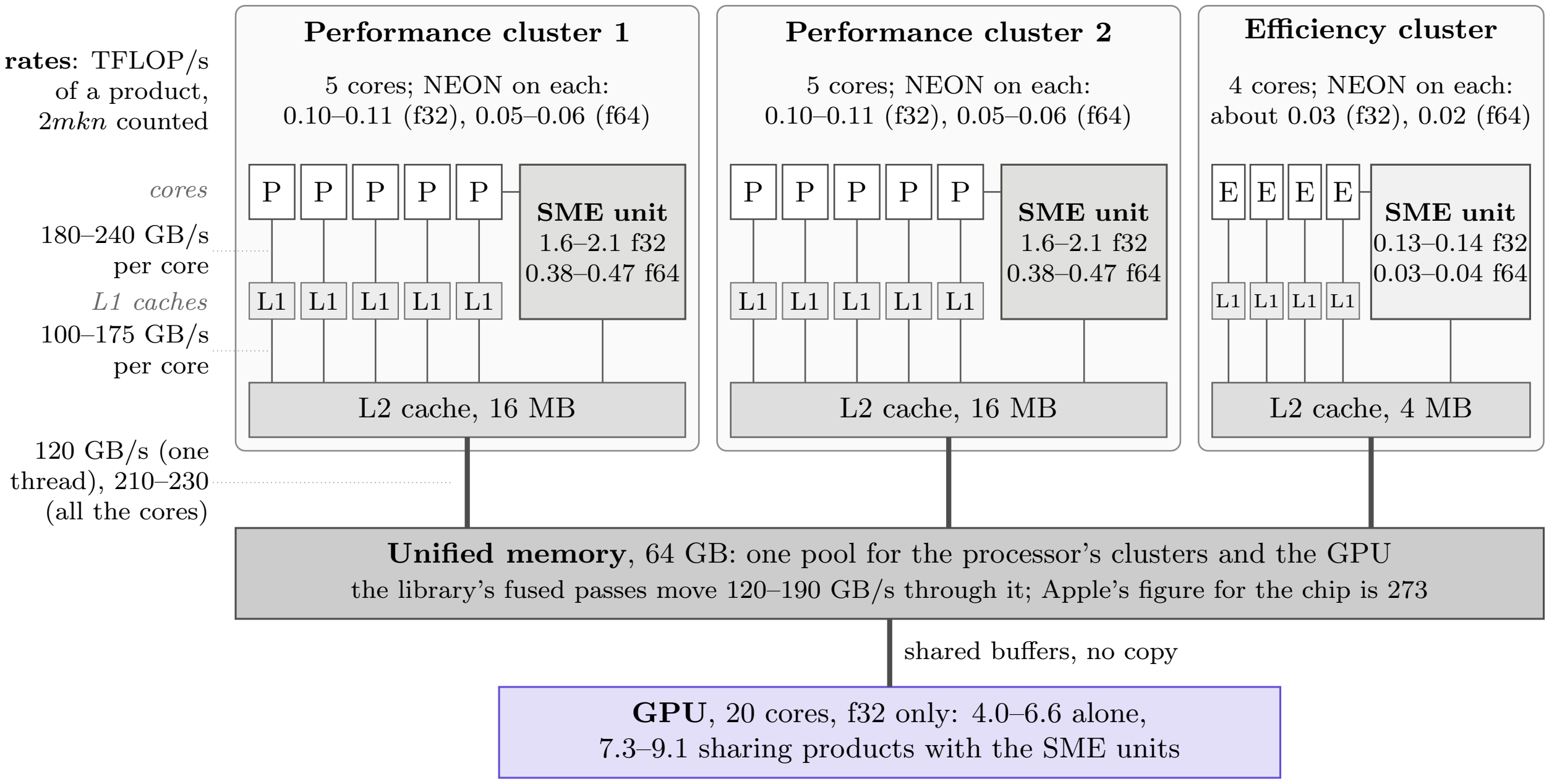

\paragraph{Setting.}
Row-major $A\in\R^{m\times k}$, $B\in\R^{k\times n}$, $C=AB$, in f32 or f64.
The machine is an Apple M4~Pro running macOS~26, drawn in
Figure~\ref{fig:machine}: two performance clusters of five cores and one
efficiency cluster of four, each cluster around its own level-2 cache and
with its own SME matrix unit, a 20-core GPU, and 64~GB of memory that all of
them share.  Accelerate's GEMM runs on the two performance clusters' SME
units, at about $3.3$~TFLOP/s in f32 and
$0.85$ in f64 on large square problems.  Times are the best of three to
seven interleaved rounds (each round runs every method once; medians are
within 3\% of the best unless stated), all from one campaign on the library
as it stands (Appendix~\ref{sec:data}).  Accelerate's own time varies between
runs more than the plans' do -- its f64 GEMM at $13824$ took $6.14$ to
$6.43$~s within the campaign, its f32 GEMM $1.55$ to $1.84$~s -- so every speedup is taken against
Accelerate's best time in the whole campaign, which understates the gains.

\paragraph{Measures.}
The \emph{error} of a product is the largest entry of
$C - C_{\mathrm{exact}}$ divided by the largest entry of
$C_{\mathrm{exact}}$, where $C_{\mathrm{exact}}$ is the f64 product of the
same inputs in f32 and a compensated double-double product in f64, on
uniform random inputs in $[-1,1)$; every method in a comparison sees the
same inputs.  (The result files print it as \texttt{err64}.)  Errors are
compared as ratios: a plan has ``$0.5$ of the GEMM's error'' when its error
is half the GEMM's on the same inputs.  The difference from Accelerate's own
result (\texttt{relerr} in the result files) is not an error measure, because
Accelerate's error is as large as a plan's; Section~\ref{sec:results} gives
an example.  Section~\ref{sec:scaled} adds a second measure, the
componentwise error, for badly scaled inputs.

\paragraph{Notation.}
Table~\ref{tab:notation} lists the codes plans are written in, from the top
of the problem down to the leaves, the levels separated by slashes.  So
\lv{R363+4/D48/G6} at $n=13824$ is a $3\times6\times3$ level with four pass
threads, whose $40$ products of $4608\times2304$ by $2304\times4608$ each get
a $4\times4$ level, whose $40\times48 = 1920$ leaf products of $1152\times576$
by $576\times1152$ go to Accelerate in six streams; the classical count would
be $54\times64 = 3456$.  Codes in typewriter type, such as \lv{R363}, name
levels and leaves; the italic letters $L$, $R$ and $P$ are the three
programs of a level (Section~\ref{sec:levels}).

\begin{table}[!tbp]
\centering\footnotesize
\begin{tabular}{@{}l p{0.57\textwidth} >{\raggedright\arraybackslash}p{0.17\textwidth}@{}}
\toprule
code & meaning & source\\
\midrule
\lv{a/b/c} & a plan: level \lv{a} on the whole problem, level \lv{b} on each of
  \lv{a}'s block products, and so on down to the leaf \lv{c} & this report\\
\addlinespace
\lv{SW} & a $2\times2$ level, $\mm{2\times2\times2:7}$: $A$, $B$, $C$ cut into
  $2\times2$ grids of blocks, $7$ block products instead of $8$; Strassen's
  scheme at the accurate point of Dumas, Pernet and Sedoglavic with
  coefficients in $\mathbb{Q}(\sqrt3)$, rounded to double (Section~\ref{sec:sw})
  & \cite{Strassen69}; point \cite[Eq.~(36)]{DPS24}\\
\lv{D48} & a $4\times4$ level of the accurate rational
  $\mm{4\times4\times4:48}$ scheme of Dumas, Pernet and Sedoglavic in its
  $262$-operation form: $4\times4$ grids, $48$ products instead of $64$
  (\lv{D} for the authors, $48$ for the rank)
  & \cite{DPS25,DPS26a}; program \cite{PLinOpt,S26b}\\
\lv{R336}, \lv{R363}, \lv{R633} & a level of the accurate
  $\mm{3\times3\times6:40}$ scheme of Dumas, Pernet and Sedoglavic, a point
  in the class of Smirnov's scheme (Tichavsk\'y's Decomposition~A), in one
  of its three orientations, named by
  the grids $g_1\times g_2\times g_3$ it cuts $A$ ($g_1\times g_2$) and $B$
  ($g_2\times g_3$) into; $40$ products instead of $54$ (\lv{R} for
  rectangular) & \cite{Smirnov13}; point \cite{Tichavsky21,DPS26b,PLinOpt}\\
\lv{T336}, \lv{T363}, \lv{T633} & the same three orientations of a ternary
  $\mm{3\times3\times6:42}$ scheme, coefficients $-1$, $0$, $1$, $42$ products
  instead of $54$ (\lv{T} for ternary; Section~\ref{sec:ternary})
  & \cite{Perminov,Perminov25}\\
\lv{X336}, \lv{X363}, \lv{X633} & in Table~\ref{tab:ternary} only: \lv{R} or
  \lv{T}, the same plan with the accurate or with the ternary levels
  & this report\\
\addlinespace
\lv{+p} & after a level, as in \lv{R363+3}: the level's passes run on $p$
  threads overlapped with the products below (Section~\ref{sec:machine})
  & this report \cite{logic}\\
\addlinespace
\lv{G} & the leaf: Accelerate's GEMM on the block products that remain, one
  call at a time with Accelerate's own threading & \cite{Accelerate}\\
\lv{G3}, \lv{G4}, \lv{G6} & the leaf as three, four or six concurrent
  single-threaded GEMM calls taking the block products in turn
  (Section~\ref{sec:machine}) & this report \cite{logic}\\
\lv{GM}, \lv{G6M} & the leaf \lv{G} or \lv{G6} with the GPU as a second
  engine (f32, GPU build; Section~\ref{sec:machine}) & this report \cite{logic}\\
\lv{K16}, \lv{K32}, \lv{K64} & no scheme: the classical product in $16$,
  $32$ or $64$ strips of rows for a shared leaf, as in \lv{K32/G6M} & this report \cite{logic}\\
\lv{B2} & the classical $2\times2$ blocked product as a level, $8$ products,
  the calibration's reference & this report \cite{logic}\\
\addlinespace
\lv{auto:}$b$ & the benchmark's planner with error budget $b$; \lv{auto:none}
  for no budget (Section~\ref{sec:planner}) & this report \cite{logic}\\
\bottomrule
\end{tabular}
\caption{The notation for plans, with the source of each scheme and of each
piece of the library.  Example: \lv{SW/R363+3/R633/G6}, read in
Figure~\ref{fig:read} and drawn in Figure~\ref{fig:plan}.}
\label{tab:notation}
\end{table}

\section{Fused levels and what they cost}\label{sec:levels}

\subsection{Schemes as linear programs}

\begin{figure}[!tbp]
\centering
\begin{tikzpicture}[x=1cm,y=1cm,scale=0.95,
  every node/.style={font=\footnotesize},
  box/.style={draw,rounded corners=3pt,inner sep=6pt,align=left},
  flow/.style={-{Stealth[length=6pt]},line width=0.8pt},
  lab/.style={font=\footnotesize,inner sep=2pt,fill=white}]
\node (A) at (-4.6,3.6) {$A=\begin{pmatrix}a_{11}&a_{12}\\a_{21}&a_{22}\end{pmatrix}$};
\node (B) at ( 4.6,3.6) {$B=\begin{pmatrix}b_{11}&b_{12}\\b_{21}&b_{22}\end{pmatrix}$};
\node[box] (L) at (-4.6,1.0) {$L$ (Left): $t = L\,\vect A$\\[3pt]
  \begin{tabular}{@{}l@{\qquad}l@{}}
  $r_1 = \tfrac12 a_{22}$ & $t_3 = a_{21}-r_1$\\
  $t_4 = a_{12}-r_1$ & $t_1 = t_3-t_4$\\
  $t_5 = a_{21}+r_1$ & $r_2 = t_3-a_{12}$\\
  $t_6 = a_{11}+\tfrac12 r_2$ & $t_2 = t_6-t_1$\\
  $t_7 = t_5-t_1$ &
  \end{tabular}};
\node[box] (R) at (4.6,1.0) {$R$ (Right): $s = R\,\vect B$\\[3pt]
  \begin{tabular}{@{}l@{\qquad}l@{}}
  $u_1 = \tfrac12 b_{12}$ & $s_2 = b_{11}+u_1$\\
  $s_3 = u_1-b_{22}$ & $u_2 = s_2-b_{22}$\\
  $s_5 = b_{22}+u_1$ & $s_1 = s_2-s_5$\\
  $s_4 = \tfrac12 u_2-b_{21}$ & $s_6 = s_1-s_3$\\
  $s_7 = s_1-s_4$ &
  \end{tabular}};
\node[box,minimum width=10.8cm] (M) at (0,-2.4) {The products, $p_i = t_i\,s_i$: the only multiplications\\[3pt]
  \begin{tabular}{@{}l@{\qquad}l@{\qquad}l@{\qquad}l@{}}
  $p_1 = t_1\,s_1$ & $p_2 = t_2\,s_2$ & $p_3 = t_3\,s_3$ & $p_4 = t_4\,s_4$\\
  $p_5 = t_5\,s_5$ & $p_6 = t_6\,s_6$ & $p_7 = t_7\,s_7$ &
  \end{tabular}};
\node[box,minimum width=10.8cm] (P) at (0,-5.0) {$P$ (Product): $\vect C = P\,p$\\[3pt]
  \begin{tabular}{@{}l@{\quad}l@{\quad}l@{\quad}l@{}}
  $c_{22} = p_5+p_3$ & $v_1 = p_1-p_6-p_3$ & $v_2 = p_7+p_6$ & $v_3 = p_4+v_1$\\
  $v_4 = \tfrac12 c_{22}$ & $c_{12} = p_2+v_1+v_4$ & $c_{21} = v_2+v_3+v_4$ & $c_{11} = \tfrac12(c_{12}+v_2-v_3)$
  \end{tabular}};
\node (C) at (0,-7.1) {$C=AB=\begin{pmatrix}c_{11}&c_{12}\\c_{21}&c_{22}\end{pmatrix}$};
\draw[flow] (A) -- (L) node[lab,midway,right] {the 4 blocks of $A$};
\draw[flow] (B) -- (R) node[lab,midway,left] {the 4 blocks of $B$};
\draw[flow] (L.south) -- (L.south |- M.north) node[lab,midway,right] {$t_1,\dots,t_7$: the 7 left factors};
\draw[flow] (R.south) -- (R.south |- M.north) node[lab,midway,left] {$s_1,\dots,s_7$: the 7 right factors};
\draw[flow] (M) -- (P) node[lab,midway,right] {$p_1,\dots,p_7$: the 7 products};
\draw[flow] (P) -- (C) node[lab,midway,right] {the 4 blocks of $C$};
\end{tikzpicture}
\caption{The three programs of a level, for the $\mm{2\times2\times2:7}$ scheme at
DPS's dyadic accurate point (the library's $2\times2$ level until
2026-11-22; it now runs their $\mathbb{Q}(\sqrt3)$ point, whose programs
have the same form with irrational constants, Section~\ref{sec:sw}): the straight-line
program of Table~4 of \cite{DPS24}, one box per program, each run on blocks
as one fused pass.  The names are Table~4's ($r_1$, $r_2$ are scratch values of the left program),
except that the factors are numbered from $1$, so that $p_i = t_i\,s_i$
(Table~4's $t_0,\dots,t_5$ and $s_0,\dots,s_5$ are our $t_1,\dots,t_6$ and
$s_1,\dots,s_6$), and the factors of $p_7$, inline there, are $t_7$ and
$s_7$; and one sign is corrected: the table prints $t_3 =
a_{12}+r_1$ (our $t_4$), which returns $c_{21}$ and $c_{22}$ but not $c_{11}$ and
$c_{12}$; with $t_4 = a_{12}-r_1$, the fourth row $(0,1,0,-\frac12)$ of the
left matrix of their Eq.~(38), the program computes $AB$ exactly, its three
matrices are Eq.~(38) row for row, and it is, term for term, PLinOpt's
\texttt{2x2x2\_7\_DPS-evenpow-12.2034}, whose matrices the library's programs
compute ($27$ additions and $6$ halvings here; $27$ and $9$ scalings in the library's cancellation-free programs).  On blocks, $t_i$, $s_i$ and $p_i$ are the factors $T_i$, $S_i$ and the products $M_i$ of Figure~\ref{fig:plan} and of the text.}
\label{fig:lrp}
\end{figure}
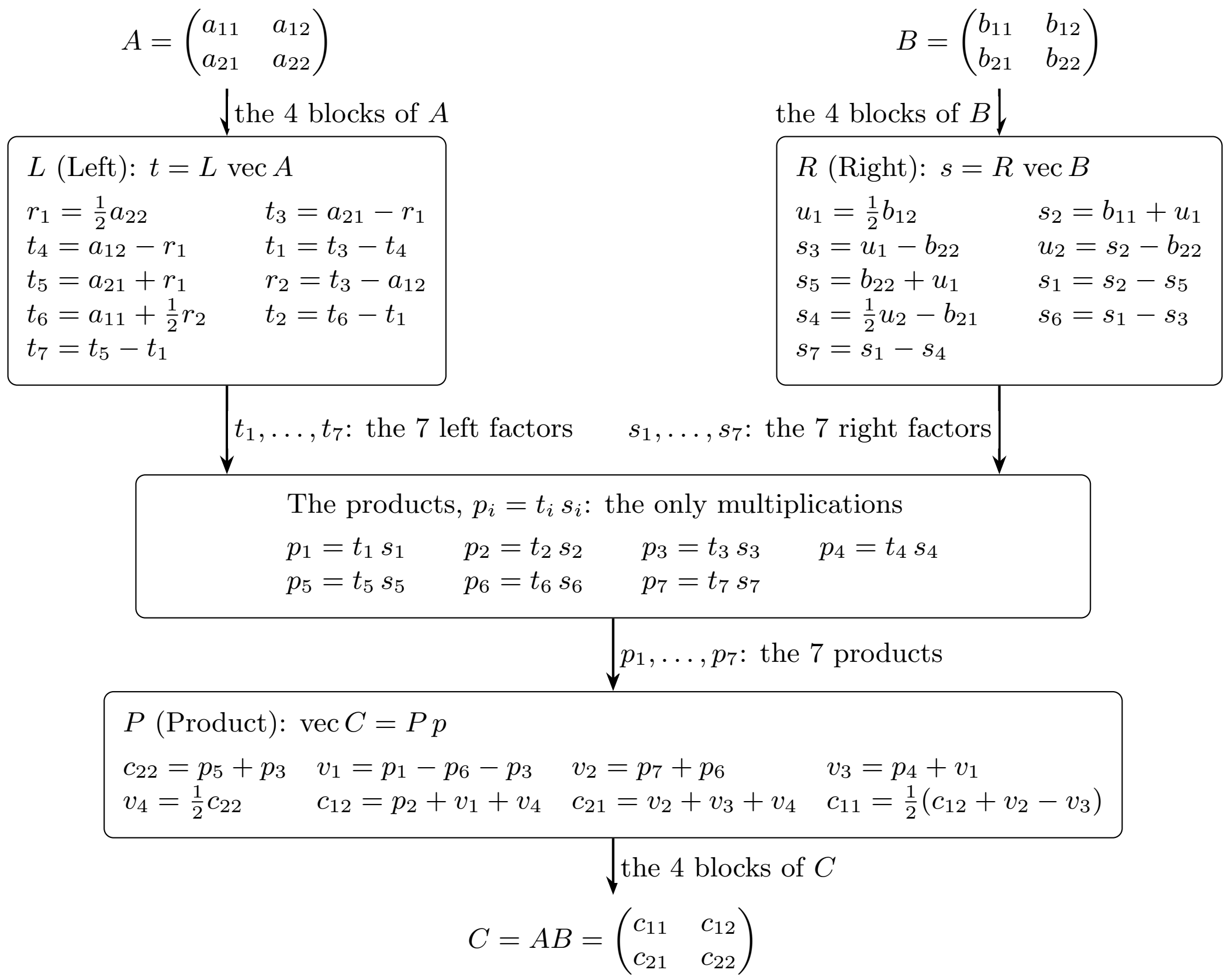

A scheme $\mm{g_1\times g_2\times g_3:r}$ is three matrices $L\in\R^{r\times
g_1g_2}$, $R\in\R^{r\times g_2g_3}$, $P\in\R^{g_1g_3\times r}$ with
\begin{equation}\label{eq:scheme}
\vect(AB) \;=\; P\bigl((L\,\vect A)\circ(R\,\vect B)\bigr)
\qquad\text{for all } A\in\R^{g_1\times g_2},\ B\in\R^{g_2\times g_3},
\end{equation}
$\circ$ the entrywise product and $\vect$ row-major.  Applied to blocks, the
same identity holds with $A$'s $g_1g_2$ blocks of size $\frac{m}{g_1}\times
\frac{k}{g_2}$, $B$'s $g_2g_3$ blocks of $\frac{k}{g_2}\times\frac{n}{g_3}$,
and $C$'s $g_1g_3$ blocks of $\frac{m}{g_1}\times\frac{n}{g_3}$: the $r$
products are $(\frac{m}{g_1}\times\frac{k}{g_2})(\frac{k}{g_2}\times
\frac{n}{g_3})$ matrix products, computed by the plan below.  A \emph{level} is
the application of \eqref{eq:scheme} once; the matrices are realized as
straight-line programs (PLinOpt's \texttt{.slp} syntax \cite{PLinOpt}), one
per side, and the three programs are what a level executes
(Figure~\ref{fig:lrp} writes them out for the $\mm{2\times2\times2:7}$ scheme):
\begin{itemize}
\item the $L$ pass reads the $g_1g_2$ blocks of $A$ and writes the $r$ left
  factors $T_i$;
\item the $R$ pass reads the $g_2g_3$ blocks of $B$ and writes the $r$ right
  factors $S_i$;
\item the products $M_i=T_iS_i$ run (GEMMs, or a level below);
\item the $P$ pass reads the $r$ products and writes the $g_1g_3$ blocks of $C$.
\end{itemize}

Each program runs as one fused pass (Section~\ref{sec:intro},
Figure~\ref{fig:fused}), so the program's operation count costs only
arithmetic, which the memory system hides.

\subsection{The cost of a level}\label{sec:levelcost}

A \emph{word} is one matrix element, an f32 or an f64; write $w$ for its
size in bytes ($4$ or $8$) and $\mathcal{B}$ for the streaming bandwidth the
passes reach, in bytes per second.  A level on an $(m\times k)(k\times n)$ problem
moves
\begin{equation}\label{eq:traffic}
W(m,k,n) \;=\; (g_1g_2+r)\,\frac{mk}{g_1g_2} \;+\; (g_2g_3+r)\,\frac{kn}{g_2g_3}
\;+\; (g_1g_3+r)\,\frac{mn}{g_1g_3}
\end{equation}
words through its three passes (each pass reads its input blocks once and
writes its output blocks once), that is $W(n,n,n) = n^2\bigl(3 +
r(\tfrac{1}{g_1g_2} + \tfrac{1}{g_2g_3} + \tfrac{1}{g_1g_3})\bigr)$ on a
square problem, and takes
\begin{equation}\label{eq:level}
\text{time}(m,k,n) \;=\; r\,\text{time}\Bigl(\frac{m}{g_1},\frac{k}{g_2},
\frac{n}{g_3}\Bigr) \;+\; \frac{w\,W(m,k,n)}{\mathcal{B}}
\end{equation}
seconds, a recurrence that ends at the leaf, where the time is the GEMM's.
Two numbers of a scheme decide the second term: its \emph{traffic},
the words it moves per $n^2$ on a square problem, $W(n,n,n)/n^2 = 3 +
r(\tfrac{1}{g_1g_2} + \tfrac{1}{g_2g_3} + \tfrac{1}{g_1g_3})$, and its
\emph{streams}, the number of blocks a pass reads and writes
at once, $g_1g_2+r$, $g_2g_3+r$ and $g_1g_3+r$.  Beyond about $64$ streams a
pass loses bandwidth: a $\mm{5\times5\times5:93}$ level's $118$-stream passes
ran at half the bandwidth of the $64$-stream passes of the $\mm{4\times4\times4:48}$ scheme.
The first term is
decided by the fraction of the classical algorithm's $g_1g_2g_3$ block products that the level \emph{keeps}, $\kappa=r/(g_1g_2g_3)$ (it \emph{saves} $1-\kappa$: an eighth for a $2\times2$ level, a quarter for the $4\times4$ level), and by the GEMM's rate on the
products' shape and size: the f32 GEMM keeps $3.2$--$3.3$~TFLOP/s down to
$2048$ and drops to $3.0$ at $1024$ and $2.6$ at $512$, where the f64 GEMM
keeps its $0.81$--$0.85$~TFLOP/s down to $1024$, so f64 rewards depth and f32
does not.  Table~\ref{tab:levels} gives $\kappa$, the streams and the traffic of the library's levels.

\begin{table}[!htb]
\centering\footnotesize
\begin{tabular}{llrrrll}
\toprule
level & scheme & kept $\kappa$ & streams $L/R/P$ & traffic ($n^2$) & amp$_{\text{eq}}$ & $\gamma_{\text{eq}}$\\
\midrule
\lv{SW} & Strassen, DPS's $\mathbb{Q}(\sqrt3)$ point (Section~\ref{sec:sw}) & 0.875 & 11/11/11 & 8.25 & 3.855 & 10.53\\
\lv{D48} & DPS accurate $\mm{4\times4\times4:48}$, 262-op.\ point & 0.750 & 64/64/64 & 12.00 & 2.828 & 5.657\\
\lv{R336} & DPS accurate $\mm{3\times3\times6:40}$ & 0.741 & 49/58/58 & 11.89 & 2.280 & 6.162\\
\lv{R363} & the same, orientation $3\times6\times3$ & 0.741 & 58/58/49 & 11.89 & 2.453 & 5.125\\
\lv{R633} & the same, orientation $6\times3\times3$ & 0.741 & 58/49/58 & 11.89 & 2.280 & 6.162\\
\midrule
\lv{T336} & ternary $\mm{3\times3\times6:42}$ (Section~\ref{sec:ternary}) & 0.778 & 51/60/60 & 12.33 & 6.407 & 11.07\\
\lv{T363} & orientation $3\times6\times3$ & 0.778 & 60/60/51 & 12.33 & 6.166 & 9.737\\
\lv{T633} & orientation $6\times3\times3$ & 0.778 & 60/51/60 & 12.33 & 8.061 & 22.84\\
\bottomrule
\end{tabular}
\caption{The levels: the fraction $\kappa=r/(g_1g_2g_3)$ of the classical products a level keeps, its streams per pass, its traffic on a square problem
($n^2$ words), and the two accuracy measures per $4\times4$-equivalent
saving, amp$_{\text{eq}}$ and $\gamma_{\text{eq}}$, which Section~\ref{sec:menu} defines.}
\label{tab:levels}
\end{table}

\subsection{Making a pass run at memory speed}\label{sec:passspeed}

Before the passes zeroed each output cache line ahead of writing it (with
Arm's \texttt{DC~ZVA}, at the end of this subsection), they reached
$120$--$175$~GB/s against the $273$ Apple states for the chip \cite{AppleM4Pro}
and the $210$--$230$ a plain pass reaches (Figure~\ref{fig:machine}).  Two
details were each worth a large factor.  With $64$ streams the compiler gave
up vectorizing the fused loop (it could not rule out overlap among $64$
pointers); a \texttt{vectorize(assume\_safety)} pragma gave $4.3\times$ on the
passes.  And scratch blocks of power-of-two size all start at the same offset
modulo the cache's set span, so element $c$ of all $64$ streams fell into one
cache set; padding each block by $1040$ elements gave another $2.5\times$.
Operation counts did not matter: our $201$-operation program for the
$\mm{4\times4\times4:48}$ scheme \cite{S26}, DPS's $315$-operation original and their
$284$-operation accurate program run at the same speed as fused passes
(totals within 1\%; the $315$-operation passes take 2--8\% longer, about
0.3\% of the total).  The exception is a wasteful program: the direct sums of
the $\mm{3\times3\times6:40}$ scheme's $R$ and $P$ matrices cost $18$ operations per
output, and their passes ran at $0.092$~s against $0.070$ for the $4\times4$
level's at $13824$ in f32 for $0.99\times$ the words.  PLinOpt's optimizer
\cite{PLinOpt}, which shares common subexpressions, shortened the programs
of all three orientations.  In the $3\times3\times6$ orientation, written as
additions${}+{}$scalings, the $L$, $R$ and $P$ programs went from $152$,
$344{+}384$ and $366{+}384$ operations as direct sums ($L$'s coefficients are
all $\pm1$, so it has no scalings) to $59$, $117{+}22$ and $148{+}18$: $39\%$,
$19\%$ and $22\%$ of the operations.  The passes got a quarter faster
(\lv{R336/G} at $13824$: $0.098$ to $0.073$~s in f32, $0.179$ to
$0.136$~s in f64, mostly in the $R$ and $P$ passes), the plans that use these
levels 1--7\% faster, and the error was unchanged
(file~\rf{r336opt}).

\paragraph{The passes paid for every output line twice.}
A plain pass $d=a+b$ moves $208$--$220$~GB/s on the machine with two
threads; the fused passes moved $121$--$130$ (the $4\times4$ level's $L$
pass: $16$ blocks in, $48$ out) to $155$--$174$ (its $P$ pass: $48$ in, $16$
out).  The more a pass writes, the slower it runs, as if every written line
were first read -- and it is: a store to a cache line that is not in the cache
fetches the line from memory before overwriting it, so a pass that reads $16$
blocks and writes $48$ moves $112$ blocks through the memory system, not $64$.
Non-temporal stores changed nothing, and strip-mining through buffers that fit
the first-level cache lost 10--20\%.  What works is Arm's \texttt{DC~ZVA}
instruction, which zeroes a whole $64$-byte line without reading it: issued for
the output lines of a $4$~KB tile just before the kernel computes that tile, it
takes the $L$ passes to $164$--$168$~GB/s, 25--37\% more, and the $P$ passes up
4--7\%, with identical results (file~\rf{probes}).
The library's passes zero their output lines this way (\texttt{FMM\_ZVA=0}
switches it off).  In whole plans, measured alternately with it switched off
(best of two rounds each, file~\rf{camp}), it
is worth 0.4--3.6\% in f32 and 0.5--1.9\% in f64, where the gains are
within the run-to-run spread: the passes are a small part of a plan's time
(Section~\ref{sec:disc}).

\subsection{Using the whole machine}\label{sec:machine}

Accelerate drives each performance cluster's SME unit from one thread
(Figure~\ref{fig:machine}), so its
GEMM keeps two or three cores busy and leaves the rest idle.  The $r$ products
of a level are independent, and a leaf \lv{G3}, \lv{G4} or \lv{G6} runs them
as three, four or six single-threaded GEMM streams taking products in turn: three streams beat
Accelerate's own threading by 10\% (f32) and 8\% (f64) at $2048$, more streams
win for the smallest products (f32 at $1024$, f64 at $512$) and lose badly
for large ones (six streams $+26$--$34$\% in f64 at $1024$).  The calibration
times each level's products with $0$, $3$, $4$ and $6$ streams and the
planner chooses per plan: 6--14\% off the best plans.

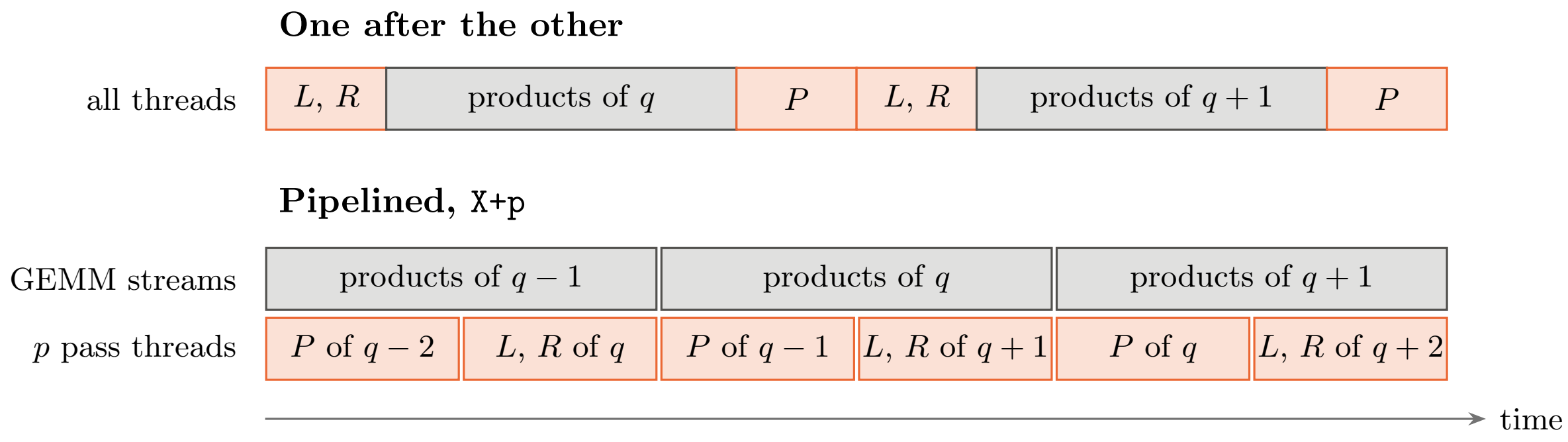
\begin{figure}[!tbp]
\centering
\begin{tikzpicture}[x=1cm,y=1cm,font=\footnotesize,
  gm/.style={draw=cG,fill=cG!18,minimum height=0.62cm,inner sep=0pt,line width=0.6pt},
  ps/.style={draw=cR,fill=cR!20,minimum height=0.62cm,inner sep=0pt,line width=0.6pt},
  feed/.style={-{Stealth[length=4pt]},black!60,line width=0.5pt}]
\node[anchor=west,font=\small\bfseries] at (0,3.75) {One after the other};
\node[anchor=east] at (-0.15,3.0) {all threads};
\foreach \x/\q in {0/{q}, 5.9/{q+1}} {
  \node[ps,minimum width=1.2cm,anchor=west] at (\x,3.0) {$L$, $R$};
  \node[gm,minimum width=3.5cm,anchor=west] at (\x+1.2,3.0) {products of $\q$};
  \node[ps,minimum width=1.2cm,anchor=west] at (\x+4.7,3.0) {$P$};
}
\node[anchor=west,font=\small\bfseries] at (0,1.95) {Pipelined, \lv{X+p}};
\node[anchor=east] at (-0.15,1.2) {GEMM streams};
\node[anchor=east] at (-0.15,0.5) {$p$ pass threads};
\foreach \x/\q in {0/{q-1}, 3.95/{q}, 7.9/{q+1}}
  \node[gm,minimum width=3.9cm,anchor=west] at (\x,1.2) {products of $\q$};
\foreach \x/\a/\b in {0/{q-2}/{q}, 3.95/{q-1}/{q+1}, 7.9/{q}/{q+2}} {
  \node[ps,minimum width=1.925cm,anchor=west] at (\x,0.5) {$P$ of $\a$};
  \node[ps,minimum width=1.925cm,anchor=west] at (\x+1.975,0.5) {$L$, $R$ of $\b$};
}
\draw[-{Stealth[length=5pt]},black!55,line width=0.6pt] (0,-0.2) -- (12.2,-0.2) node[right,text=black] {time};
\end{tikzpicture}
\caption{Pipelining a level whose sub-problems $q-1$, $q$, $q+1$, \dots\
are themselves levels over GEMM streams (schematic).  One after the other,
the passes and the products of a sub-problem take turns.  Pipelined, the
pass threads work beside the streams: while the streams compute the
products of sub-problem $q$, the pass threads finish sub-problem $q-1$ (its
$P$ pass) and prepare $q+1$ (its $L$ and $R$ passes), so the streams never
wait for a pass.}
\label{fig:pipeline}
\end{figure}

A streaming pass beside the SME streams slows them, but less than the pass
costs run alone: at f64 blocks of $512$ two pass threads slow six streams by
14\% while the pass runs at $49$~GB/s, about 35\% of its stand-alone cost.  So
a level whose products are themselves levels over GEMM streams can
\emph{pipeline} them (written \lv{X+p}, $p$ pass threads; Figure~\ref{fig:pipeline}): while the streams
compute sub-problem $q$, the pass threads write sub-problem $q-1$'s result
and the factors of $q+1$, the scratch below double-buffered.  Pipelining
takes 8--22\% off the plans over six-stream leaves; one pass thread is too
few, and sub-problems under $1$~ms are not pipelined.  The calibration times
a pipeline of $2$ and of $10$ synthetic sub-problems per level, block size,
stream count ($3$, $6$) and pass threads ($2$, $3$, $4$): the slope is the
steady time per sub-problem, the rest the fill and drain.  NEON GEMMs on the
idle cores were tried and not adopted: at most 1--3\%, fragile, and the SME
units lose most of what the vector units add.  Our own SME kernel was tried
too (outer-product accumulation with the left operand kept transposed, so
that nothing is packed): it reproduces Accelerate's results bit for bit and
is faster than Accelerate in cache on blocks of $128$, but as a level's
products through memory it runs at $0.81$--$0.94\times$ Accelerate's rate on
blocks of $384$ and up and equals it below, where both are bound by
bandwidth; an open kernel does not unlock deeper plans, and the library keeps
Accelerate (file~\rf{smeleaf}).

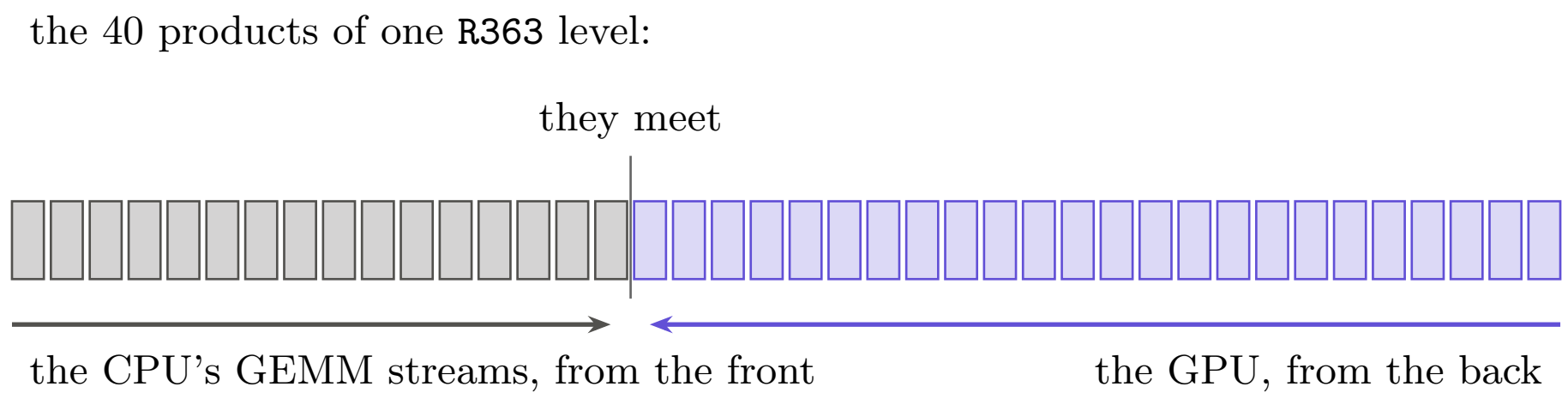
\begin{figure}[!tbp]
\centering
\begin{tikzpicture}[x=0.3cm,y=1cm,font=\footnotesize]
\foreach \i in {0,...,15} \draw[cG,fill=cG!25,line width=0.5pt] (\i,0) rectangle ++(0.82,0.6);
\foreach \i in {16,...,39} \draw[cM,fill=cM!22,line width=0.5pt] (\i,0) rectangle ++(0.82,0.6);
\draw[-{Stealth[length=5pt]},cG,line width=1pt] (0,-0.35) -- (15.4,-0.35);
\draw[-{Stealth[length=5pt]},cM,line width=1pt] (39.82,-0.35) -- (16.4,-0.35);
\node[anchor=north west] at (0,-0.45) {the CPU's GEMM streams, from the front};
\node[anchor=north east] at (39.82,-0.45) {the GPU, from the back};
\draw[black!60,line width=0.5pt] (15.91,-0.15) -- (15.91,0.95);
\node[anchor=south] at (15.91,0.95) {they meet};
\node[anchor=south west] at (0,1.62) {the $40$ products of one \lv{R363} level:};
\end{tikzpicture}
\caption{A shared leaf, \lv{G6M}: the two engines take a level's products
from the two ends of the list until they meet (the split drawn is
schematic).  Both compute the same bits, so the result does not depend on
where they meet.}
\label{fig:share}
\end{figure}

\paragraph{A second engine: the GPU.}
In f32 the machine has a second matrix engine, the GPU (Metal has no f64).
Three facts make it usable \emph{inside} a level rather than instead of the
library.  Its f32 product is bit for bit Accelerate's (Section~\ref{sec:gpu});
the memory is unified, so it reads a level's factors and writes the level's
products in the library's own scratch blocks without a copy; and it runs
beside the SME units without slowing them.  The GPU build adds two kinds of
code to the notation (Table~\ref{tab:notation}).  A leaf with the suffix
\lv{M}, \lv{GM} or \lv{G6M}, is the leaf \lv{G} or \lv{G6} with the GPU
added: it shares the $r$ products of the level above between the two
engines.  Accelerate's GEMM (one call at a time in \lv{GM}, six streams in
\lv{G6M}) takes products from the front of the list, and one thread submitting to the GPU
(Metal Performance Shaders' matrix multiplication on the scratch wrapped as
shared Metal buffers, a few milliseconds of products per command buffer, at
most eight) takes them from the back, until they meet (Figure~\ref{fig:share}).  On blocks of $1152$ and up the two together run at
$7.3$--$9.1$~TFLOP/s, against $3.3$--$3.8$ for the streams alone and
$4.0$--$6.6$ for the GPU alone (file~\rf{gpushare}).
Since both engines compute the same bits, it does not matter which one takes
which product: a plan's result, and so its error, is that of the same plan
without the GPU.  Products smaller than $576$ are left to the streams (below
that a command buffer's overhead is the product's time).  To separate the
hardware's share of the gain from the scheme's, the same sharing is offered
without any scheme: the levels \lv{K16}, \lv{K32} and \lv{K64} are the
classical product cut into $16$, $32$ or $64$ strips of rows
of $A$ and $C$, each strip one product for a shared leaf below
(\lv{K32/G6M}: $32$ strips shared between six GEMM streams and the GPU).
They have no passes and save no products, and their result is the GEMM's
own.  The GPU
leaves live in a second build of the C library (\texttt{libfmm\_gpu.a}) and
behind a feature of the Rust port (\texttt{gpu}), with their own calibration
file beside the cache, so the default builds never see them.

\section{The menu: which schemes can help}\label{sec:menu}

\begin{figure}[!tbp]
\centering
\begin{tikzpicture}
\begin{axis}[report, width=14.2cm, height=8.8cm,
  xmin=0, xmax=205, ymode=log, ymin=1.9, ymax=21,
  log ticks with fixed point, ytick={2,3,5,8,12}, xtick={0,32,64,96,128,160,192},
  xlabel={streams: blocks the widest pass reads and writes at once},
  ylabel={amp$_{\text{eq}}$: error amplification per $4\times4$-equivalent saving},
  clip=false]
\fill[cR!7] (axis cs:0,1.9) rectangle (axis cs:64,2.828);
\draw[black!45,line width=0.5pt] (axis cs:64,1.9) -- (axis cs:64,21);
\draw[black!45,line width=0.5pt] (axis cs:0,2.828) -- (axis cs:205,2.828);
\node[anchor=north east,font=\footnotesize,text=black!70,align=right] at (axis cs:204,20.3) {beyond about $64$ streams\\ a pass loses bandwidth};
\node[anchor=south east,font=\footnotesize,text=black!70,align=right] at (axis cs:204,2.88) {below this line: no worse\\ than the $4\times4$ scheme};
\addplot[only marks,mark=o,mark size=2.8pt,mark options={draw=cG,line width=0.9pt}]
  coordinates {(78,4.637) (91,5.734) (118,8.56) (140,10.38) (189,9.229) (52,4.962) (11,6.89) (11,10.66) (11,4.87) (32,12.7) (58,9.754)};
\addplot[only marks,mark=*,mark size=3.2pt,mark options={fill=cR,draw=white,line width=0.4pt}] coordinates {(58,2.280) (58,2.453)};
\addplot[only marks,mark=square*,mark size=3.1pt,mark options={fill=cD,draw=white,line width=0.4pt}] coordinates {(64,2.828)};
\addplot[only marks,mark=triangle*,mark size=3.6pt,mark options={fill=cS,draw=white,line width=0.4pt}] coordinates {(11,3.855)};
\node[anchor=west,font=\footnotesize] at (axis cs:67,2.3) {$\mm{3\times3\times6:40}$ accurate, three orientations};
\node[anchor=south west,font=\footnotesize] at (axis cs:66,2.86) {$\mm{4\times4\times4:48}$ accurate};
\node[anchor=south west,font=\footnotesize] at (axis cs:5,14.6) {$\mm{2\times2\times2:7}$:};
\draw[black!45,line width=0.5pt] (axis cs:5,14.4) -- (axis cs:5,3.5);
\node[anchor=west,font=\footnotesize,xshift=5pt] at (axis cs:11,10.66) {Winograd};
\node[anchor=west,font=\footnotesize,xshift=5pt] at (axis cs:11,6.89) {Strassen};
\node[anchor=west,font=\footnotesize,xshift=5pt] at (axis cs:11,4.87) {dyadic point};
\node[anchor=west,font=\footnotesize,xshift=5pt] at (axis cs:11,3.855) {$\mathbb{Q}(\sqrt3)$ point};
\node[anchor=west,font=\footnotesize,xshift=5pt] at (axis cs:32,12.7) {$\mm{3\times3\times3:23}$};
\node[anchor=west,font=\footnotesize,xshift=5pt,fill=white,inner sep=1pt] at (axis cs:58,9.754) {Smirnov's original point};
\node[anchor=south east,font=\footnotesize,yshift=2pt] at (axis cs:62,4.962) {$\mm{2\times4\times5:32}$};
\node[anchor=west,font=\footnotesize,xshift=5pt] at (axis cs:78,4.637) {$\mm{3\times4\times6:54}$};
\node[anchor=west,font=\footnotesize,xshift=5pt] at (axis cs:91,5.734) {$\mm{3\times4\times7:63}$};
\node[anchor=west,font=\footnotesize,xshift=5pt] at (axis cs:118,8.56) {$\mm{5\times5\times5:93}$};
\node[anchor=west,font=\footnotesize,xshift=5pt] at (axis cs:140,10.38) {$\mm{5\times5\times6:110}$};
\node[anchor=north east,font=\footnotesize,yshift=-4pt] at (axis cs:204,9.229) {$\mm{6\times6\times6:153}$};
\end{axis}
\end{tikzpicture}
\caption{The menu screen of Table~\ref{tab:menu} as a picture: every scheme
by the streams of its widest pass and by its error amplification per
$4\times4$-equivalent saving.  Filled marks are the levels of the menu, open
marks the schemes screened out.  Only the accurate $\mm{3\times3\times6:40}$
scheme lies in the shaded corner, where a level can replace the $4\times4$
one; the $\mathbb{Q}(\sqrt3)$ point of the $\mm{2\times2\times2:7}$ class is kept for the top of a
plan, where its $11$ streams and low traffic count.}
\label{fig:menu}
\end{figure}
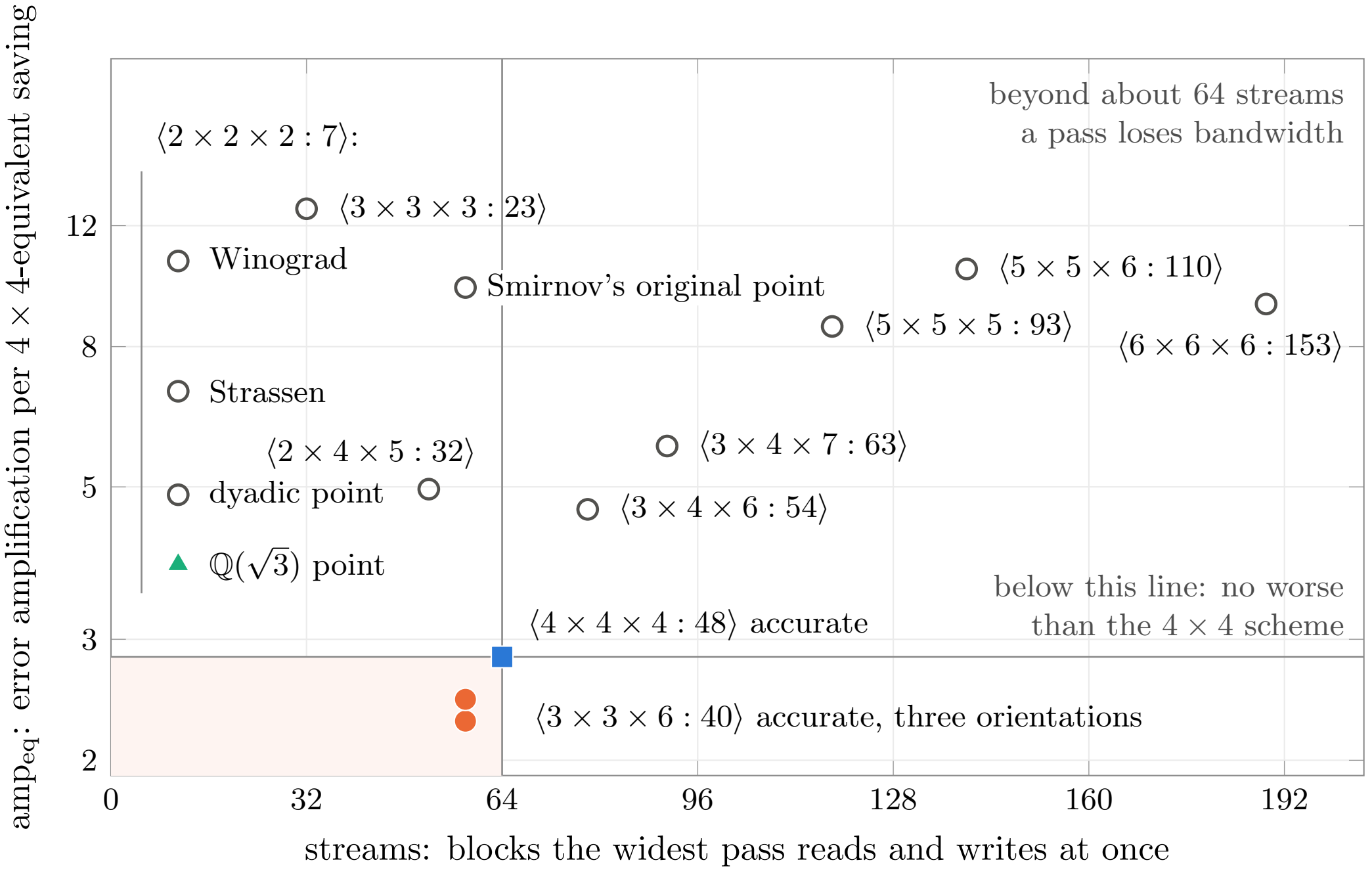

In a fused-pass library a plan already is a combination of sub-schemes, one
per level on the current $(m,k,n)$ sub-problem, so the question of mixing is
the question of the menu.  We scored $71$ schemes, $65$ from public
collections, each given there with explicit coefficients, and six composites of them: the $13$ of
AlphaEvolve's $15$ shapes whose coefficients are real
\cite{Novikov25}; the $29$ rectangular schemes of Kauers and Wood \cite{KW}
and the $5\times5\times5$ and $6\times6\times6$ schemes of Moosbauer, Poole
and Kemper \cite{MP25}, both as distributed with the addition-reduction
library of M\r{a}rtensson and Stankovski Wagner \cite{MSW25}; $21$ small and
rectangular schemes from PLinOpt's data \cite{PLinOpt}, DPS's accurate
variants and Smirnov's $\mm{3\times3\times6:40}$ \cite{Smirnov13} among
them; and the six Kronecker products of pairs of the accurate pieces.  (Perminov's
larger catalogue is screened by rank and streams at the end of this
section.)  Each scheme is scored by the
quantities that decide a level's cost and error in \eqref{eq:level}:
\begin{itemize}
\item the fraction of products kept $\kappa=r/(g_1g_2g_3)$, streams $g_1g_2+r$, $g_2g_3+r$, $g_1g_3+r$, and
  traffic $3 + r\sum 1/(g_ig_j)$;
\item the amplification of the leaf error,
\begin{equation}\label{eq:amp}
\amp \;=\; \max_j\ \tfrac12\sqrt{\textstyle\sum_i P_{ji}^2\,\|L_i\|_2^2\,\|R_i\|_2^2},
\end{equation}
  the root-sum-square growth of independent rounding errors in the $r$
  products through output $j$, which divided by the classical algorithm's $\tfrac12\sqrt{g_2}$ is the scheme's \emph{amplification ratio}; and DPS's growth factor $\ginf=\max_j\sum_i
  |P_{ji}|\,\|L_i\|_2\|R_i\|_2$ \cite{DPS24,DPS26a}, normalized by $g_2$.
\end{itemize}
Schemes of different shapes save different fractions per level, so the amplification ratio and the normalized growth factor are raised to the power $\ln0.75/\ln\kappa$, the
growth per $4\times4$-equivalent saving, and are then directly comparable with
the $\mm{4\times4\times4:48}$ scheme's $2.828$ and $5.657$ (the columns amp$_{\text{eq}}$ and
$\gamma_{\text{eq}}$ of Table~\ref{tab:levels}).  The complete table is
\texttt{matmul/levels/levels.csv} \cite{logic}; Table~\ref{tab:menu} shows the
candidates, and Figure~\ref{fig:menu} places them by the two quantities that
screen them out.

\begin{table}[t]
\centering\footnotesize
\begin{tabular}{lrrrrl}
\toprule
scheme & kept $\kappa$ & streams & traffic & amp$_{\text{eq}}$ & $\gamma_{\text{eq}}$\\
\midrule
$\mm{3\times3\times6:40}$ DPS accurate (and $6\times3\times3$) & \textbf{0.741} & 49/58/58 & \textbf{11.89} & \textbf{2.280} & 6.162\\
$\mm{3\times6\times3:40}$ DPS accurate & 0.741 & 58/58/49 & 11.89 & 2.453 & \textbf{5.125}\\
$\mm{4\times4\times4:48}$ accurate, 262-op.\ point & 0.750 & 64/64/64 & 12.00 & 2.828 & 5.657\\
$\mm{3\times4\times6:54}$ AlphaEvolve ($\pm1$ point) & 0.750 & 66/78/72 & 12.75 & 4.637 & 6.784\\
$\mm{3\times4\times7:63}$ rational & 0.750 & 75/91/84 & 13.50 & 5.734 & 11.13\\
$\mm{5\times5\times5:93}$ AlphaEvolve & 0.744 & 118/118/118 & 14.16 & 8.560 & 17.46\\
$\mm{5\times5\times6:110}$ Kemper & 0.733 & 135/140/140 & 14.73 & 10.38 & 22.9\\
$\mm{6\times6\times6:153}$ Moosbauer--Poole & 0.708 & 189/189/189 & 15.75 & 9.229 & 18.3\\
$\mm{2\times4\times5:32}$ AlphaEvolve & 0.800 & 40/52/42 & 11.80 & 4.962 & 10.95\\
$\mm{2\times2\times2:7}$ DPS accurate (in $\mathbb{Q}(\sqrt3)$): \lv{SW} & 0.875 & 11/11/11 & 8.25 & 3.855 & 10.53\\
$\mm{2\times2\times2:7}$ DPS dyadic, Eq.~(38) of \cite{DPS24} & 0.875 & 11/11/11 & 8.25 & 4.87 & 10.84\\
$\mm{2\times2\times2:7}$ Strassen / Winograd & 0.875 & 11/11/11 & 8.25 & 6.89 / 10.66 & 14.1 / 19.8\\
$\mm{3\times3\times3:23}$ best of four & 0.852 & 32/32/32 & 10.67 & 12.7 & 34.1\\
$\mm{3\times3\times6:40}$ Smirnov's original (same class) & 0.741 & 49/58/58 & 11.89 & 9.754 & 26.1\\
\bottomrule
\end{tabular}
\caption{The menu screen (\texttt{matmul/levels/levels.py}): the schemes that
could matter, at their published points.  The accuracy columns are per
$4\times4$-equivalent saving.}
\label{tab:menu}
\end{table}

Three conclusions.  Every scheme that saves more per level than the
$\mm{4\times4\times4:48}$ scheme ($5\times5\times5$ and larger) needs $118$--$225$ streams per
pass, where the passes run at half bandwidth, and has $8.6$--$107$ times the
classical amplification per $4\times4$-equivalent saving at its published
point; the other shapes larger than $4\times4$ save no more and still need
$78$--$120$ streams.
Kronecker products add nothing: a composite keeps the per-saving
amplification of its pieces ($\mm{3\times3\times6}\otimes\mm{3\times3\times6}$
keeps $2.280$, the accurate $2\times2$ squared keeps $3.855$) and multiplies
their stream counts, so it is never better than its two levels run one after
the other and the stream explosion makes it worse.  And the one scheme that
beats the $\mm{4\times4\times4:48}$ scheme on every count but $\gamma_{\text{eq}}$ is DPS's
accurate $\mm{3\times3\times6:40}$: as a level it keeps $0.741$ of the
products, moves $11.89\,n^2$ words, stays within $58$ streams, and amplifies
the leaf error by $2.28$--$2.45$ per $4\times4$-equivalent saving against
$2.83$.  Its three orientations differ in which dimension the level cuts into
six parts, so for each sub-problem the planner can use the orientation whose
passes suit its shape.  Perminov's catalogue of
$680$ shapes up to $16\times16\times16$ \cite{Perminov} confirms the menu: only
$\mm{3\times3\times6:40}$, $\mm{4\times4\times4:48}$ and $\mm{3\times4\times6:54}$
save a quarter of the products within $80$ streams per pass, and the new
shapes that save a quarter, $\mm{3\times6\times8:108}$ and $\mm{4\times7\times9:187}$,
need $126$--$215$ streams.

The paper cost model \eqref{eq:level} with the GEMM at the lone $8192^3$ f32
rate and the passes at the $4\times4$ level's bandwidth puts the
$3\times3\times6$ levels 1--3 points ahead of the $4\times4$ plans at one and
at two levels (\lv{R336/G} 17.8\% off the
GEMM against \lv{D48/G} 16.8\%; \lv{R336/R363/G} 13.9\% against 10.8\%), and
the assumption behind it holds: Accelerate on the rectangular leaves at equal
flops runs within 5\% of the square GEMM after one level and 13\% ahead after
two, because its block products are then larger matrix products.

\section{Accuracy}\label{sec:accuracy}

Five quantities describe accuracy in this report, three computed from a
scheme's matrices and two measured (Table~\ref{tab:accnames}).

\begin{table}[H]
\centering\footnotesize
\begin{tabular}{@{}l l p{0.57\textwidth}@{}}
\toprule
name & symbol & what it is\\
\midrule
amplification ratio & $\amp/(\tfrac12\sqrt{g_2})$, amp$_{\text{eq}}$ & computed from a scheme by
  \eqref{eq:amp}: the root-sum-square growth of the leaf errors through one
  level, over the classical algorithm's; amp$_{\text{eq}}$ is the same per
  $4\times4$-equivalent saving (Section~\ref{sec:menu})\\
growth factor & $\ginf$, $\gamma_{\text{eq}}$ & computed likewise: DPS's worst-case bound on that growth\\
participation number & $\nu_X$ & computed likewise: how many products an output of the level effectively averages (Section~\ref{sec:errmodel})\\
level factor & & measured: the error of one level $X$ over the GEMM, as a multiple of the GEMM's own (Section~\ref{sec:errmodel})\\
error growth & & a whole plan's error as a multiple of the GEMM's: predicted by the error model \eqref{eq:errmodel}, or
  measured; the planner's error budget bounds it (Section~\ref{sec:planner})\\
\bottomrule
\end{tabular}
\caption{The five accuracy quantities and the names the report keeps for them.}
\label{tab:accnames}
\end{table}

Of the five, only the growth factor is the literature's.  It is the constant
of the worst-case error bounds for fast multiplication, from Brent
\cite{Brent70} and Bini and Lotti \cite{BL80} through Higham \cite{Higham02}
to the stability factor of Ballard et al.\ \cite{BBDLS16}; $\ginf$ is DPS's
form of it with $2$-norms, their $\gamma_{2,1,\infty}$
\cite[Eq.~(32)]{DPS24}, a relaxation of which they minimize over the orbit.
The amplification ratio was introduced with the $262$-operation programs
\cite{S26b}.  It applies to a scheme the usual model of rounding errors as
independent random variables, which add in quadrature (Higham and Mary
\cite{HM19}): the sum over the products in $\ginf$ becomes a root-sum-square.
It is a model, not a bound.  The quantities of \cite{DPS24} and
\cite{BBDLS16} are all worst-case sums and this one is not among them; we
have not searched the literature beyond those two papers.  The participation
number, the level factor and the error growth are this report's, as is the
error model of Section~\ref{sec:errmodel} that uses them.

\subsection{What a level does to the error}

The rounding errors of the $r$ leaf products pass through $P$ into the
outputs; the roundings of the passes themselves add to them.  Taking the leaf
errors as independent and proportional to the sizes of the factors gives
\eqref{eq:amp}: a level multiplies the leaf error by $\amp$, against
$\tfrac12\sqrt{g_2}$ for the classical algorithm on the same blocks.  This
root-sum-square measure, not the worst-case sum $\ginf$, is what the
measurements follow: the $4\times4$ orbit's $\ginf$ minimum ($21.41$, in
$\mathbb{Q}(\sqrt2)$) has $\amp=3.14$ and measures no better than DPS's
Table~1 point, while the $262$-operation point, with $\amp=2\sqrt2$ the
orbit's minimum and the smallest $\ginf$ among dyadic points, measures
$0.75$--$0.77$ of it \cite{S26b}.

DPS's accurate $\mm{3\times3\times6:40}$ point is Decomposition~A of
Tichavsk\'y \cite{Tichavsky21}: the same $40$ rank-one terms, in each of the
three orientations of PLinOpt's files, with only the scaling of the three
factors different (his coefficients are all $\pm\tfrac12$; checked in exact
arithmetic against the matrices of his appendix).  He obtained it from
Smirnov's decomposition, whose coefficients are $\pm1$ and $\pm\tfrac18$, by
minimising the $\ell_1$ norm of the factors over the orbit; DPS reach the
same point from Smirnov's by minimising the growth factor, which gives the
diagonal transformation $\operatorname{diag}(\tfrac14,2,2,2,\tfrac14,2)$ of
the six-sided grid \cite[Eq.~(97)]{DPS26b}.  His signature, an
invariant of the class, takes the values $\tfrac32$ and $\tfrac54$ on it, so
the class has no point with coefficients in $\{-1,0,1\}$.  His sparser
Decomposition~B ($648$ nonzero coefficients against $960$) has amplification
ratio $3.2$ in the $3\times3\times6$ orientation and $2.8$ in the
$3\times6\times3$ one, against $2.36$ and $2.55$; we did not measure it as a
level.

Over its whole orbit under $GL_3\times GL_3\times GL_6$ the
$\mm{3\times3\times6:40}$ class has, as far as we could find, no point with
amplification ratio below $2.242$, in any orientation.  That floor is the smallest value found by L-BFGS, the
limited-memory Broyden--Fletcher--Goldfarb--Shanno method \cite{LN89}: a
quasi-Newton minimizer, which builds its estimate of the curvature from the
last few gradients and so needs only the function and its gradient.  It is run
from random starting points of the orbit (\texttt{orbit\_rect.py}), so the
floor is a numerical minimum, not a proved bound.  DPS's
$3\times3\times6$ point is 5\% above that floor and 5\% above its $\ginf$
floor, their $3\times6\times3$ point 14\% above (and 41\% above that
orientation's $\ginf$ floor of $3.90$, far below the $4\times4$ class's
$5.66$).  Smirnov's original point of the same class (coefficients $\pm1$
and $\pm\tfrac18$, amplification ratio $10.8$) reaches the same floors, so the class carries them, not the point.
Per $4\times4$-equivalent saving the floor is $2.17$ against the $4\times4$
class's $2.83$, which is its orbit's minimum and, as far as continuous search
from many starts can tell, a local minimum of the whole variety of rank-48
decompositions.

\paragraph{Other points without denser programs.}\label{par:mixed}
A point of the orbit is the scheme run on transformed grids
(Section~\ref{sec:intro}): with $A'=U^{-1}AV$ and $B'=V^{-1}BW$ it computes
$C'=U^{-1}(AB)W$, and $C=UC'W^{-1}$.  So a level at another point needs no
new programs: its blocks are mixed before the $L$ and $R$ programs and after
the $P$ program, inside the same fused passes, however dense the transformed
scheme is.  For the $3\times6\times3$ orientation the floor is where all nine
outputs amplify alike; at DPS's point six are at $2.55$ and three at $1.44$.
Two $3\times3$ rotations reach it, $U$ and $W$; the $6\times6$ side $V$ does
not help, and diagonal rescalings gain nothing.  A rational matrix taken for
both, $U=W=\tfrac23J-I$ ($I$ the $3\times3$ identity, $J$ the all-ones
matrix: entries $-\tfrac13$ on the diagonal and $\tfrac23$ off it, the
reflection through the line of $(1,1,1)$, orthogonal and its own inverse),
comes within $1.6\%$ of
it: amplification ratio $2.28$ against DPS's $2.55$ and the floor's $2.24$,
and $\ginf/g_2$ $4.22$ against $5.50$.  The programs grow from $383$
operations to $491$ with the reflection and $668$ with the rotations
(\path{orbit_points.py}); written out, the minimizer has $1{,}800$ nonzero
coefficients where DPS's point has $960$.

Measured as the level, in three binaries otherwise identical
(Table~\ref{tab:r363pts}, file~\rf{r363pts}), the two points take the same
time as DPS's ($0.98$--$1.01$) and divide the plans.  Where the level sits
above another level, which is where the planner uses it at $6912$ and
$13824$, the error is $0.83$--$0.97$ of that at DPS's point with the
reflection and $0.82$--$0.92$ with the rotations.  Where it is the only level
over the GEMM the error is no lower, and up to $15$--$18\%$ higher.  With
the f32 passes computed in f64 the single-level penalty disappears ($1.00$
and $1.02$ of DPS's point) and the two-level gain stays ($0.91$ and $0.90$;
file~\rf{r363f64}): the penalty is the rounding of the mixing itself, and in
a single level the amplified leaf error is not what bounds the error, so
lowering the amplification gains nothing there.  The library keeps DPS's
point; the mixed points would pay as a second form of the level, taken above
another level.

\begin{table}[!tbp]
\centering\footnotesize
\begin{tabular}{llrr}
\toprule
 & & reflection $\tfrac23J-I$ & a minimizer's rotations\\
position of the level & plan & ($\amp=2.28$) & ($\amp=2.24$)\\
\midrule
alone & \lv{R363/G6}, $3456$ f32 & $1.15$ & $1.13$\\
 & \lv{R363/G6}, $3456$ f64 & $1.00$ & $1.05$\\
 & \lv{R363/G}, $6912$ f32 & $1.06$ & $1.05$\\
 & \lv{R363/G}, $6912$ f64 & $1.07$ & $1.06$\\
 & \lv{R363/G}, $13824$ f32 & $1.15$ & $1.07$\\
 & \lv{R363/G}, $13824$ f64 & $1.08$ & $1.18$\\
\midrule
above another level & \lv{R363+4/R633/G6}, $6912$ f32 & $0.90$ & $0.89$\\
 & \lv{R363+4/D48/G6}, $6912$ f32 & $0.86$ & $0.83$\\
 & \lv{R363+2/R336/G6}, $6912$ f64 & $0.86$ & $0.88$\\
 & \lv{R363+3/R633/G6}, $6912$ f64 & $0.97$ & $0.92$\\
 & \lv{R363+3/R336/G6}, $13824$ f32 & $0.88$ & $0.82$\\
 & \lv{R363+4/R363/G6}, $13824$ f32 & $0.83$ & $0.82$\\
 & \lv{R363+4/R336/G3}, $13824$ f64 & $0.91$ & $0.91$\\
\midrule
below a $2\times2$ level & \lv{SW+4/R363/G6}, $3456$ f64 & $0.99$ & $0.92$\\
\bottomrule
\end{tabular}
\caption{The $3\times6\times3$ level at two other points of its orbit, run
as DPS's programs with the blocks mixed around them: the error against the
exact product over the error at DPS's point (amplification ratio
$\amp=2.55$), geometric means over four random instances (two at $13824$).
The first point mixes the two $3\times3$ grids by $U=W=\tfrac23J-I$, with $I$
the identity and $J$ the all-ones matrix; the second by the two rotations a
minimizer found, which reach the orbit's floor.  Under each is its
amplification ratio.  The times are within $2\%$ of each other on every
plan.}
\label{tab:r363pts}
\end{table}

\subsection{The error model and its validation}\label{sec:errmodel}

The planner needs a plan's error before running it.  A first model took the
GEMM's error times one measured factor per level, with a depth factor for
every level after the first, the order of the levels immaterial.  It served
for two-level plans without a $2\times2$ level and failed elsewhere: it
over-predicted the f32 plans with a $2\times2$ level by $1.5$--$1.8\times$,
and it could not tell \lv{R363/R336} from \lv{R336/R363}, which the planner
had to choose between.  So we measured what the order does
(file~\rf{emdata}): every level alone, all $25$ ordered pairs and $17$
three-level plans, at $3456$ and $6912$ in both precisions, each on four
random instances, against the exact product.  The instances matter: the
max-norm error of one plan varies by $10$--$20\%$ from one instance to the
next (up to $1.5\times$ between the extremes of four), so a single
measurement cannot separate two plans that differ by less.

\begin{table}[!tbp]
\centering\footnotesize
\begin{tabular}{lrrrrrr}
\toprule
 & & \multicolumn{5}{c}{level below}\\
\cmidrule(lr){3-7}
level on top & alone & \lv{SW} & \lv{D48} & \lv{R336} & \lv{R363} & \lv{R633}\\
\midrule
\lv{SW}   & 1.03 & 0.98 & 0.81 & 0.79 & 0.59 & 0.76\\
\lv{D48}  & 0.74 & 0.82 & 0.84 & \textbf{0.89} & 0.66 & 0.83\\
\lv{R336} & 0.65 & 0.77 & 0.78 & 0.82 & 0.62 & 0.75\\
\lv{R363} & 0.48 & 0.58 & 0.60 & \textbf{0.72} & 0.50 & 0.61\\
\lv{R633} & 0.65 & 0.76 & 0.76 & \textbf{0.82} & 0.60 & 0.76\\
\bottomrule
\end{tabular}
\caption{The error of a level alone and of every ordered pair of levels over
the GEMM, as a multiple of the GEMM's: geometric means over four instances,
two sizes ($3456$, $6912$) and both precisions.  The table is not symmetric:
\lv{R336} below another level (in bold) costs $9$--$16\%$ more than above
it, and a $2\times2$ level commutes with everything.}
\label{tab:pairs}
\end{table}

Table~\ref{tab:pairs} has the answer for two levels.  The order matters
where \lv{R336} is involved: \lv{R363/R336} has $1.13$--$1.20$ times the
error of \lv{R336/R363}, and \lv{D48/R336} $1.09$--$1.29$ times that of
\lv{R336/D48}, in all four cases.  At three levels the effect is large: the
same three levels measure $0.85$--$0.99$ of the GEMM's error as
\lv{R363/R336/D48} and $1.35$--$1.86$ as \lv{D48/R363/R336}, and
\lv{R336/R336/R336} has $1.6$ times the error of \lv{R363/R363/R363}.  A
level at the bottom of a plan does something that the levels above it then
multiply.

\paragraph{The model.}
For a plan with the fused levels $X_1$ (on top) to $X_d$ over a GEMM leaf, on
a problem of inner dimension $k$,
\begin{equation}\label{eq:errmodel}
\text{err}^2 \;=\; \underbrace{c^2\Bigl(1+\frac{\beta}{N^{3/4}}\Bigr)\,k_{\text{leaf}}\;A_d^{\,2}}_{\text{the leaf GEMM's error, amplified}}
 \;+\; \underbrace{\sum_{i=1}^{d}\pi_{X_i}^{\,2}\,A_{i-1}^{\,2}}_{\text{the levels' own rounding}},
\qquad A_i=\prod_{j\le i}a_{X_j},\quad N=\prod_{i}\nu_{X_i},
\end{equation}
in units of the unit roundoff $u$ ($2^{-24}$ in f32, $2^{-53}$ in f64), with
$k_{\text{leaf}}=k/\prod_i g_2(X_i)$ the leaves' inner dimension.  Its pieces:
\begin{itemize}
\item \emph{The GEMM's error grows like $\sqrt{k}$.}  Measured at
  $1152\times k\times1152$ for $k$ from $48$ to $6912$, Accelerate's error is
  about $0.7\,u\sqrt{k}$ ($0.63$--$0.77$ over that range), the same in both precisions.
\item \emph{A level multiplies the relative error of its products by its
  amplification ratio} $a_X$ (Table~\ref{tab:accnames}), computed from the
  scheme: $1.87$, $2.83$, $2.36$, $2.55$ and $2.36$ for \lv{SW}, \lv{D48},
  \lv{R336}, \lv{R363} and \lv{R633}.  Nothing is fitted here; fitted freely
  instead, the five came out at $1.84$, $2.98$, $2.40$, $2.66$ and $2.45$.
\item \emph{The tail.}  By that count alone one level of the $4\times4$
  scheme would have $a/\sqrt{g_2}=1.41$ times the GEMM's error, and it
  measures $0.74$.  The error of an entry of the GEMM's product is far from
  normal: its size follows that entry's own partial sums, so a few entries
  carry errors several times the typical one, and the max-norm error is
  theirs.  An output of a level is a weighted sum of many products, which
  averages those tails away.  The \emph{participation number} $\nu_X$ of a
  scheme, $(\sum_r w_r)^2/\sum_r w_r^2$ over the weights
  $w_r=P_{jr}^2\|L_r\|^2\|R_r\|^2$ of the products in its worst output, counts
  how many products an output effectively averages: $3.2$, $16$, $23.5$,
  $27.7$ and $23.5$ for the five levels.  The factor $1+\beta/N^{3/4}$
  interpolates between a single product ($N=1$, the GEMM alone) and the fully
  averaged limit; $c$ and $\beta$ are fitted, the exponent was fitted once
  ($0.76$) and is fixed.
\item \emph{A level's own rounding} $\pi_X$, in its three passes, is added to
  its output and multiplied by the amplification of every level above it,
  $A_{i-1}$.  This is the term that knows the order.  On top of a plan a
  level's rounding counts once; at the bottom of a three-level plan it is
  multiplied by two amplification ratios, about $7$.
\end{itemize}
The fitted constants are in Table~\ref{tab:errconst}.  \lv{R336} has the
largest own rounding, $9$--$10\,u$ against $4$--$7$ for the other schemes'
levels, which is the asymmetry of Table~\ref{tab:pairs}; alone it does not
show, because a single level's rounding of $10\,u$ is small beside a leaf
error of $30\,u$ and up.  We have not traced it to a statement of its
programs.  The model has no term for the size of the blocks, and none is
needed at these sizes.

\begin{table}[!tbp]
\centering\footnotesize
\begin{tabular}{lrrrrrrrr}
\toprule
 & GEMM & leaf & tail & \multicolumn{5}{c}{own rounding $\pi_X$ ($u$)}\\
\cmidrule(lr){5-9}
 & $c_G$ & $c$ & $\beta$ & \lv{SW} & \lv{D48} & \lv{R336} & \lv{R363} & \lv{R633}\\
\midrule
f32 & 0.746 & 0.241 & 11.4 & 3.5 & 5.7 & 10.2 & 6.5 & 6.8\\
f64 & 0.724 & 0.271 & 8.2 & 0.0 & 4.3 & 9.2 & 5.9 & 6.0\\
\midrule
from the scheme: $a_X$ & & & & 1.871 & 2.828 & 2.363 & 2.550 & 2.363\\
from the scheme: $\nu_X$ & & & & 3.18 & 16.0 & 23.5 & 27.7 & 23.5\\
\bottomrule
\end{tabular}
\caption{The error model's constants on the machine.  $c_G$: the GEMM alone
has error $c_G\,u\sqrt{k}$.  The fitted constants $c$, $\beta$ and $\pi_X$
are the calibration's, from $2304$; the amplification ratio and the
participation number are computed from each scheme's programs by the kernel
generator.  (\lv{SW}'s own rounding in f64 was fitted negative and set to
zero: it is too small to separate from the leaf term at one size.)}
\label{tab:errconst}
\end{table}

\paragraph{Calibration.}
The calibration measures $31$ plans at $N_E=2304$ (divisible by $16$ and by
$36$) against the exact product -- the f64 product of the f32 inputs, the
compensated product of the f64 inputs: the GEMM alone, the five levels alone
and their $25$ ordered pairs, each on four random instances, and keeps the
root mean square of each.  Squared, \eqref{eq:errmodel} is linear in $c^2$,
$c^2\beta$ and the $\pi_X^2$, so the fit is a weighted linear least squares
on the squared errors, seven unknowns from $30$ equations, with an unknown
that comes out negative set to zero; it is the same arithmetic in C and in
Rust, and takes a few seconds.  The \emph{level factor} of
Table~\ref{tab:accnames}, one level's error over the GEMM's, is among these
measurements: \lv{SW} $1.05$, \lv{D48} $0.69$, \lv{R336} $0.69$, \lv{R363}
$0.50$, \lv{R633} $0.65$ in f32, and $1.03$, $0.78$, $0.70$, $0.55$, $0.67$
in f64.  The accurate levels are below one although they amplify, because
their leaves accumulate $k/g_2$ products instead of $k$ and their outputs
average the GEMM's tails.

\paragraph{Validation.}
Calibrated at $2304$ and applied to the plans measured at $3456$ and $6912$
(file~\rf{empicks}), the model predicts all $188$ errors, at one, two and
three levels, within $0.86$--$1.23$ of the measured geometric means
(root-mean-square deviation $7\%$), and the ratio between two plans of one
size to $4$--$10\%$ (root mean square; nine in ten pairs within
$7$--$18\%$).  It has the order right: \lv{R363/R336} over \lv{R336/R363}
is predicted at $1.11$--$1.22$ and measures $1.13$--$1.20$;
\lv{D48/R363/R336} over \lv{R363/R336/D48} is predicted at $1.4$--$1.6$ and
measures $1.6$--$1.9$.  On the same $188$ plans the first model, calibrated
at each size, is off by $0.24$--$2.88$.  Against the campaign
(Section~\ref{sec:results}; one instance, sizes up to $13824$, streams and
pipelines), the predicted error of the $53$ plans is within $0.80$--$1.20$
of the measured, and their growth over the GEMM, which the budget uses,
within $0.90$--$1.26$ (Figure~\ref{fig:growth}), where the first model's
growth was within $0.80$--$2.01$.  The growth is over-predicted at $13824$,
by up to $1.26$, because the GEMM's own error there exceeds its $\sqrt{k}$
extrapolation from $2304$ by $1.2$--$1.3\times$ while the plans' exceed
theirs by less; that is the safe side.  What the model does not have: the
$3\times6\times3$ level's other orbit points of
Section~\ref{par:mixed}, which trade a lower amplification for more rounding
of their own and so fit its two terms, but are not calibrated; and inputs
that are not well scaled (Section~\ref{sec:scaled}).

\subsection{Programs and precision}

For a fixed scheme the program still matters, through the roundings of the
$P$ pass: a first-order rounding model with exact second moments of every
intermediate shows that DPS's $262$-operation programs collect $15$ roundings
of RMS size $0.80$ of the output per output where our $445$-operation
programs for the same point collect $36$ of size $1.39$, and among four
programs of equal amplification the model's worst-output error orders the
measured errors exactly at three and five levels \cite{S26b}.  The lever is
nearly exhausted for the $4\times4$ class: a 10\% smaller modelled pass
error was invisible at one to three levels and worth about 5\% at depth.  The
precision lever remains: running the passes of the f32 path in f64 (the
blocks stay f32, the arithmetic is f64, one rounding at the GEMM interface)
costs $\le2$\% below $16384$ and cuts the error by 36\% at five levels, nothing
at one to three, where the leaf GEMMs' error dominates.  It is a switch in the
library (\texttt{FMM\_F64PASS}), not the default.

\subsection{The $2\times2$ level's programs}\label{sec:sw}

The $2\times2$ level runs Strassen's scheme \cite{Strassen69} at points of
its orbit that Dumas, Pernet and Sedoglavic found by minimizing a growth
factor \cite{DPS24}.  Their Eq.~(38) is a point whose coefficients are $1$,
$\frac12$ and $\frac14$ only, so its scalings are exact in floating point,
with amplification ratio ($\amp$ of \eqref{eq:amp} over the classical
algorithm's, the per-level measure that Table~\ref{tab:menu} raises to a
$4\times4$-equivalent saving) $2.08$ per level against $3.00$ for Winograd's
variant \cite{Winograd71}, the usual choice for its $15$ additions, and
$2.45$ for Strassen's original, and $\ginf$ $6.05$ against $8$ and $6.83$
($3.03$, $4$ and $3.42$ normalized by $g_2=2$).

The dyadic point is not the most accurate point of the orbit.  DPS's optimum,
their Eq.~(36), has coefficients in $\mathbb{Q}(\sqrt3)$ \cite{DPS24}: its amplification
ratio is $1.87$ per level, $10\%$ below the dyadic point's $2.08$, and its
$\ginf$ is $5.97$ against $6.05$.  (Table~\ref{tab:menu} shows the same gap
as $3.86$ against $4.87$, because it raises both to a $4\times4$-equivalent
saving, a power of $2.15$.)  The library first ran the dyadic point, taken
for its exact scalings, and runs the $\mathbb{Q}(\sqrt3)$ point since
2026-11-22, for the measurement that follows.  Run as the level, the $\mathbb{Q}(\sqrt3)$ point is the more accurate
by about what the amplification ratios say (Table~\ref{tab:q3},
file~\rf{swq3}): its error is $0.83$--$0.94$ of the dyadic point's at one
level and $0.62$--$0.68$ at four, against $0.90$ and $0.66$ predicted, and
the five plans with a $2\times2$ level on top have $0.88$--$0.97$ of their
error, below one on every instance, in the same time (within $2.5\%$, the
run-to-run spread).  The comparison is of three binaries that differ in
this level's programs only: the dyadic point's optimized programs, the
library's at the time, and each point written as direct
sums, the $\mathbb{Q}(\sqrt3)$ point with its constants rounded to double.
The dyadic point as direct sums has $0.93$--$1.10$ of the error of its
optimized programs, so the form of the program is not what
separates the two points.  The irrational constants, which all round, cost
nothing that shows, as in the rescaling test of Section~\ref{sec:disc}; DPS's
own measurements order the two points the same way \cite[Figure~1]{DPS24}.

The library was therefore moved to the $\mathbb{Q}(\sqrt3)$ point
(its programs are \path{schemes/sw/q3_{L,R,P}.slp}, direct sums with the
constants rounded to double; the kernel generator's
option \path{--sw schemes/opt/dps7} builds the dyadic kernels), recalibrated,
and the campaign repeated on it (file~\rf{camp}).  Every table and figure of
this report is from that campaign and from runs on that library, except the
comparisons of this subsection that name the dyadic point or Winograd's
programs.  Against the campaign before the move (file~\rf{campold}), the
level factor of \lv{SW}, on the one instance the calibration then used, fell from $1.05$ to $0.94$ in f32 and from
$0.97$ to $0.96$ in f64; the plans with a $2\times2$ level have $0.85$--$0.95$
of the errors they had, and the plans
without one have the same errors to the digits printed.  The times are within the
run-to-run spread, and the recalibration moved the planner's picks among
plans within $2\%$ of each other, as any recalibration does
(Section~\ref{sec:planner}).  One of the moves cost accuracy for nothing: at $3456$ f64 the
new calibration predicted \lv{SW+4/R363/G6} $2.1\%$ faster than
\lv{R363/G4}, just outside the $2\%$ window in which the planner prefers a
clearly more accurate plan, and the planner took it ($0.0706$~s, $0.69$ of the
GEMM's error) in place of \lv{R363/G4} ($0.0715$~s, $0.58$).  The window is
now $3\%$ and the choice \lv{R363/G4} again (Section~\ref{sec:planner}).

\begin{table}[!tbp]
\centering\footnotesize
\begin{tabular}{lrrr}
\toprule
 & $\mathbb{Q}(\sqrt3)$ point & predicted, & dyadic as direct sums\\
plan & over dyadic & $0.90$ per level & over its optimized programs\\
\midrule
\lv{SW/G} & $0.83$--$0.94$ & $0.90$ & $0.99$--$1.01$\\
\lv{SW/SW/G} & $0.76$--$0.83$ & $0.81$ & $0.98$--$0.99$\\
\lv{SW/SW/SW/G} & $0.64$--$0.76$ & $0.73$ & $0.93$--$1.02$\\
\lv{SW/SW/SW/SW/G} & $0.62$--$0.68$ & $0.66$ & $0.96$--$1.10$\\
\midrule
\lv{SW+3/R363/G6}, $3456$ f64 & $0.97$ & & $1.01$\\
\lv{SW/R363/G6}, $6912$ f64 & $0.91$ & & $0.99$\\
\lv{SW/R363+3/R633/G6}, $13824$ f64 & $0.92$ & & $1.01$\\
\lv{SW/D48+2/D48/G6}, $6912$ f32 & $0.90$ & & $0.99$\\
\lv{SW/D48+2/D48/G6}, $13824$ f32 & $0.88$ & & $0.98$\\
\bottomrule
\end{tabular}
\caption{The $2\times2$ level at DPS's $\mathbb{Q}(\sqrt3)$ point against
the dyadic point, the library's until this measurement: ratios of the errors against the exact
product, geometric means over four random instances (two at $13824$).  The
first four rows give the range over $4096$ and $8192$ in both precisions; the
last column is the control for the form of the programs.}
\label{tab:q3}
\end{table}

Before either accurate point, the comparison was with Winograd's programs.
The dyadic point's programs,
\path{schemes/opt/dps7_{L,R,P}.slp}, are cancellation-free programs from
PLinOpt's optimizer: $8+3$, $8+3$ and $11+3$ additions and scalings.  Against
Winograd's programs, with the two binaries otherwise identical, run
alternately and measured against the exact product
(file~\rf{swdps7}; \texttt{gen\_kernels.py --sw
winograd} builds the kernels with Winograd's programs):
\begin{itemize}
\item One level alone has $0.88$--$1.35\times$ the GEMM's error at $4096$ and
  $8192$ in both precisions where Winograd's programs have $1.44$--$1.95\times$;
  at two, three and four recursive levels Winograd's programs have $2.5$--$3.6$, $3.7$--$4.0$ and $5.1$--$6.3$ times the dyadic point's error.  The
  calibration's factor for \lv{SW} at the dyadic point was $1.05$ in f32 and $0.97$ in f64 against
  Winograd's $1.41$ and $1.51$; the other levels' factors
  agree within $1\%$.
\item The passes cost the same: at $8192$ f64 the three passes of one level take
  $0.010$, $0.010$ and $0.009$~s with either program, and every plan's time is
  within the run-to-run noise of the other binary's ($\pm1\%$).
\item The plans with a $2\times2$ level on top have $0.69$--$0.75$ of the
  error they have with Winograd's programs, in the same time
  (Table~\ref{tab:sw}; \lv{SW/R363+3/R633/G6} at $13824$ f64 takes $3.25$~s
  either way).  With Winograd's programs these plans have $0.96$--$1.55$
  times the GEMM's error, with the dyadic point's $0.66$--$1.14$, and with
  the $\mathbb{Q}(\sqrt3)$ point's, in the campaign, $0.59$--$0.97$; the one
  that was above the GEMM, \lv{SW/D48+2/D48/G6} at $6912$ f32, is a three-level plan
  twice as slow as the best at its size.
\end{itemize}
\begin{table}[!tbp]
\centering\footnotesize
\setlength{\tabcolsep}{3.6pt}
\begin{tabular}{llrrrrrr}
\toprule
 & & \multicolumn{2}{c}{$\mathbb{Q}(\sqrt3)$ point} & \multicolumn{2}{c}{dyadic point} & \multicolumn{2}{c}{Winograd's programs}\\
\cmidrule(lr){3-4}\cmidrule(lr){5-6}\cmidrule(lr){7-8}
$n$, prec. & plan & error & of GEMM's & error & of GEMM's & error & of GEMM's\\
\midrule
3456 f64 & \lv{SW+3/R363/G6} & $3.29\cdot10^{-15}$ & $0.69$ & $3.47\cdot10^{-15}$ & $0.73$ & $4.62\cdot10^{-15}$ & $0.97$\\
6912 f64 & \lv{SW/R363/G6} & $3.94\cdot10^{-15}$ & $0.59$ & $4.43\cdot10^{-15}$ & $0.66$ & $6.40\cdot10^{-15}$ & $0.96$\\
13824 f64 & \lv{SW/R363+3/R633/G6} & $8.03\cdot10^{-15}$ & $0.65$ & $8.66\cdot10^{-15}$ & $0.70$ & $1.25\cdot10^{-14}$ & $1.02$\\
6912 f32 & \lv{SW/D48+2/D48/G6} & $3.74\cdot10^{-6}$ & $0.97$ & $4.40\cdot10^{-6}$ & $1.14$ & $5.95\cdot10^{-6}$ & $1.55$\\
13824 f32 & \lv{SW/D48+2/D48/G6} & $5.50\cdot10^{-6}$ & $0.86$ & $6.30\cdot10^{-6}$ & $0.98$ & $9.13\cdot10^{-6}$ & $1.43$\\
\bottomrule
\end{tabular}
\caption{Plans with a $2\times2$ level on top: the error with the
$\mathbb{Q}(\sqrt3)$ point (the library's, from the campaign), with the dyadic
point's programs and with Winograd's (those two binaries otherwise identical,
run alternately); the times are equal within the noise.}
\label{tab:sw}
\end{table}

So a $2\times2$ level on top costs little accuracy and stays within the
default budget: \lv{SW+4/R363/G6} is the fastest plan measured at $3456$
f64 and the planner's choice there without a budget ($0.0706$~s, $1.38\times$ over Accelerate, $0.69\times$ its error), and
\lv{SW/R363+3/R633/G6} is within $1\%$ of the fastest at $13824$ f64 ($3.269$~s, $1.88\times$,
$0.65\times$).  The error model of Section~\ref{sec:errmodel} predicts
these plans as it does the others: \lv{SW+4/R363/G6} at $6912$ f32 at
$0.56\times$ the GEMM's error, where it measures $0.53\times$.  (The model
before it charged a $2\times2$ level on top as if it were an accurate
level's second, and predicted $0.92$.)

\subsection{Badly scaled inputs}\label{sec:scaled}

Every error so far is measured on uniform random inputs in $[-1,1)$ and in
the max norm, relative to the largest entry of $C_{\mathrm{exact}}$.  Both
choices favour a fast scheme.  The classical algorithm has a componentwise
bound: the error in $c_{ij}$ is at most a small multiple of
$(|A||B|)_{ij}=\sum_l |a_{il}||b_{lj}|$, the size of the terms that make up
$c_{ij}$ itself.  The Strassen-type schemes have only a normwise bound
\cite{Higham02}: a level adds blocks from different rows and columns of $A$
and of $B$, so one product can carry rounding errors at the scale of the
largest block into an output entry whose own terms are far smaller.  When
all the entries are of one size the two bounds say the same thing, and a
max-norm error cannot tell them apart; this is the case ESSL's warning about
inputs of divergent scale is about \cite{ESSL}.  We therefore measured the
plans on inputs with structured magnitudes, by the max-norm error of Section~\ref{sec:setup} (``max'' in the tables) and by the componentwise error
$$
\text{cw} \;=\; \max_{i,j}\ \frac{|C-C_{\mathrm{exact}}|_{ij}}{(|A||B|)_{ij}},
$$
which is about $4\cdot10^{-7}$ for Accelerate's GEMM in f32 and
$7\cdot10^{-16}$ in f64 on uniform inputs.  Scaling the whole of $A$ or $B$
changes nothing (by a power of two, not a bit), so only uneven scalings
matter.  Each class below modifies the uniform inputs (\texttt{FMM\_INPUT}
in the benchmark):
\begin{description}
\item[nonneg:] $|A|$ and $|B|$, so that $C$ has no cancellation.
\item[outer:$s$:] row $i$ of $A$ multiplied by $10^{-s\,i/(m-1)}$ and
  column $j$ of $B$ by $10^{-s\,j/(n-1)}$: graded rows and columns, $C$'s
  entries spanning $2s$ decades.  \textbf{outerp:$s$} applies the same factors
  to the rows and columns in a scrambled order.
\item[inner:$s$:] column $l$ of $A$ multiplied by $10^{s(l/(k-1)-1/2)}$ and
  row $l$ of $B$ divided by the same factor: $C$ and every term
  $a_{il}b_{lj}$ are unchanged, but the columns of $A$ span $s$ decades.
  \textbf{innerp:$s$}: scrambled.
\item[cross:$s$:] the off-diagonal halves of $A$ and of $B$ (the entries
  whose row lies in one half and column in the other) multiplied by
  $10^{-s}$; no scaling of rows and columns evens these.
\item[entry:$s$:] every entry multiplied by its own $10^{-sU}$, $U$ uniform
  in $[0,1)$.
\end{description}

\begin{table}[!tbp]
\centering\footnotesize
\newcommand{\E}[2]{$#1{\cdot}10^{#2}$}
\begin{tabular}{llrrrrrr}
\toprule
 & & \multicolumn{2}{c}{GEMM} & \multicolumn{2}{c}{default plan} & \multicolumn{2}{c}{plan, prescaled}\\
\cmidrule(lr){3-4}\cmidrule(lr){5-6}\cmidrule(lr){7-8}
prec. & inputs & max & cw & max & cw & max & cw\\
\midrule
f32 & uniform  & \E{3.85}{-6} & \E{3.74}{-7} & \E{2.20}{-6} & \E{2.18}{-7} & \E{3.02}{-6} & \E{2.96}{-7}\\
    & nonneg   & \E{6.23}{-6} & \E{6.63}{-6} & \E{4.37}{-6} & \E{4.63}{-6} & \E{5.60}{-6} & \E{5.94}{-6}\\
    & outer:6  & \E{2.59}{-6} & \E{3.97}{-7} & \E{2.57}{-7} & \cellcolor{black!10}{\boldmath\E{6.01}{2}} & \E{1.63}{-6} & \E{3.27}{-7}\\
    & outerp:6 & \E{2.71}{-6} & \E{3.87}{-7} & \E{3.11}{-7} & \cellcolor{black!10}{\boldmath\E{1.69}{3}} & \E{1.62}{-6} & \E{3.65}{-7}\\
    & inner:6  & \E{3.81}{-6} & \E{3.70}{-7} & \cellcolor{black!10}{\boldmath\E{6.51}{-2}} & \cellcolor{black!10}{\boldmath\E{6.40}{-3}} & \E{2.71}{-6} & \E{2.63}{-7}\\
    & innerp:6 & \E{3.80}{-6} & \E{3.69}{-7} & \cellcolor{black!10}{\boldmath\E{6.84}{-2}} & \cellcolor{black!10}{\boldmath\E{6.49}{-3}} & \E{2.70}{-6} & \E{2.62}{-7}\\
    & cross:6  & \E{2.48}{-6} & \E{3.98}{-7} & \E{1.46}{-6} & \cellcolor{black!10}{\boldmath\E{9.96}{-2}} & \E{1.79}{-6} & \cellcolor{black!10}{\boldmath\E{1.23}{-1}}\\
    & entry:6  & \E{3.15}{-6} & \E{1.94}{-6} & \E{2.12}{-6} & \E{1.63}{-6} & \E{2.12}{-6} & \E{1.63}{-6}\\
\midrule
f64 & uniform  & \E{6.69}{-15} & \E{6.60}{-16} & \E{4.07}{-15} & \E{3.92}{-16} & \E{4.07}{-15} & \E{3.92}{-16}\\
    & nonneg   & \E{1.18}{-14} & \E{1.25}{-14} & \E{8.96}{-15} & \E{9.56}{-15} & \E{8.96}{-15} & \E{9.56}{-15}\\
    & outer:6  & \E{4.42}{-15} & \E{6.60}{-16} & \E{4.41}{-16} & \cellcolor{black!10}{\boldmath\E{1.16}{-6}} & \E{2.76}{-15} & \E{6.42}{-16}\\
    & outerp:6 & \E{5.15}{-15} & \E{5.89}{-16} & \E{5.28}{-16} & \cellcolor{black!10}{\boldmath\E{3.23}{-6}} & \E{3.00}{-15} & \E{6.78}{-16}\\
    & inner:6  & \E{6.86}{-15} & \E{6.77}{-16} & \cellcolor{black!10}{\boldmath\E{1.37}{-10}} & \cellcolor{black!10}{\boldmath\E{1.35}{-11}} & \E{4.89}{-15} & \E{4.78}{-16}\\
    & innerp:6 & \E{6.69}{-15} & \E{6.60}{-16} & \cellcolor{black!10}{\boldmath\E{1.25}{-10}} & \cellcolor{black!10}{\boldmath\E{1.19}{-11}} & \E{5.11}{-15} & \E{4.98}{-16}\\
    & cross:6  & \E{1.20}{-14} & \E{1.62}{-15} & \E{2.61}{-15} & \cellcolor{black!10}{\boldmath\E{1.73}{-10}} & \E{2.61}{-15} & \cellcolor{black!10}{\boldmath\E{1.73}{-10}}\\
    & entry:6  & \E{6.87}{-15} & \E{4.31}{-15} & \E{3.80}{-15} & \E{2.96}{-15} & \E{3.80}{-15} & \E{2.96}{-15}\\
\bottomrule
\end{tabular}
\caption{Errors on badly scaled inputs at $n=6912$: Accelerate's GEMM, the
planner's default plan (\lv{R363+4/R633/G6} in f32, \lv{R633+3/R363/G6} in
f64), and the same plan prescaled; in bold on grey, errors more than $100$ times
the GEMM's.  File~\rf{scaled} also has the
single levels, \lv{D48/D48/G}, \lv{R363/R336/G}, $s=3$, and $n=13824$; the f64 rows here and the f32 cases at $13824$, whose default plans changed since, are from file~\rf{emreruns}.}
\label{tab:scaled}
\end{table}

\newcommand{\icon}[1]{\raisebox{-3pt}{\tikz[x=0.42cm,y=0.42cm]{#1\draw[black!60,line width=0.4pt] (0,0) rectangle (1,1);}}}
\newcommand{\iuni}{\icon{\fill[black!45] (0,0) rectangle (1,1);}}
\newcommand{\iout}{\icon{\shade[top color=black!75,bottom color=black!4] (0,0) rectangle (1,1);}}
\newcommand{\iinn}{\icon{\shade[left color=black!4,right color=black!75] (0,0) rectangle (1,1);}}
\newcommand{\icro}{\icon{\fill[black!6] (0,0) rectangle (1,1);\fill[black!65] (0,0.5) rectangle (0.5,1);\fill[black!65] (0.5,0) rectangle (1,0.5);}}
\newcommand{\ient}{\icon{\pgfmathsetseed{11}\foreach \i in {0,...,4} \foreach \j in {0,...,4} {\pgfmathtruncatemacro{\g}{5+70*rnd}\fill[black!\g] (0.2*\i,0.2*\j) rectangle ++(0.2,0.2);}}}
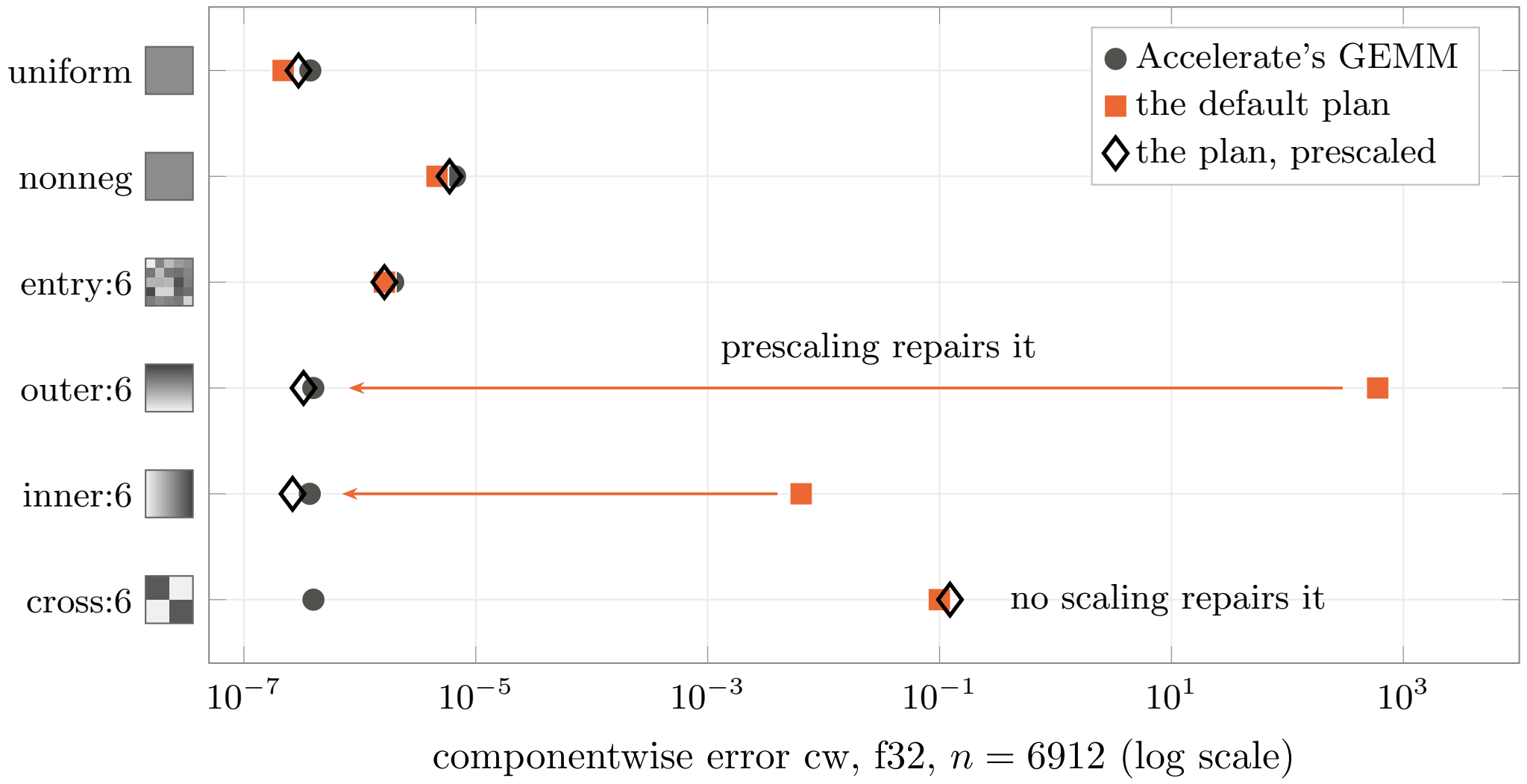
\begin{figure}[!tbp]
\centering
\begin{tikzpicture}
\begin{axis}[report, width=13.2cm, height=7.4cm,
  xmode=log, xmin=5e-8, xmax=1e4, ymin=0.4, ymax=6.6, y dir=reverse,
  xtick={1e-7,1e-5,1e-3,1e-1,1e1,1e3},
  ytick={1,2,3,4,5,6},
  yticklabels={{uniform \iuni},{nonneg \iuni},{entry:6 \ient},{outer:6 \iout},{inner:6 \iinn},{cross:6 \icro}},
  ymajorgrids=true, xmajorgrids=true,
  xlabel={componentwise error cw, f32, $n=6912$ (log scale)},
  legend pos=north east, clip=false]
\addplot[only marks,mark=*,mark size=3pt,mark options={fill=cG,draw=white,line width=0.4pt}]
  coordinates {(3.74e-7,1) (6.63e-6,2) (1.94e-6,3) (3.97e-7,4) (3.70e-7,5) (3.98e-7,6)};
\addlegendentry{Accelerate's GEMM}
\addplot[only marks,mark=square*,mark size=2.9pt,mark options={fill=cR,draw=white,line width=0.4pt}]
  coordinates {(2.18e-7,1) (4.63e-6,2) (1.63e-6,3) (6.01e2,4) (6.40e-3,5) (9.96e-2,6)};
\addlegendentry{the default plan}
\addplot[only marks,mark=diamond,mark size=4pt,mark options={draw=black,line width=1pt}]
  coordinates {(2.96e-7,1) (5.94e-6,2) (1.63e-6,3) (3.27e-7,4) (2.63e-7,5) (1.23e-1,6)};
\addlegendentry{the plan, prescaled}
\draw[cR,line width=0.7pt,-{Stealth[length=4pt]}] (axis cs:3e2,4) -- (axis cs:8e-7,4);
\draw[cR,line width=0.7pt,-{Stealth[length=4pt]}] (axis cs:4e-3,5) -- (axis cs:7e-7,5);
\node[font=\footnotesize,anchor=south] at (axis cs:3e-2,3.92) {prescaling repairs it};
\node[font=\footnotesize,anchor=west] at (axis cs:3e-1,6) {no scaling repairs it};
\end{axis}
\end{tikzpicture}
\caption{The f32 componentwise errors of Table~\ref{tab:scaled} as a
picture.  The small squares show the pattern of magnitudes in $A$ for each
class of inputs (darker is larger): even, graded by row (outer), graded by
column (inner), and in crossed blocks.  On graded inputs the default plan
loses four to nine orders of magnitude against the GEMM and prescaling
brings it back; on cross-structured inputs nothing does.}
\label{fig:scaled}
\end{figure}

What Table~\ref{tab:scaled} says, with the rest of the run (Figure~\ref{fig:scaled} draws its f32 componentwise columns):
\begin{itemize}
\item \emph{Graded rows and columns are invisible to the max-norm error and ruinous componentwise.}  On outer:6 the default plans' max-norm error is a tenth to a fourteenth
  of the GEMM's, because it is set by the largest entries, but their
  componentwise error is $6\cdot10^2$ in f32 ($182$ at $13824$) and $1.2\cdot10^{-6}$
  in f64 ($1.2\cdot10^{-6}$), $10^9$ times the GEMM's: the smallest entries
  of $C$, $10^{-12}$ of the largest, have no correct digit in f32 and about
  six in f64.  At $s=3$ the f32 plan's cw is $3.2\cdot10^{-3}$, and
  \lv{D48/D48/G} reaches $3.9\cdot10^3$ at $s=6$.  The scrambled order makes
  no difference: it is the spread, not its pattern.
\item \emph{Inner scaling shows in the max-norm error as well.}  inner:6 leaves $C$ as it
  was, yet the default plans' max-norm error is $0.065$ in f32 ($0.090$ at $13824$) and
  $1.4\cdot10^{-10}$ in f64 ($2.0\cdot10^{-10}$), about $2\cdot10^4$ times the
  GEMM's; at $s=3$ already $40$ times.  A level adds column blocks of $A$
  whose sizes differ by up to $10^s$, and the sum is rounded at the scale of
  the larger.
\item \emph{Nonnegative inputs cost the plans much of their margin.}  The
  default plans keep $0.70$ (f32) and $0.76$ (f64) of the GEMM's error, where
  they have $0.57$ and $0.61$ on uniform inputs, and single levels are above the GEMM in both precisions (\lv{R336/G} $1.41\times$ in
  f32, \lv{R633/G} $1.47\times$ in f64), as is a plan with an \lv{R336} level
  at the bottom: \lv{R363+2/R336/G6}, the planner's choice at $6912$ f64 before
  the error model of Section~\ref{sec:errmodel}, has $1.16$ times the GEMM's error there.  The amplification of
  \eqref{eq:amp} assumes independent zero-mean blocks; with nonnegative ones
  a scheme's sums of blocks do not partly cancel and its products are larger
  against $C$, while the GEMM's relative error, with nothing cancelling in
  $C$, is at its smallest.
\item \emph{Cross structure defeats the plans componentwise}, by $10^5$ at
  $s=6$ ($0.10$ in f32, $1.7\cdot10^{-10}$ in f64), with the max-norm error still below the
  GEMM's: the off-diagonal blocks of $C$ are $10^s$ smaller than the diagonal
  ones and receive their rounding errors.
\item \emph{Random per-entry scaling is harmless}: $0.52$--$0.84$ of
  the GEMM's componentwise error at $s=3$ and $6$.
\end{itemize}

\paragraph{Prescaling.}
Ballard et al.\ \cite{BBDLS16} repair much of this by diagonal scaling
before the product and the inverse scaling after it.  We scale by powers of
two, so that both are exact: \emph{outside}, row $i$ of $A$ by $2^{-e_i}$ and
column $j$ of $B$ by $2^{-f_j}$, where $2^{e_i}$ and $2^{f_j}$ are the
largest powers of two not above the row's and the column's largest
magnitude; then \emph{inside}, column $l$ of the scaled $A$ by $2^{t_l}$ and
row $l$ of the scaled $B$ by $2^{-t_l}$, with $t_l$ half the difference of
the exponents of their largest magnitudes, rounded toward zero; and row $i$
and column $j$ of the computed $C$ by $2^{e_i}$ and $2^{f_j}$ at the end.
The GEMM's result is unchanged bit for bit.  For the plans
(Table~\ref{tab:scaled}) it removes the graded and the inner failures
completely: the default plans' errors on every class but cross are
$0.44$--$1.15$ of the GEMM's in both measures at $6912$ and $13824$ (the
worst, outerp:6 in f64, $1.15$ componentwise), and every single level and
two-level plan of the run stays within $1.85\times$ (\lv{R336/G} on nonneg
f32; the \lv{SW} level within $1.53\times$).
Cross structure it leaves as it was, and no scaling of rows and columns can
do better; such a product would have to be split into its blocks.  It costs
some accuracy on well-scaled f32 inputs: the default plan's max-norm error there rises
from $0.57$ to $0.78$ of the GEMM's at $6912$ and from $0.38$ to $0.54$ at
$13824$, because rounding uniform inputs to f32 leaves a few rows whose
largest entry is exactly $1$, which get the factor $1$ while the others get
$2$, and the level then adds blocks whose scales differ by two.  Skipping the
factors when the inputs' range is small would avoid that; the library does
not do it yet.

\begin{table}[!tbp]
\centering\footnotesize
\begin{tabular}{llrrrr}
\toprule
$n$, prec. & GEMM (s) & default plan (s) & separate passes & folded & speedup: none / separate / folded\\
\midrule
6912 f32  & 0.194 & 0.137 & $+5.8\%$ & $+3.6\%$ & $1.42$ / $1.34$ / $1.37\times$\\
6912 f64  & 0.758 & 0.447 & $+4.3\%$ & $+2.0\%$ & $1.70$ / $1.63$ / $1.66\times$\\
13824 f32 & 1.544 & 0.901 & $+4.9\%$ & $+2.5\%$ & $1.71$ / $1.63$ / $1.67\times$\\
13824 f64 & 6.176 & 3.255 & $+2.2\%$ & $+0.3\%$ & $1.90$ / $1.86$ / $1.89\times$\\
\bottomrule
\end{tabular}
\caption{The time prescaling costs the default plans: the best of two
alternating rounds of seven (6912) or three (13824) runs, with the factors
computed inside the timed call in both forms.  The two forms give the same
result bit for bit.  File~\rf{prescale}; the $6912$ f64 and $13824$ f32 rows, whose default plans changed since, from file~\rf{emreruns}.}
\label{tab:prescale}
\end{table}

The library does the prescaling in one of two forms
(\texttt{exec\_prescaled} in \path{fmm_impl.h}).  As \emph{separate passes},
it scales copies of $A$ and $B$, runs the plan on them and scales $C$: three
more reads and writes of a matrix.  \emph{Folded} into a fused top level, the
level's $L$ and $R$ passes scale each tile of their input as they read it,
and its $P$ pass scales each tile of $C$ right after writing it, while the
tile is still in cache; what is left is computing the factors, which reads
$A$ and $B$ twice each.  Table~\ref{tab:prescale} gives the price: folded,
$0.3$--$3.6\%$ of the plan's time, a sixth to two thirds of what the separate passes cost, and the
plans stay $1.37$--$1.89\times$ faster than the GEMM.  (With the default plans of the campaign before, two of
them with a $3\times3\times6$ level on top, folding cost $1.2$--$5.3\%$.)  Most of the folded
cost is the factors; estimating the inside factors from the unscaled
columns would need one read of each input and should roughly halve it.  The
prescaled path is in the C library but not yet in its public interface, in
the planner's default, or in the Rust port.

\section{The planner}\label{sec:planner}

\begin{figure}[!tbp]
\centering
\begin{tikzpicture}[font=\footnotesize,
  stp/.style={draw=black!70,rounded corners=3pt,fill=white,align=center,inner sep=4pt,line width=0.5pt},
  io/.style={draw=black!45,fill=black!6,align=center,inner sep=4pt,line width=0.4pt},
  opt/.style={draw=black!60,rounded corners=2pt,align=center,inner sep=3pt,line width=0.4pt,text width=2.05cm,minimum height=1.25cm},
  ar/.style={-{Stealth[length=4pt]},black!65,line width=0.6pt},
  ln/.style={black!55,line width=0.5pt},
  pt/.style={circle,fill=cG!70,inner sep=0pt,minimum size=4.2pt}]
\node[io,text width=2.45cm] (call) at (1.35,9.2) {a call:\\ shape $(m,k,n)$,\\ precision,\\ error budget $b$};
\node[stp,text width=2.85cm] (pad) at (4.75,9.2) {\textbf{1\ \ pad}\\ round the shape up\\ to one the levels\\ divide};
\node[stp,text width=2.85cm] (price) at (8.3,9.2) {\textbf{2\ \ price}\\ predicted time and\\ error growth of the\\ candidate plans};
\node[stp,text width=2.5cm] (choose) at (11.68,9.2) {\textbf{3\ \ choose}\\ apply the error\\ budget, then\\ the tie rule};
\node[io,text width=2.85cm] (plan) at (14.8,9.2) {a plan, cached:\\ \lv{R363+4/R336/G3}};
\draw[ar] (call) -- (pad); \draw[ar] (pad) -- (price); \draw[ar] (price) -- (choose); \draw[ar] (choose) -- (plan);
\node[io,text width=11.6cm] (cal) at (8.3,7.45) {the calibration, measured once per machine and precision:\\
  every level's passes; the GEMM per product with $0$, $3$, $4$ and $6$ streams;\\ the pipelines; the error model's factors};
\draw[ar] (cal.north) -- (price.south);
\node[anchor=west,font=\small\bfseries] at (0,6.3) {2\ \ Price: the recursion of Eq.~\eqref{eq:dp}};
\node[stp,text width=4.7cm] (root) at (4.475,5.45) {the time for shape $\sigma=(m,k,n)$\\ is the least of};
\node[opt,fill=cG!18] (o1) at (1.1,3.9) {a leaf\\ \lv{G}, \lv{G3}, \lv{G4},\\ \lv{G6} or \lv{B2}};
\node[opt,fill=cS!16] (o2) at (3.35,3.9) {\lv{SW}\\ passes $+\ 7\,\times$\\ shape $\sigma/2$};
\node[opt,fill=cD!14] (o3) at (5.6,3.9) {\lv{D48}\\ passes $+\ 48\,\times$\\ shape $\sigma/4$};
\node[opt,fill=cR!16] (o4) at (7.85,3.9) {an \lv{R} level\\ passes $+\ 40\,\times$\\ shape $\sigma/g$};
\foreach \o in {o1,o2,o3,o4} \draw[ln] (root.south) -- (\o.north);
\node[stp,text width=4.7cm] (sub) at (4.475,2.35) {the time for one of a level's\\ products is the least of};
\draw[ar] (o2.south) -- (sub.north -| o2.south);
\draw[ar] (o3.south) -- (sub.north -| o3.south);
\draw[ar] (o4.south) |- (sub.east);
\node[opt,fill=cG!18,text width=2.6cm,minimum height=1.55cm] (s1) at (1.45,0.62) {a GEMM leaf,\\ timed \emph{inside} the\\ level: $0$, $3$, $4$\\ or $6$ streams};
\node[opt,text width=2.6cm,minimum height=1.55cm] (s2) at (4.475,0.62) {a further level:\\ the same question\\ one level down};
\node[opt,text width=2.6cm,minimum height=1.55cm] (s3) at (7.5,0.62) {a level over\\ GEMM streams,\\ pipelined (\lv{+}$p$)};
\foreach \o in {s1,s2,s3} \draw[ln] (sub.south) -- (\o.north);
\begin{scope}[xshift=10.1cm,yshift=0.35cm]
\node[anchor=west,font=\small\bfseries] at (-0.45,5.95) {3\ \ Choose};
\fill[black!7] (0,3.75) rectangle (6.1,5.35);
\draw[black!60,line width=0.5pt,-{Stealth[length=4pt]}] (0,0) -- (6.25,0);
\node[anchor=north east,inner sep=2pt] at (6.2,-0.08) {predicted time};
\draw[black!60,line width=0.5pt,-{Stealth[length=4pt]}] (0,0) -- (0,5.55);
\node[rotate=90,anchor=south] at (-0.05,2.7) {predicted error growth};
\draw[black!70,dashed,line width=0.6pt] (0,3.75) -- (6.1,3.75);
\node[anchor=south east,inner sep=2pt] at (6.1,3.75) {the budget $b$};
\node[anchor=north east,inner sep=3pt,text=black!65] at (6.1,5.35) {over budget: not admitted};
\node[pt] at (0.45,4.6) {}; \node[pt] at (1.1,4.2) {}; \node[pt] at (2.3,4.75) {};
\node[pt] at (1.6,2.95) {}; \node[pt] at (2.3,2.2) {}; \node[pt] at (3.2,2.45) {};
\node[pt] at (4.1,1.9) {}; \node[pt] at (5.4,3.3) {}; \node[pt] at (5.0,2.35) {};
\node[pt] at (5.75,1.8) {};
\node[circle,draw=black,line width=0.8pt,fill=cG!70,inner sep=0pt,minimum size=5.4pt] (F) at (0.9,3.1) {};
\draw[black!70,dotted,line width=0.7pt] (0.9,0) -- (0.9,3.75);
\draw[black!70,dotted,line width=0.7pt] (1.85,0) -- (1.85,3.75);
\draw[black,line width=0.7pt,dashed] (0.9,0) rectangle (1.85,2.79);
\node[rectangle,rotate=45,fill=black,inner sep=0pt,minimum size=5.6pt] (W) at (1.42,1.7) {};
\draw[black!70,line width=0.4pt,{Stealth[length=3pt]}-{Stealth[length=3pt]}] (0.9,-0.28) -- (1.85,-0.28);
\node[anchor=north,inner sep=1pt] at (1.375,-0.34) {$3\%$};
\node[anchor=west,align=left,inner sep=1.5pt] (Fl) at (2.2,3.3) {the fastest plan\\ within the budget};
\draw[black!60,line width=0.4pt] (F.east) -- (Fl.west);
\node[anchor=west,align=left,inner sep=1.5pt] (Wl) at (2.35,0.78) {a plan within $3\%$ of its\\ time and at least $10\%$\\ below its error is\\ chosen instead};
\draw[black!60,line width=0.4pt] (W.east) -- (Wl.west);
\end{scope}
\end{tikzpicture}
\caption{The planner.  Top: a call is padded, priced and decided from one
calibration; the plan is cached.  Left: the pricing recursion
\eqref{eq:dp}, where ``passes $+\ 7\,\times$ shape $\sigma/2$'' means the
time of the passes of the level plus seven times the time for a product of
that shape; a product's leaf is timed inside the level that issues it,
where back-to-back calls run slower than a lone one.  Right: the choice,
drawn schematically (the points are not data): plans whose predicted error
growth exceeds the budget are not admitted, and among the rest the fastest
wins unless another within $3\%$ of its time is predicted at least $10\%$
more accurate.}
\label{fig:planner}
\end{figure}
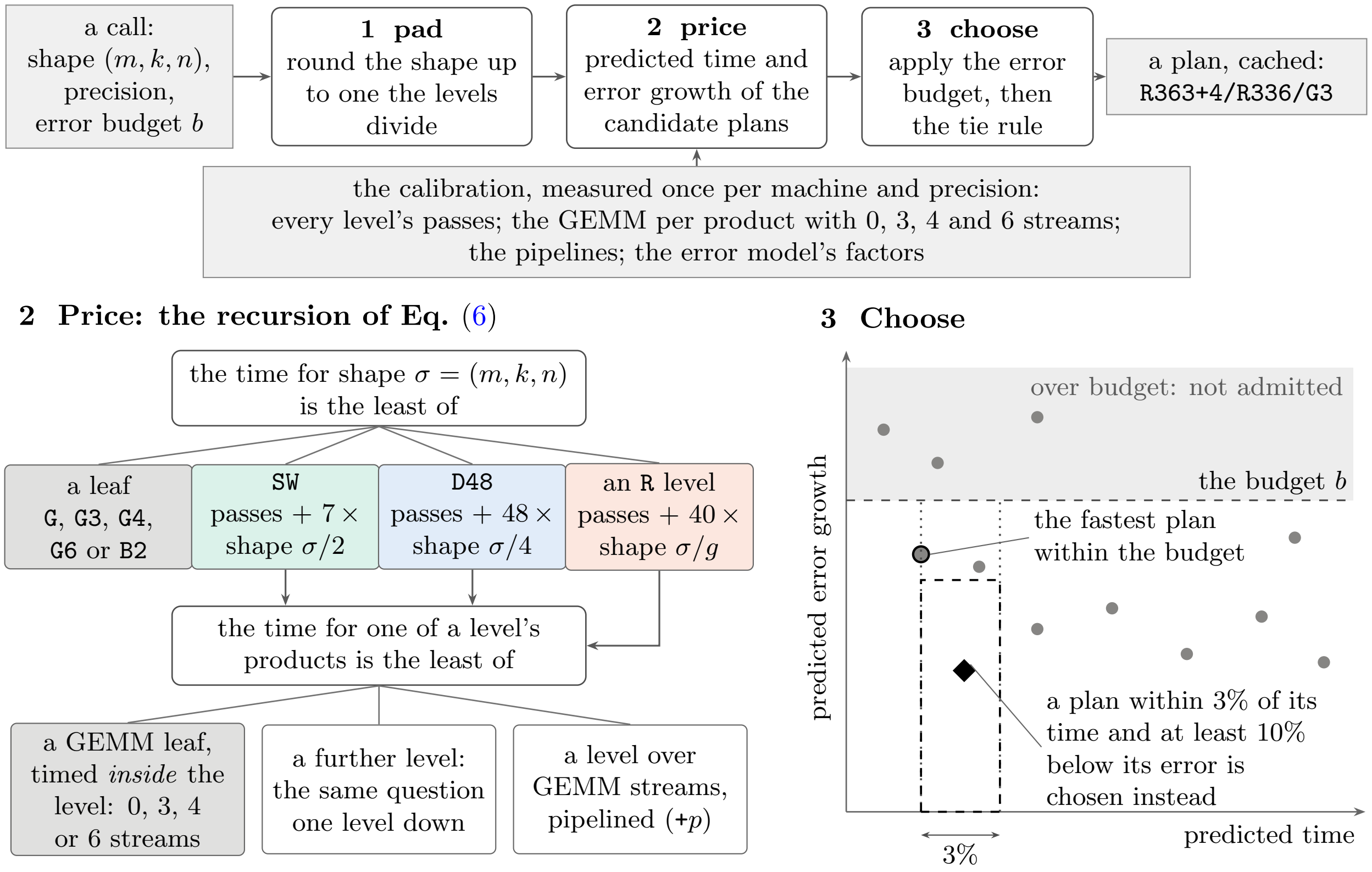

A plan is a sequence of levels from the top, ending in a leaf: \lv{G} (one
GEMM, Accelerate's threading), \lv{G3}, \lv{G4} or \lv{G6} (three, four or six
GEMM streams), or the classical
blocked GEMM \lv{B2} on its own.  A call goes through three steps
(Figure~\ref{fig:planner}): the shape is padded to shapes the levels divide,
candidate plans are priced from the calibration, and one is chosen under the
error budget.  The pricing is a dynamic program over the
first level:
\begin{equation}\label{eq:dp}
\text{cost}(\sigma) \;=\; \min\Bigl(\text{leaf}(\sigma),\ \min_X\bigl[\text{passes}_X(\sigma)
+ r_X\,\text{cost}_X(\sigma/g_X)\bigr]\Bigr),
\end{equation}
where $\sigma=(m,k,n)$ is the shape, $\sigma/g_X$ the shape of $X$'s products, and $\text{cost}_X(\tau)$ is the cheaper of a GEMM leaf \emph{priced inside level
$X$} (with $0$, $3$, $4$ or $6$ streams) and a further level on $\tau$.  Pricing
the leaf inside the level matters: a GEMM shorter than about $1$~ms takes
13--27\% longer called back to back, as inside a plan, than called alone, and
the first calibration, which timed single calls, picked a two-level plan at
$8192$ f64 that ran 12\% slower than the right one.  A level whose
sub-problems are a level over GEMM streams may instead be pipelined
(Section~\ref{sec:machine}), priced from the calibrated steady time per
sub-problem plus fill and drain.

\begin{figure}[!tbp]
\centering
\begin{tikzpicture}
\begin{axis}[report, width=9.6cm, height=9.6cm,
  xmode=log, ymode=log, xmin=0.3, xmax=1.4, ymin=0.3, ymax=1.4,
  log ticks with fixed point, xtick={0.3,0.5,0.7,1}, ytick={0.3,0.5,0.7,1},
  xlabel={predicted error, as a multiple of the GEMM's},
  ylabel={measured error, as a multiple of the GEMM's},
  legend pos=north west, clip=false]
\fill[black!6] (axis cs:0.3,0.375) -- (axis cs:1.12,1.4) -- (axis cs:1.4,1.4) -- (axis cs:1.4,1.05) -- (axis cs:0.4,0.3) -- (axis cs:0.3,0.3) -- cycle;
\draw[black!45,line width=0.5pt] (axis cs:0.3,0.3) -- (axis cs:1.4,1.4);
\addplot[only marks,mark=*,mark size=2.6pt,mark options={fill=cR,draw=white,line width=0.4pt}]
  coordinates {(0.49,0.421) (0.721,0.661) (0.674,0.553) (0.523,0.581) (0.76,0.779) (0.563,0.571) (0.715,0.73) (0.653,0.59) (0.479,0.457) (0.752,0.764) (0.564,0.587) (0.68,0.646) (0.757,0.731) (0.696,0.677) (0.514,0.428) (0.614,0.547) (0.823,0.837) (0.612,0.561) (0.61,0.646) (0.567,0.552) (0.713,0.65) (0.643,0.594) (0.473,0.427) (0.522,0.522) (0.727,0.68) (0.534,0.545) (0.415,0.384) (0.617,0.502) (0.755,0.634) (0.688,0.586) (0.51,0.404) (0.582,0.546) (0.809,0.673) (0.592,0.472)};
\addlegendentry{no $2\times2$ level}
\addplot[only marks,mark=triangle*,mark size=3.4pt,mark options={fill=cS,draw=white,line width=0.4pt}]
  coordinates {(0.639,0.693) (0.612,0.589) (1.092,1.057) (1.055,0.902) (0.798,0.653)};
\addlegendentry{a $2\times2$ level on top, f64}
\addplot[only marks,mark=triangle,mark size=3.4pt,mark options={draw=cS,line width=1pt}]
  coordinates {(0.598,0.559) (0.562,0.532) (1.026,0.971) (0.958,0.859)};
\addlegendentry{a $2\times2$ level on top, f32}
\node[anchor=north west,font=\footnotesize,align=left] at (axis cs:0.78,0.5) {predicted too high:\\ the safe side};
\end{axis}
\end{tikzpicture}
\caption{The error model against the campaign: the predicted and measured
error of each of its $53$ plans (points that coincide drawn once), as multiples of the GEMM's, at $3456$ to $13824$ in both
precisions.  On the diagonal the model is exact; the band is $\pm25\%$.
Every plan is in the band; the plans with a $2\times2$ level on top, which the first
model over-predicted in f32 by up to $2\times$, lie with the others.}
\label{fig:growth}
\end{figure}
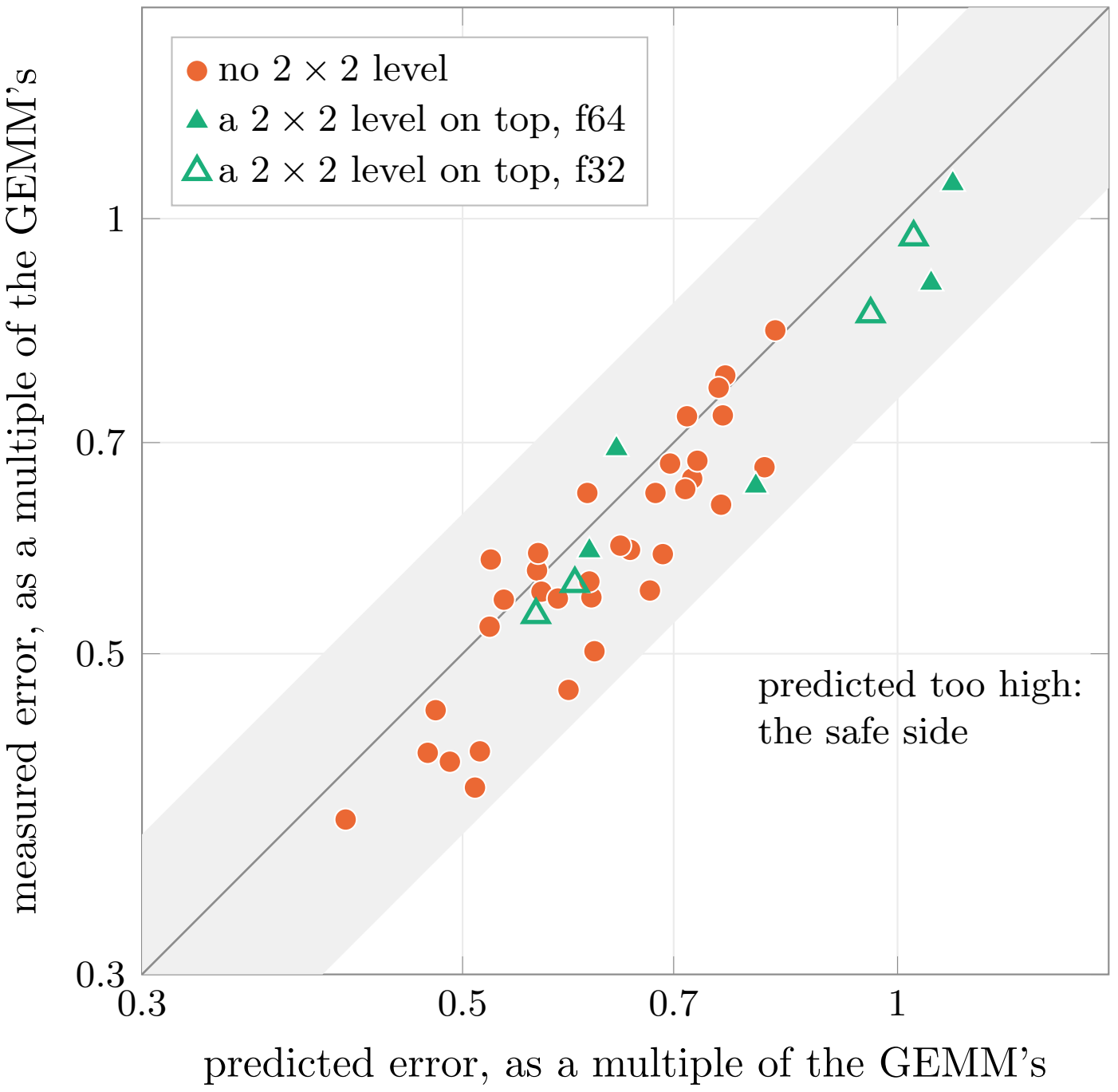

\paragraph{The error budget.}\label{par:budget}
Every call takes a number $b$: \lv{max\_growth} in the C and Rust
interfaces, \lv{auto:}$b$ in the benchmark.  For each candidate plan the
planner computes its predicted \emph{error growth}: the plan's
error by the error model \eqref{eq:errmodel}
divided by the classical GEMM's predicted error on the same problem,
so that a growth of $1$ means as accurate as the GEMM, $0.5$ twice as
accurate, $2$ twice the error.  With $b>0$ the planner returns the fastest
plan whose growth is at most $b$: $b=1$ asks for a plan no less accurate than
the GEMM it replaces, $b<1$ for a more accurate one, and $b>1$ admits plans
up to $b$ times the GEMM's error, which is where the deep plans live.
$b=0$ means the default budget, $1$; a negative $b$ means no budget, the
fastest plan whatever its error (\lv{auto:none} in the benchmark).  On the
square problems measured the fastest plan the model sees is always predicted
to be more accurate than the GEMM, so the budget's bound excludes none of
them; the two differ only where the default's tie rule (below) takes a plan
predicted more accurate, which happens at three of the six square sizes and
precisions (Table~\ref{tab:budget}).  The growth is the model's estimate (Figure~\ref{fig:growth}), within $0.90$--$1.26$ of the measured growth on the campaign's plans (Section~\ref{sec:errmodel}), so a
budget is a target rather than a guarantee, and a plan predicted just under
it can measure just over.

\paragraph{Faster first, more accurate second.}
The best plans at a size are predicted within a few percent of each other,
less than the calibration can resolve, and they can differ in error by more
than that.  So among the plans predicted within 3\% of the fastest, one whose
predicted error is at least 10\% lower wins, the most accurate of those.  The window is the calibration's
resolution.  It was $2\%$ until this campaign: at $3456$ f64 the plan
\lv{SW+4/R363/G6} was predicted $2.1\%$ faster than \lv{R363/G4}, the planner
took it, and it measured $1.3\%$ faster with $1.19$ times the error; measured
again, alternately, the two tie ($0.0703$~s for \lv{R363/G4} in both rounds
against $0.0688$ and $0.0705$, medians $0.0705$ against $0.0713$ and $0.0716$;
file~\rf{tie3}).  The threshold on the error is the error
model's resolution, which is why it follows the model.  With the first model
it had to be $25\%$: at $10\%$ that model decided three picks in an early
campaign and was wrong at two of them.  The present model resolves the
ratio between two plans of one size to about $10\%$
(Section~\ref{sec:errmodel}), and with the threshold there the rule decides
three of the six square cases, each time for a plan that measures more
accurate at the same speed (file~\rf{empicks}, two alternating rounds):
\begin{itemize}
\item $3456$ f64: \lv{R363/G4} ($0.0698$~s, $0.58$ of the GEMM's error) over
  \lv{SW+4/R363/G6} ($0.0702$~s, $0.69$);
\item $6912$ f64: \lv{R633+3/R363/G6} ($0.4451$~s, $0.61$) over
  \lv{R363+2/R336/G6} ($0.4506$~s, $0.65$), the pair of orders that the first
  model predicted equal;
\item $13824$ f32: \lv{R363+4/R363/G6} ($0.8994$~s, $0.38$) over
  \lv{R363+3/R336/G6} ($0.8942$~s, $0.55$).
\end{itemize}
The rule is part of the default budget; \lv{auto:none} returns the
plan predicted fastest, whatever its error.

\paragraph{The GPU's leaves.}
In the build with the GPU (Section~\ref{sec:machine}) the planner adds the
shared leaves \lv{GM} and \lv{G6M} on products of $576$ and up, pipelines over
them with three or four pass threads, and the \lv{K} levels on top for shapes
calibrated exactly.  Two rules came from the measurements.  A plan with the
GPU repeats only to 10--20\% (the same plan is slower right after CPU-only
work), so the window of the rule above is 8\% there instead of 3\%, and the
gain it asks for $25\%$ instead of $10\%$: at $10\%$ it took \lv{R363/GM} at
$13824$, predicted $2.5\%$ slower than \lv{R363+4/R336/G6M} and measured
$16$--$20\%$ slower (file~\rf{emreruns}).  And the
plans with the GPU are held to the predicted error of the fastest plan
without it: otherwise \lv{K32/G6M}, the classical product on both engines
with the GEMM's own error, replaces \lv{R363/G6M} at $3456$ for a tie in
time.  \texttt{FMM\_GPU=0} plans without the GPU.

\paragraph{Calibration.}
Everything the model needs is measured once per machine and precision and
cached (format~5, \texttt{calib5\_f32.txt}, \texttt{calib5\_f64.txt}, keyed by
the machine, the padding, the file format and the programs the kernels came
from): the lone GEMM and the blocked GEMM at each size; for each level and
product shape its three passes and the GEMM time per product at $0$, $3$, $4$
and $6$ streams, measured together by running the level; the pipelines; and
the $31$ error-model plans at $N_E=2304$, against the exact product on four instances.  For $n$ up to $13824$ this is $433$
entries per precision, measured in $5$ minutes in f32 and $13$ in f64.  Square shapes interpolate (log-log
between the measured sizes); rectangular shapes do not, and are either
calibrated for the shape ($118$ entries, a few seconds) or priced by
\emph{shape estimates} from the square calibration: GEMM-like entries at the
square size of equal flops, $\sqrt[3]{mkn}$ (Accelerate runs rectangular shapes
within 5\% of the square rate), and a level's passes at the square size of
equal traffic,
\begin{equation}\label{eq:equiv}
n_{\text{eq}} \;=\; \sqrt{\frac{(g_1g_2+r)\,mk + (g_2g_3+r)\,kn + (g_1g_3+r)\,mn}
{g_1g_2+g_2g_3+g_1g_3+3r}},
\end{equation}
the size at which a square problem moves the same words through the level's
passes \eqref{eq:traffic}.  On the rectangular shapes measured the estimates
are within 3.5\% of the measured times (Table~\ref{tab:rect}).

\paragraph{Sizes and shapes the levels do not divide.}
A plan runs at the smallest shape its levels divide: every dimension is
rounded up to the same multiple of $2^i3^j$ ($i\le10$, $j\le3$; 44 candidates),
the operands copied into zeroed matrices and the result copied back, priced at
$2\times$ the words at $80$~GB/s.  The planner takes the cheapest padded
candidate, so an $n$ that $3$ does not divide gets a $4\times4$ plan unless
padding to a multiple of $54$ pays.  Planning a new shape costs $0.07$~s in C
and $0.010$~s in Rust and is cached for $64$ shapes.

\paragraph{Validation.}
In the campaign of Section~\ref{sec:results}, of the $76$ measurements
without the GPU that the planner can price, $56$ are within 2\% of the
prediction and $72$ within 5.5\%; the only ones beyond 7\% are three
measurements at $13824$ f32 taken right after GPU runs, the lone GEMM's (9--11\%) and one plan's (8\%).  Measured alternately against its nearest competitors (file~\rf{empicks}), the planner's choice was the fastest of them at
four of the six square sizes and precisions and within $0.7\%$ of the fastest at the other two ($6912$ and $13824$ f32, the second by the tie rule).  The best plans at a size lie within 2--3\% of each
other, about the run-to-run spread (a plan measured twice in the campaign
repeats within 1.3\%), so the choice among them is not robust to the
calibration's noise, and the tie rule is what keeps it from flipping between
plans of different accuracy: the campaign before this one, on another calibration and with the first error model, chose a different plan at four of the six.  The GEMM itself drifts between runs
($13824$ f32: $1.55$--$1.84$~s in one campaign), which is why the speedups
are taken against its best time.

\section{Results}\label{sec:results}

\begin{table}[tp]
\centering\small
\begin{tabular}{llrrrr}
\toprule
$n$, precision & plan & time (s) & over GEMM & error & of the GEMM's\\
\midrule
6912 f32 & \lv{G} (Accelerate) & 0.1940 & 1.00 & $3.85\cdot10^{-6}$ & $1.00$\\
 & \lv{D48/G} & 0.1528 & 1.27 & $2.81\cdot10^{-6}$ & $0.73$\\
 & \lv{R336/G} & 0.1628 & 1.19 & $2.27\cdot10^{-6}$ & $0.59$\\
 & \lv{R363/G} & 0.1534 & 1.26 & $1.76\cdot10^{-6}$ & $0.46$\\
 & \lv{D48+4/D48/G6} & 0.1458 & 1.33 & $2.94\cdot10^{-6}$ & $0.76$\\
 & \lv{R363+4/D48/G6} & 0.1395 & 1.39 & $2.26\cdot10^{-6}$ & $0.59$\\
 & \lv{SW+4/R363/G6} & 0.1403 & 1.38 & $2.05\cdot10^{-6}$ & $0.53$\\
\rowcolor{cR!12} & \lv{R363+4/R633/G6} (auto) & \textbf{0.1374} & \textbf{1.41} & $2.20\cdot10^{-6}$ & $0.57$\\
\midrule
13824 f32 & \lv{G} (Accelerate) & 1.551 & 1.00 & $6.40\cdot10^{-6}$ & $1.00$\\
 & \lv{D48/G} & 1.282 & 1.21 & $4.16\cdot10^{-6}$ & $0.65$\\
 & \lv{R336/G} & 1.235 & 1.26 & $3.80\cdot10^{-6}$ & $0.59$\\
 & \lv{R363/G} & 1.250 & 1.24 & $2.73\cdot10^{-6}$ & $0.43$\\
 & \lv{R336/R363/G} & 1.171 & 1.32 & $3.34\cdot10^{-6}$ & $0.52$\\
 & \lv{D48+4/D48/G6} & 0.9517 & 1.63 & $4.35\cdot10^{-6}$ & $0.68$\\
 & \lv{R363+3/D48/G6} & 0.9178 & 1.69 & $3.49\cdot10^{-6}$ & $0.55$\\
 & \lv{R336+2/R363/G6} & 0.9179 & 1.69 & $3.34\cdot10^{-6}$ & $0.52$\\
 & \lv{R363+4/R336/G6} & 0.9084 & 1.71 & $3.53\cdot10^{-6}$ & $0.55$\\
 & \lv{SW/D48+2/D48/G6} & 1.298 & 1.19 & $5.50\cdot10^{-6}$ & $0.86$\\
 & \lv{R363+3/R336/G6} (no budget) & \textbf{0.9011} & \textbf{1.72} & $3.53\cdot10^{-6}$ & $0.55$\\
\rowcolor{cR!12} & \lv{R363+4/R363/G6} (auto) & 0.9069 & 1.71 & $2.46\cdot10^{-6}$ & $0.38$\\
\midrule
6912 f64 & \lv{G} (Accelerate) & 0.7648 & 1.00 & $6.69\cdot10^{-15}$ & $1.00$\\
 & \lv{D48/G} & 0.6078 & 1.26 & $4.89\cdot10^{-15}$ & $0.73$\\
 & \lv{R336/G} & 0.5950 & 1.29 & $4.53\cdot10^{-15}$ & $0.68$\\
 & \lv{R363/G} & 0.6440 & 1.19 & $2.86\cdot10^{-15}$ & $0.43$\\
 & \lv{R336/R363/G} & 0.5824 & 1.31 & $3.66\cdot10^{-15}$ & $0.55$\\
 & \lv{D48+4/D48/G6} & 0.4735 & 1.62 & $5.60\cdot10^{-15}$ & $0.84$\\
 & \lv{R363+4/D48/G6} & 0.4534 & 1.69 & $4.32\cdot10^{-15}$ & $0.65$\\
 & \lv{R363+3/R633/G6} & 0.4542 & 1.68 & $3.75\cdot10^{-15}$ & $0.56$\\
 & \lv{R363+2/R633/G6} & 0.4572 & 1.67 & $3.75\cdot10^{-15}$ & $0.56$\\
 & \lv{R336+2/R363/G6} & 0.4527 & 1.69 & $3.66\cdot10^{-15}$ & $0.55$\\
 & \lv{SW/R363/G6} & 0.5228 & 1.46 & $3.94\cdot10^{-15}$ & $0.59$\\
 & \lv{R363+2/R336/G6} (no budget) & \textbf{0.4490} & \textbf{1.70} & $4.32\cdot10^{-15}$ & $0.65$\\
\rowcolor{cR!12} & \lv{R633+3/R363/G6} (auto)$^\dagger$ & 0.4451 & 1.72 & $4.07\cdot10^{-15}$ & $0.61$\\
\midrule
13824 f64 & \lv{G} (Accelerate) & 6.138 & 1.00 & $1.23\cdot10^{-14}$ & $1.00$\\
 & \lv{D48/G} & 5.026 & 1.22 & $7.80\cdot10^{-15}$ & $0.63$\\
 & \lv{R336/G} & 4.982 & 1.23 & $7.21\cdot10^{-15}$ & $0.59$\\
 & \lv{R363/G} & 5.102 & 1.20 & $4.97\cdot10^{-15}$ & $0.40$\\
 & \lv{R336/R363/G} & 3.624 & 1.69 & $6.72\cdot10^{-15}$ & $0.55$\\
 & \lv{D48+4/D48/G6} & 3.784 & 1.62 & $8.28\cdot10^{-15}$ & $0.67$\\
 & \lv{R363+3/D48/G6} & 3.503 & 1.75 & $5.81\cdot10^{-15}$ & $0.47$\\
 & \lv{R363+3/R336/G3} & 3.281 & 1.87 & $6.18\cdot10^{-15}$ & $0.50$\\
 & \lv{R363+2/R336/G3} & \textbf{3.238} & \textbf{1.90} & $6.18\cdot10^{-15}$ & $0.50$\\
 & \lv{R336+3/R363/G3} & 3.286 & 1.87 & $6.72\cdot10^{-15}$ & $0.55$\\
 & \lv{SW/D48+2/D48/G6} & 3.404 & 1.80 & $1.11\cdot10^{-14}$ & $0.90$\\
 & \lv{SW/R363+3/R633/G6} & 3.269 & 1.88 & $8.03\cdot10^{-15}$ & $0.65$\\
\rowcolor{cR!12} & \lv{R363+4/R336/G3} (auto) & 3.265 & 1.88 & $6.18\cdot10^{-15}$ & $0.50$\\
\bottomrule
\end{tabular}
\caption{Square problems on the machine, best of five (f32 at $6912$) or
three interleaved rounds (file~\rf{camp}); ``over GEMM'' is Accelerate's best time in the campaign divided by the plan's, and the last column the plan's error as a multiple of the GEMM's.  ``auto'' (shaded) is the planner's choice under its default budget as the library stands, ``no budget'' its choice
without one where the two differ; in bold, the fastest plan the campaign measured at each
size.  $^\dagger$The campaign did not have this plan: its row is from two later alternating rounds against the plans around it (file~\rf{empicks}), where \lv{R363+2/R336/G6} took $0.4506$~s and \lv{D48+4/D48/G6} $0.4707$~s.}
\label{tab:results}
\end{table}

Table~\ref{tab:results} compares, at two sizes and in both precisions, the
single levels, the best plan the $4\times4$ menu offered, and the planner's
mixed plan.  What the table says:

\begin{itemize}
\item As a single level the $\mm{3\times3\times6:40}$ scheme ties the $\mm{4\times4\times4:48}$ scheme
  in time at $13824$ (from 4\% faster to 1.5\% slower; at $6912$ the orientations range from 2\%
  faster to 7\% slower, because the passes are a larger share there and they
  move different amounts per orientation) and has less error in every orientation: \lv{R336} $0.81$--$0.93$ of the $4\times4$ level's, \lv{R363} $0.58$--$0.66$, in both precisions
  ($1.76$ against $2.81\cdot10^{-6}$ at $6912$ f32, $4.97$ against
  $7.80\cdot10^{-15}$ at $13824$ f64).  The planner knows the pass costs and
  uses the orientation whose cheapest pass faces the largest operand.
\item At matched time the mixed plans are the more accurate ones.  In each of
  the four cases the planner's choice is faster than the best
  $4\times4$-only plan, \lv{D48+4/D48/G6}, by 5--16\%, and has $0.57$--$0.75$ of its error: \lv{R363+4/R633/G6} at $6912$ f32 ($2.20$
  against $2.94\cdot10^{-6}$), \lv{R363+4/R363/G6} at $13824$ f32 ($2.46$
  against $4.35\cdot10^{-6}$), \lv{R633+3/R363/G6} at $6912$ f64 ($4.07$
  against $5.60\cdot10^{-15}$) and \lv{R363+4/R336/G3} at $13824$ f64 ($6.18$
  against $8.28\cdot10^{-15}$).  These plans, at $1.4$--$1.9\times$ over the
  GEMM, also have less error than a single $4\times4$ level at
  $1.21$--$1.27\times$ (\lv{D48/G}).  Two of the four choices are the tie rule's (Section~\ref{sec:planner}).  At $13824$ f32 the fastest plan is \lv{R363+3/R336/G6}; the rule takes \lv{R363+4/R363/G6}, the most accurate of the
  fast plans, 0.6\% slower with $0.70$ of its error.  At $6912$ f64 it takes \lv{R633+3/R363/G6} over \lv{R363+2/R336/G6}, which has an \lv{R336} level at the bottom (Section~\ref{sec:errmodel}); \lv{R336+2/R363/G6} and \lv{R363+3/R633/G6} measure $8$--$10\%$ more accurate still, a difference the error model does not resolve.
\item A $2\times2$ level on top, at DPS's $\mathbb{Q}(\sqrt3)$ point
  (Section~\ref{sec:sw}), costs little accuracy.  At $13824$ f64 the plan
  \lv{SW/R363+3/R633/G6} is within $1\%$ of the fastest measured ($3.269$~s,
  $1.88\times$; $3.22$~s in the ternary trial's run of the same campaign,
  Table~\ref{tab:ternary}, which would make it the fastest) with $0.65$ of
  the GEMM's error.  At $6912$ f32 the plan \lv{SW+4/R363/G6} is $2\%$ behind
  the fastest ($0.140$~s, $1.38\times$) with the lowest error of the fast
  plans there, $0.53$ of the GEMM's, and at $3456$ f64 it is the fastest
  ($0.0706$~s, $1.38\times$, $0.69$) and the planner's choice without a
  budget; under the default budget the tie rule takes \lv{R363/G4},
  predicted $2.1\%$ slower and $18\%$ more accurate, which measured $1.3\%$
  slower with $0.84$ of the error.  Only the three-level \lv{SW/D48+2/D48/G6}
  stays near the GEMM's error ($0.86$--$1.06\times$ over $6912$ and $13824$; above it only at $6912$ f64), and it is slow except
  at $13824$ f64.
\item Every plan in the table has less error than the GEMM's own
  in both precisions (for the planner's choices $2.20$ against
  $3.85\cdot10^{-6}$ at $6912$ f32, $2.46$ against $6.40\cdot10^{-6}$ at
  $13824$ f32, $4.07$ against $6.69\cdot10^{-15}$ at $6912$ f64, $6.18$
  against $1.23\cdot10^{-14}$ at $13824$ f64): a fast multiplication that is
  also the more accurate one (Section~\ref{sec:accuracy} says why), on
  these well-scaled inputs; on nonnegative inputs the default plans keep $0.70$--$0.76$ of the GEMM's error,
  and on graded ones they need prescaling
  (Section~\ref{sec:scaled}).  The
  difference from Accelerate's product hides all of this: it mixes the GEMM's error with the plan's, so that \lv{R336/G} and \lv{R363/G} at $6912$ f64 differ from Accelerate's product by the same $6.8\cdot10^{-15}$ while their errors are $4.53$ and $2.86\cdot10^{-15}$.
\end{itemize}

\begin{table}[!htb]
\centering\small
\begin{tabular}{lllrrrr}
\toprule
$n$, prec. & plan & & time (s) & \multicolumn{2}{c}{growth over the GEMM} & error\\
\cmidrule(lr){5-6}
 & & & & predicted & measured & \\
\midrule
3456 f32 & \lv{R363/G6} & auto & 0.0204 & 0.49 & 0.42 & $1.28\cdot10^{-6}$\\
 & \lv{SW+3/R363/G6} & & 0.0241 & 0.60 & 0.56 & $1.70\cdot10^{-6}$\\
3456 f64 & \lv{R363/G4} & auto & 0.0715 & 0.52 & 0.58 & $2.76\cdot10^{-15}$\\
 & \lv{SW+4/R363/G6} & none & 0.0706 & 0.64 & 0.69 & $3.29\cdot10^{-15}$\\
6912 f32 & \lv{R363+4/R633/G6} & auto & 0.137 & 0.56 & 0.57 & $2.20\cdot10^{-6}$\\
 & \lv{SW+4/R363/G6} & & 0.140 & 0.56 & 0.53 & $2.05\cdot10^{-6}$\\
6912 f64 & \lv{R633+3/R363/G6}$^\dagger$ & auto & 0.445 & 0.60 & 0.61 & $4.07\cdot10^{-15}$\\
 & \lv{R363+2/R336/G6} & none & 0.449 & 0.68 & 0.65 & $4.32\cdot10^{-15}$\\
 & \lv{SW/R363/G6} & & 0.523 & 0.61 & 0.59 & $3.94\cdot10^{-15}$\\
13824 f32 & \lv{R363+4/R363/G6} & auto & 0.907 & 0.41 & 0.38 & $2.46\cdot10^{-6}$\\
 & \lv{R363+3/R336/G6} & none & 0.901 & 0.57 & 0.55 & $3.53\cdot10^{-6}$\\
 & \lv{SW/D48+2/D48/G6} & & 1.298 & 0.96 & 0.86 & $5.50\cdot10^{-6}$\\
13824 f64 & \lv{R363+4/R336/G3} & auto & 3.265 & 0.62 & 0.50 & $6.18\cdot10^{-15}$\\
 & \lv{SW/R363+3/R633/G6} & & 3.269 & 0.80 & 0.65 & $8.03\cdot10^{-15}$\\
 & \lv{SW/D48+2/D48/G6} & & 3.404 & 1.05 & 0.90 & $1.11\cdot10^{-14}$\\
\bottomrule
\end{tabular}
\caption{Error growth over the GEMM (the plan's error divided by the GEMM's),
predicted by the error model \eqref{eq:errmodel} and measured, for the planner's choices (auto: the default budget,
predicted growth at most $1$; none: no budget, where the choice differs)
and the plans with a $2\times2$ level on top, from the campaign of Table~\ref{tab:results} ($^\dagger$as there).}
\label{tab:budget}
\end{table}

The planner's choices by size under the default budget, from the calibration
at $13824$: f32 \lv{G} at $1728$, \lv{R363/G6} at $3456$,
\lv{R363+4/R633/G6} at $6912$, \lv{R363+4/R363/G6} at $13824$; f64
\lv{R363/G6} at $1728$, \lv{R363/G4} at $3456$, \lv{R633+3/R363/G6} at
$6912$, \lv{R363+4/R336/G3} at $13824$.  Sizes that $3$ does not divide keep
their $4\times4$ plans (at $8192$ \lv{D48+4/D48/G6} in f32, $1.44\times$, and
\lv{D48+3/D48/G6} in f64, $1.60\times$; at $9000$ padded to $9008$, $1.42\times$ and
$1.62\times$).

\begin{table}[!b]
\centering\small
\begin{tabular}{llllrr}
\toprule
$m\times k\times n$ & calibration & prec. & plan & predicted / measured (s) & over GEMM\\
\midrule
$3456\times3456\times1728$ & for the shape & f32 & \lv{R363/G6} & 0.0112 / 0.0111 & 1.09\\
 & & f64 & \lv{R363/G6} & 0.0359 / 0.0356 & 1.33\\
$1000\times3000\times2000$ & estimated & f32 & \lv{G} & -- & 1.00\\
 & & f64 & \lv{G} & -- & 1.00\\
$2048\times4096\times1024$ & estimated & f32 & \lv{G} & -- & 1.00\\
 & & f64 & \lv{D48/G6} & 0.0177 / 0.0171 & 1.14\\
$6912\times6912\times3456$ & estimated & f32 & \lv{SW+4/D48/G6} & 0.0778 / 0.0778 & 1.28\\
 & & f64 & \lv{D48+3/R363/G6} & 0.2578 / 0.2518 & 1.53\\
\bottomrule
\end{tabular}
\caption{Rectangular problems through the library's entry points
(\texttt{test\_fmm} in file~\rf{emreruns}): the planner's plan under the default budget, its predicted
and measured time, and the speedup over Accelerate's GEMM on the same shape.
``Estimated'' shapes are priced from the square calibration by
\eqref{eq:equiv}.}
\label{tab:rect}
\end{table}

Rectangular problems (Table~\ref{tab:rect}) go through the same planner over
$(m,k,n)$: a level applies when its grid divides the current shape, so a
$6912\times6912\times3456$ product gets a $4\times4$ level whose products are
$1728\times1728\times864$ and then a $3\times6\times3$ level in f64.  Where a
level pays, the shape estimates are within 3.5\% of the measurements; where none does, the planner correctly returns the GEMM (at $1000\times3000\times2000$ in f64 a lone $2\times2$ level would be $4\%$ faster, but the error model puts it just above the GEMM's error, and the default budget excludes it; a
version of the estimates without that case padded $1000\times3000\times2000$
to $3072^3$, 30 times slower than Accelerate).

\subsection{Against ESSL's Strassen--Winograd}\label{sec:essl}

The one fast matrix multiplication a vendor ships as a library routine is
IBM's ESSL \texttt{\_GEMMS} \cite{ESSL}: ``Winograd's variation of the
Strassen's algorithm'', applied ``up to four times'', each step halving the
dimensions, over ESSL's own GEMM, with the number of steps chosen from the
size by an unpublished rule.  ESSL is closed and runs on IBM's Power systems,
not on the machine, so we implemented the published algorithm over
Accelerate's GEMM (\path{gemms_impl.h}, benchmark method
\texttt{gemms:}$d$): Winograd's variant with its $15$ block additions, in the
usual schedule with two temporaries per level (as in Douglas et al.'s GEMMW
\cite{DHSS94} and in Boyer et al.\ \cite{BDPZ09}), each addition a
separate parallel pass over memory and each product one call to the GEMM;
at the last level two additions ride in the GEMM's $\beta$, in place of
ESSL's unpublished ``minor changes for tuning''.  It is a conventional
implementation of the algorithm, not ESSL's code, and its leaves are
Accelerate's, not ESSL's.  Since the depth rule is unpublished we measured
one to four levels and credit the algorithm with its fastest depth at each
size.  All methods ran interleaved in one run, on the same inputs, against
the exact product (\path{run_essl_gemms.sh}, file~\rf{essl}).

\begin{table}[!tbp]
\centering\footnotesize
\newcommand{\E}[2]{$#1{\cdot}10^{#2}$}
\setlength{\tabcolsep}{3.2pt}
\begin{tabular}{llrrlrrlrrr}
\toprule
 & & \multicolumn{2}{c}{GEMM} & \multicolumn{3}{c}{\texttt{\_GEMMS}, fastest depth} & \multicolumn{3}{c}{the planner's plan} & ours vs\\
\cmidrule(lr){3-4}\cmidrule(lr){5-7}\cmidrule(lr){8-10}
$n$ & prec. & time (s) & error & $d$ & time (s) & error & plan & time (s) & error & \texttt{\_GEMMS}\\
\midrule
3456  & f32 & 0.0245 & \E{3.04}{-6}  & 1 & 0.0248 & \E{5.43}{-6}  & \lv{R363/G6}        & 0.0203 & \E{1.28}{-6}  & $1.22\times$, $4.2\times$\\
4096  & f32 & 0.0416 & \E{3.00}{-6}  & 1 & 0.0416 & \E{5.35}{-6}  & \lv{D48/G6}         & 0.0345 & \E{2.05}{-6}  & $1.21\times$, $2.6\times$\\
6912  & f32 & 0.1949 & \E{3.85}{-6}  & 1 & 0.1875 & \E{6.83}{-6}  & \lv{R363+4/R633/G6} & 0.1361 & \E{2.20}{-6}  & $1.38\times$, $3.1\times$\\
8192  & f32 & 0.3264 & \E{4.61}{-6}  & 1 & 0.3123 & \E{8.07}{-6}  & \lv{D48+4/D48/G6}   & 0.2279 & \E{3.39}{-6}  & $1.37\times$, $2.4\times$\\
13824 & f32 & 1.5509 & \E{6.40}{-6}  & 2 & 1.3790 & \E{1.89}{-5}  & \lv{R363+4/R363/G6} & 0.8939 & \E{2.46}{-6}  & $1.54\times$, $7.7\times$\\
\midrule
3456  & f64 & 0.0964 & \E{4.75}{-15} & 1 & 0.0923 & \E{9.95}{-15} & \lv{R363/G4}        & 0.0701 & \E{2.76}{-15} & $1.32\times$, $3.6\times$\\
4096  & f64 & 0.1614 & \E{6.75}{-15} & 1 & 0.1552 & \E{9.69}{-15} & \lv{D48+4/D48/G6}   & 0.1158 & \E{5.15}{-15} & $1.34\times$, $1.9\times$\\
6912  & f64 & 0.7716 & \E{6.69}{-15} & 2 & 0.6709 & \E{2.12}{-14} & \lv{R633+3/R363/G6} & 0.4453 & \E{4.07}{-15} & $1.51\times$, $5.2\times$\\
8192  & f64 & 1.2829 & \E{8.29}{-15} & 2 & 1.1240 & \E{2.34}{-14} & \lv{D48+3/D48/G6}   & 0.7940 & \E{6.54}{-15} & $1.42\times$, $3.6\times$\\
13824 & f64 & 6.2008 & \E{1.23}{-14} & 3 & 4.8106 & \E{6.01}{-14} & \lv{R363+4/R336/G3} & 3.2649 & \E{6.18}{-15} & $1.47\times$, $9.7\times$\\
\bottomrule
\end{tabular}
\caption{The planner's default plan against ESSL's \texttt{\_GEMMS}
algorithm at its fastest depth $d$ and Accelerate's GEMM, uniform inputs,
best of seven (3456, 4096), five (6912, 8192) or three (13824) rounds.  Last
column: how much faster our plan is, and the algorithm's error as a multiple of the plan's.  The \texttt{\_GEMMS} columns are our implementation of ESSL's published algorithm, not ESSL's code.  The
GEMM's times are within 2\% of its best in the campaign of
Table~\ref{tab:results}; ratios to it are taken within each row's run.  The
$3456$ f64, $6912$ f64 and $13824$ f32 rows were measured again after the planner's tie rule changed its choice there
(Section~\ref{sec:planner}; file~\rf{emreruns}).}
\label{tab:essl}
\end{table}

On well-scaled inputs (Table~\ref{tab:essl}) the planner's plan is
$1.21$--$1.54\times$ faster than the algorithm at its fastest depth, and the algorithm has $1.9$--$9.7$ times the plan's error, at every size and in both precisions.  At its
fastest depth the algorithm is $0.99$--$1.29\times$ as fast as the GEMM in
this run with $1.4$--$4.9\times$ its error: every Winograd level roughly
doubles the error (one to four levels at $13824$ f64:
$2.1$, $3.8$, $6.0$ and $11.5\cdot10^{-14}$), where an accurate level lowers
it (Section~\ref{sec:accuracy}), and the deeper levels lose time, because
their $7^d$ leaf products run one after another and are small for the GEMM
(four levels at $13824$ f32 take $2.24$~s, more than the GEMM's $1.56$),
where the plans run their leaves in streams (Section~\ref{sec:machine}).
The best depth grows with the size, from one level at $3456$ to three at
$13824$ f64, which fits ESSL's choosing it from the size.

\begin{table}[!tbp]
\centering\footnotesize
\begin{tabular}{llrrrrrrrrr}
\toprule
 & & \multicolumn{3}{c}{one level} & \multicolumn{3}{c}{two levels} & \multicolumn{3}{c}{three levels}\\
\cmidrule(lr){3-5}\cmidrule(lr){6-8}\cmidrule(lr){9-11}
$n$ & prec. & fused & separate & ratio & fused & separate & ratio & fused & separate & ratio\\
\midrule
8192  & f32 & 0.309 & 0.316 & $1.02$ & 0.293 & 0.322 & $1.10$ & 0.353 & 0.466 & $1.32$\\
13824 & f32 & 1.447 & 1.446 & $1.00$ & 1.353 & 1.368 & $1.01$ & 1.244 & 1.387 & $1.12$\\
8192  & f64 & 1.171 & 1.183 & $1.01$ & 1.103 & 1.125 & $1.02$ & 1.079 & 1.215 & $1.13$\\
13824 & f64 & 5.582 & 5.576 & $1.00$ & 4.971 & 5.038 & $1.01$ & 4.613 & 4.781 & $1.04$\\
\bottomrule
\end{tabular}
\caption{What fusing alone is worth: times in seconds of fused $2\times2$
levels (\lv{SW/G}, \lv{SW/SW/G}, \lv{SW/SW/SW/G}) and of the
one-operation-at-a-time Strassen--Winograd implementation at the same depth,
both over one GEMM call per product, and the ratio of the separate time to
the fused one.  Best of five ($8192$) or three ($13824$) rounds, file~\rf{fused}.}
\label{tab:fused}
\end{table}

\paragraph{How much of the gap is fusing.}
Little.  Table~\ref{tab:fused} holds everything else fixed: fused $2\times2$
levels against the one-at-a-time implementation at the same depth, the same
class of scheme, and the same leaf.  At one level the two are equal within
about 2\%; at two levels the fused plans are 1--10\% faster and at three
4--32\%, the f32 cases the larger.  Best depth against best depth the fused
$2\times2$ plans are 4--10\% faster.  The reason is that Winograd's variant
is already frugal: its $15$ additions, one at a time, move $45$ blocks per
level where the three fused passes move $33$, and the passes are a small
part of a shallow plan.  So of the $1.37$--$1.54\times$ that the planner's
plans have over the algorithm at these two sizes (Table~\ref{tab:essl}),
fusing accounts for $1.04$--$1.10\times$; the rest comes from the schemes
(a $4\times4$ or $3\times3\times6$ level saves a quarter of the products in
one step, which a $2\times2$ level needs two steps and twice the passes to
match) and from the streams and pipelines.  What fusing buys is those
schemes.  One at a time, the $262$ operations of a $4\times4$ level's
programs would be up to $786$ block moves against the $192$ of its three
fused passes, four times the traffic, on passes that are already 2--5\% of
a one-level plan at $13824$; we have not built that implementation, so this
is a count, not a measurement.

\paragraph{What the comparison does not say.}
The $1.21$--$1.54\times$ is between the planner's plans and a conventional
implementation of Strassen--Winograd: one block operation at a time, one GEMM
call per product.  It is not the advantage of the
$\mm{4\times4\times4:48}$ scheme over Winograd's variant at twice the depth.
That advantage is small, as the product counts say: one $4\times4$ level
makes $48$ products where two $2\times2$ levels make $49$.  In this library,
at $13824$, one \lv{D48} level over one GEMM call per product takes
$1.282$~s in f32 and $5.026$~s in f64 (\lv{D48/G}, Table~\ref{tab:results})
and two fused $2\times2$ levels take $1.353$~s and $4.971$~s (\lv{SW/SW/G},
Table~\ref{tab:fused}): $5\%$ apart in f32 and $1\%$ in f64, from two
different runs.  The exact-arithmetic measurements of \cite{DPSST26} on two
x86 processors find the same (Figures~1 and~2 of its current source):
Winograd's variant at two, four and six levels is within a few percent of
one, two and three levels of $\mm{4\times4\times4:48}$.  So the two papers
agree, and the plans' margin over the conventional algorithm comes from
elsewhere: the rectangular levels ($40$ products where the classical
algorithm makes $54$), the leaves run in streams and the passes in pipelines,
the fused $2\times2$ levels the planner itself uses where they are fastest,
and, for the error, the accurate points of every scheme.

\begin{table}[!tbp]
\centering\footnotesize
\newcommand{\E}[2]{$#1{\cdot}10^{#2}$}
\begin{tabular}{lllrrrrr}
\toprule
 & & & & \multicolumn{2}{c}{unscaled} & \multicolumn{2}{c}{prescaled}\\
\cmidrule(lr){5-6}\cmidrule(lr){7-8}
prec. & inputs & measure & GEMM & \texttt{\_GEMMS} & plan & \texttt{\_GEMMS} & plan\\
\midrule
f32 & nonneg  & max   & \E{6.23}{-6} & \E{6.63}{-6} & \E{4.37}{-6} & \E{6.63}{-6} & \E{5.60}{-6}\\
    & outer:6 & cw    & \E{3.97}{-7} & \E{2.54}{-1} & \E{6.01}{2}  & \E{7.00}{-7} & \E{3.27}{-7}\\
    & inner:6 & max   & \E{3.81}{-6} & \E{4.06}{-3} & \E{6.51}{-2} & \E{7.13}{-6} & \E{2.71}{-6}\\
    & cross:6 & cw    & \E{3.98}{-7} & \E{4.19}{-1} & \E{9.96}{-2} & \E{4.19}{-1} & \E{1.23}{-1}\\
\midrule
f64 & nonneg  & max   & \E{1.18}{-14} & \E{6.10}{-15} & \E{8.96}{-15} & \E{6.10}{-15} & \E{8.96}{-15}\\
    & outer:6 & cw    & \E{6.60}{-16} & \E{3.41}{-7}  & \E{1.16}{-6}  & \E{2.64}{-15} & \E{6.42}{-16}\\
    & inner:6 & max   & \E{6.86}{-15} & \E{1.98}{-10} & \E{1.37}{-10} & \E{3.20}{-14} & \E{4.89}{-15}\\
    & cross:6 & cw    & \E{1.62}{-15} & \E{1.50}{-9}  & \E{1.73}{-10} & \E{1.50}{-9}  & \E{1.73}{-10}\\
\bottomrule
\end{tabular}
\caption{ESSL's \texttt{\_GEMMS} algorithm at its fastest depth at $6912$
(one level in f32, two in f64) against the planner's default plan on the
badly scaled inputs of Section~\ref{sec:scaled}, without and with the same
power-of-two prescaling for both; for each class the measure that shows its effect (max: the max-norm error; cw: the componentwise error).}
\label{tab:essl-scaled}
\end{table}

On the badly scaled inputs of Section~\ref{sec:scaled}
(Table~\ref{tab:essl-scaled}) neither is safe without prescaling, and the
comparison is mixed.  On graded rows and columns one Winograd level loses
less than the plan's two accurate levels do, componentwise $0.25$ against
$6\cdot10^2$ in f32 (at two levels $142$, at three $3.6\cdot10^3$), and on inner
scaling one level has $4\cdot10^{-3}$ against $7\cdot10^{-2}$ (two levels,
$0.11$, no better than the plan): the damage grows with the depth more than with
the scheme.  On cross-structured inputs the plan is better than the
algorithm at every depth ($0.10$ against $0.42$--$2.5$ in f32).  On
nonnegative inputs the algorithm wins in f64: two Winograd levels have
$6.1\cdot10^{-15}$, half the GEMM's $1.18\cdot10^{-14}$ and below the plan's
$8.96\cdot10^{-15}$, and four levels $2.6\cdot10^{-15}$ (in f32
$1.2\cdot10^{-6}$ at four levels, against the plan's $4.4\cdot10^{-6}$); we
have not explained this.  With prescaling, which repairs the graded and inner
cases for both, the plan has less error than the algorithm on those
inputs in both precisions, and less than the GEMM, and the comparison returns to that of
Table~\ref{tab:essl}.

\subsection{The GPU, alone and as a second engine}\label{sec:gpu}

The other engine on the machine is the 20-core GPU, in f32 only: Metal, MPS
and MPSGraph have no 64-bit floating type.  We measured Apple's own GEMM paths
on it (\path{gpu_gemm.m}, Appendix~\ref{sec:glossary} for the names):
MPSGraph's matrix multiplication, the faster of the two, and the older
\lv{MPSMatrixMultiplication}, within 3\% of it at $13824$ and 10\% slower at
$3456$.  The inputs are the benchmark's, wrapped as shared Metal buffers
without a copy (the memory is unified), and the time is the wall clock from
encoding to completion.  Neither MLX nor PyTorch was installed, so these are
the vendor's GEMMs, as Accelerate is on the CPU side.  The library uses the
GPU differently, as a second engine inside a level
(Section~\ref{sec:machine}): the leaf \lv{G6M} shares a level's products
between the CPU's GEMM streams and the GPU, and the levels \lv{K32} and
\lv{K64} share the classical product the same way with no scheme at all,
which measures what the hardware alone buys.

\begin{table}[t]
\centering\footnotesize
\begin{tabular}{lrrrrlrr}
\toprule
 & Accelerate & \multicolumn{2}{c}{GPU alone} & classical shared & \multicolumn{3}{c}{the planner's plan with the GPU}\\
\cmidrule(lr){3-4}\cmidrule(lr){6-8}
$n$ (f32) & time (s) & time (s) & TFLOP/s & time (s) & plan & time (s) & error\\
\midrule
3456 & 0.0244 & 0.013 & 6.2 & 0.0136 & \lv{R363/G6M} & 0.0130 & $1.28\cdot10^{-6}$\\
6912 & 0.194 & 0.112 & 5.9 & 0.0879 & \lv{R363/G6M} & 0.0670 & $1.76\cdot10^{-6}$\\
13824 & 1.551 & 0.873 & 6.1 & 0.801 & \lv{R363+4/R336/G6M} & 0.505 & $3.53\cdot10^{-6}$\\
\bottomrule
\end{tabular}
\caption{The GPU in f32 (file~\rf{camp},
step (2g)): best of two rounds of five to seven (for the GPU alone, the
better median of the two); Accelerate's time is its
best in the campaign.  The GPU alone and the classical product shared between
the two engines (\lv{K32/G6M}, \lv{K64/G6M}, \lv{K32/GM}) have the GEMM's
error, $3.04$, $3.85$ and $6.40\cdot10^{-6}$; the plans with the GPU have
that of the same plans on the CPU.}
\label{tab:gpu}
\end{table}

What Table~\ref{tab:gpu} says:
\begin{itemize}
\item The GPU's f32 product is bit for bit Accelerate's at every size.  Both
  accumulate each output in increasing $k$ with fused multiply-adds and
  neither splits the inner dimension, so both equal the sequential FMA chain
  (64 sampled outputs recomputed as that chain agree with both), and the
  GPU's error is the GEMM's.  This is what lets the library share a level's
  products between the two engines without changing a bit of the result.
\item Alone, the GPU is $1.6\times$ ahead of the CPU plan at $3456$ and
  $1.2\times$ at $6912$, and ties it at $13824$ ($0.873$ against $0.901$~s),
  at $5.9$--$6.2$~TFLOP/s; it has $1.7$--$2.4\times$ the
  CPU plan's error.
\item As a second engine it pays at every size: the planner's plans with the
  GPU run $1.9$, $2.9$ and $3.1\times$ faster than Accelerate, and $1.0$,
  $1.7$ and $1.7\times$ faster than the GPU alone; each of those two has $1.8$--$2.4$ times the plans' error.  The hardware alone, the classical product shared
  between the two engines, gives $1.8$--$2.2\times$ over Accelerate with the
  GEMM's error; the accurate level on top adds $1.05$, $1.31$ and $1.59\times$
  to that and halves the error.  With a leaf this fast the best plans have
  one level or two, and the single level \lv{R363/G6M} at $13824$, $0.52$~s
  with $2.73\cdot10^{-6}$, is within 3\% of the fastest.
\item Plans with the GPU repeat only to 10--20\%: the two engines share the
  package's power and thermal budget, and the same plan runs slower right
  after CPU-only work (at $13824$ the planner's choice measured $0.505$ to
  $0.557$~s within the campaign, the shared classical product $0.80$~s where the campaign before measured $0.74$, and Accelerate's GEMM $1.55$~s alone but
  $1.80$--$1.84$~s in median after GPU runs).  The planner's window for ties is wider there
  (Section~\ref{sec:planner}).
\item The GPU's f16-input path (f32 result) is only 14\% faster than its f32
  at $13824$, and its error, $2.5$--$2.8\cdot10^{-4}$, is the rounding of the
  inputs to 11 bits (file~\rf{gpugemm}).
\item In f64 the GPU does not compete, and the plans' $1.70$--$1.88\times$
  over Accelerate's dgemm at $6912$ and $13824$ stands alone.
\end{itemize}

\section{The ternary trial: cheap passes against accurate ones}\label{sec:ternary}

Perminov's catalogue \cite{Perminov,Perminov25} lists, beside the ranks, the
schemes with coefficients in $\{-1,0,1\}$ whose additive complexity his
parallel common-subexpression search reduced; for $3\times3\times6$ it has four
rank-$42$ ternary schemes reduced to $139$, $140$, $145$ and $150$ additions
for the three linear forms together (file
\texttt{3x3x6\_m42\_cr139\_cn287\_ZT\_reduced.json} and its siblings; the
naive counts are $287$--$369$).  Against DPS's accurate rank-$40$ scheme,
whose nine optimized programs take $374$, $383$ and $353$ operations per
orientation, this is the opposite design: two products more ($\kappa=0.778$ against $0.741$), two more streams per pass ($51/60/60$), and a third of the
pass arithmetic.  Is the arithmetic worth anything once the passes are fused,
and what does the accuracy cost?

\paragraph{The scheme.}
The reduced JSON is a straight-line program already (fresh variables for the
common subexpressions, then the forms); each of the four schemes was expanded
to its matrices and checked exactly against the Brent equations.  The
$139$-addition scheme is the most accurate of the four by far (amplification
ratio $5.07$ in the $3\times3\times6$ orientation against $7.6$--$9.5$ for the
others), so it is the one tried; Table~\ref{tab:levels} gives its three
orientations.  Its programs are Perminov's forms for the three maps the
$3\times3\times6$ orientation needs ($L$: $33$ additions, $R$: $39$, $P$: $67$),
and for the other orientations the same forms re-indexed where an
orientation reuses a map and transposed by PLinOpt's \texttt{transpozer}
where it needs the transposed map ($3\times6\times3$: $39$, $43{+}6$ negations,
$66$; $6\times3\times3$: $43{+}6$, $33$, $63$), all nine checked by
\texttt{SLPchecker} against the matrices.  PLinOpt's optimizer (the fork's
\texttt{-s 1 -Q -O 500}) on the same nine matrices finds $34$, $41$, $71$,
$41$, $45$, $67$, $45$, $34$, $66$ additions: one to four more than Perminov's
reduction on every matrix, in $0.2$~s each.  The levels \lv{T336}, \lv{T363},
\lv{T633} went into the C library beside the R levels, outside the planner's
menu, and every plan containing them computes the product (3456, both
precisions, including pipelined plans).

\begin{table}[t]
\centering\footnotesize
\begin{tabular}{llrrlrrl}
\toprule
$n$, prec. & plan & R (s) & T (s) & T/R & R error & T error & T/R\\
\midrule
13824 f32 & \lv{X336/G} & 1.221 & 1.292 & 1.06 & 3.80e-6 & 1.28e-5 & 3.4\\
 & \lv{X363/G} & 1.269 & 1.311 & 1.03 & 2.73e-6 & 6.25e-6 & 2.3\\
 & \lv{X633/G} & 1.264 & 1.331 & 1.05 & 3.95e-6 & 1.01e-5 & 2.6\\
 & \lv{X363+4/D48/G6} & 0.912 & 0.950 & 1.04 & 3.49e-6 & 6.23e-6 & 1.8\\
 & \lv{X363+4/X363/G6} & 0.902 & 0.987 & 1.09 & 2.46e-6 & 9.04e-6 & 3.7\\
\midrule
13824 f64 & \lv{X336/G} & 4.783 & 5.207 & 1.09 & 7.21e-15 & 2.49e-14 & 3.5\\
 & \lv{X363/G} & 4.902 & 5.094 & 1.04 & 4.97e-15 & 1.25e-14 & 2.5\\
 & \lv{SW/X363+3/X633/G6} & 3.223 & 3.543 & 1.10 & 8.03e-15 & 3.55e-14 & 4.4\\
 & \lv{X363+3/X633/G6} & 5.624 & 6.237 & 1.11 & 6.37e-15 & 2.51e-14 & 3.9\\
\midrule
6912 f32 & \lv{X336/G} & 0.159 & 0.165 & 1.04 & 2.27e-6 & 8.00e-6 & 3.5\\
 & \lv{X363/G} & 0.149 & 0.163 & 1.10 & 1.76e-6 & 3.87e-6 & 2.2\\
 & \lv{X363+4/D48/G6} & 0.136 & 0.142 & 1.04 & 2.26e-6 & 4.07e-6 & 1.8\\
 & \lv{X363+3/X633/G6} & 0.148 & 0.164 & 1.11 & 2.20e-6 & 8.72e-6 & 4.0\\
\midrule
6912 f64 & \lv{X336/G} & 0.585 & 0.612 & 1.05 & 4.53e-15 & 1.44e-14 & 3.2\\
 & \lv{X363+4/D48/G6} & 0.450 & 0.470 & 1.04 & 4.32e-15 & 7.13e-15 & 1.7\\
 & \lv{X363+3/X633/G6} & 0.449 & 0.492 & 1.10 & 3.75e-15 & 1.60e-14 & 4.3\\
\bottomrule
\end{tabular}
\caption{The ternary $\mm{3\times3\times6:42}$ levels (T) against the accurate
$\mm{3\times3\times6:40}$ levels (R) on the same plans
(file~\rf{camp}, step (3); \lv{X} stands for \lv{R} or
\lv{T}): best of five (f32) or three (f64) interleaved rounds; the f64
errors from a second run on the same inputs
(file~\rf{t42}).  The GEMM's own errors are
$6.40\cdot10^{-6}$ and $3.85\cdot10^{-6}$ in f32, $1.23\cdot10^{-14}$ and
$6.69\cdot10^{-15}$ in f64.}
\label{tab:ternary}
\end{table}

\paragraph{Measured.}
Table~\ref{tab:ternary} answers both questions.  The pass arithmetic is worth
nothing: the T levels' passes take the R levels' times (at $13824$ f32,
$L/R/P$ $0.025/0.015/0.016$~s against $0.024/0.015/0.017$; at $6912$,
$0.006/0.004/0.004$ against $0.006/0.004/0.005$), because a fused pass is bound by the blocks it
moves and the T level moves two more per pass.  The two extra products cost
what the rank says: the GEMM part of a single level is 4--8\% longer
($42/40=1.05$), a single T level is 3--10\% slower than its R counterpart, a
plan with one T level 4\% slower, and a plan with two T levels 9--11\%
slower, at every size and precision.  The accuracy cost is the
amplification: against the exact product, $2.2$--$3.5\times$ the R level's
error as a single level in both precisions, $3.7$--$4.4\times$ in plans with
two T levels and $1.7$--$1.8\times$ with one T level over a $4\times4$ level,
in line with the ratio of the amplification ratios
($5.07/2.36 = 2.1$ for the $3\times3\times6$ orientation, the measured
$2.2$--$3.5$ including the passes' own roundings).  In f32 the T plans'
errors are at or above the GEMM's own (\lv{T336/G} $1.28\cdot10^{-5}$ against
$6.40\cdot10^{-6}$ at $13824$; the best, \lv{T363/G} and \lv{T363+4/D48/G6},
level with it at $6.2\cdot10^{-6}$), and in f64 every T plan's error is
above the GEMM's, where every R plan's is well below it in both precisions.
The planner's menu was left as it was: a level that is slower and less
accurate than one already there cannot be chosen.

\paragraph{What it shows.}
The trial is the cleanest confirmation of the cost model's premise.  Two
schemes of the same shape, with the same pass structure, differ in rank by
two and in pass arithmetic by a factor of $2.7$; the measured difference is
the rank's, to the percent, and none of the arithmetic's.  For a fused-pass
library the objectives of a scheme search are rank and amplification, in
that order, and additive complexity is not among them.  The caveat is the
one stated in Section~\ref{sec:levels}: arithmetic shows when a program is
wasteful enough to make a pass compute-bound, as the $18$-operation-per-output
direct sums were; Perminov's $139$ additions and DPS's optimized programs are
both well inside the bandwidth-bound regime.  Whether the class of the
ternary scheme contains an accurate point is a separate question, and the
answer is no: over $GL_3\times GL_3\times GL_6$ the class's amplification
ratio is at least
$2.825$ in the $3\times3\times6$ orientation and $2.831$ in $3\times6\times3$
(eight starts each), the $4\times4$ class's $2\sqrt2$ and 26\% above the
accurate $\mm{3\times3\times6:40}$ class's $2.242$; the $139$-addition point
sits $1.8\times$ above its own floor, and its $\ginf$ floors ($5.12$ and $4.82$)
are likewise above the accurate class's ($6.37$ and $3.90$) in the orientation
that matters.  Even at its best point the class would match the
$\mm{4\times4\times4:48}$ scheme's accuracy with two products more than the accurate
$\mm{3\times3\times6:40}$ scheme.

\section{Discussion}\label{sec:disc}

The mixed approach rests on two measurements that contradict the usual
accounting.  A fused level's cost is the blocks it moves, so the right
measure of a scheme is its saving per block moved and its stream count, not
its rank or the length of its programs; and the accuracy of a level is set by
the point in its orbit and, at depth, by the roundings of its $P$ program,
not by the operation count.  Under these measures the menu is short: two
accurate schemes and an accurate $\mm{2\times2\times2:7}$ scheme for the top of a plan.  A
scheme enters the menu only if a plan containing it beats every plan without
it on the machine, and the planner decides that per shape and precision from
measurements, which is why the library can take a new level (or a new machine) without
any change but a calibration.  Its default objective is time at the GEMM's
own accuracy or better, and on the machine that objective costs no time:
with DPS's accurate point in the $2\times2$ level the fastest plans are
within it (Section~\ref{sec:sw}), and where plans tie in time the default
takes the one predicted clearly more accurate.  That accuracy is the
max-norm accuracy on well-scaled inputs, which is what the calibration
measures; componentwise, on graded inputs, the plans need prescaling, which
costs up to $4\%$, and on cross-structured inputs nothing we have repairs them
(Section~\ref{sec:scaled}).

The limits are also measured.  The GEMM is 85--95\% of every plan, so no menu
can take more than the products it saves, and below $n\approx3000$ nothing
pays in f32 ($1728$ in f64).  A level's passes must not fall behind the
streams it feeds: six streams of $1024$-blocks in f64 are the most sensitive
to pass traffic, and pipelining gains nothing there.  Shapes above
$4\times4$ cannot help because of the stream count, whatever their rank, and
Kronecker composites because they inherit their pieces' amplification and
multiply their streams.  And the ternary trial (Section~\ref{sec:ternary})
closes the remaining door on the cheap side: when the passes are already
bandwidth-bound, a program with a third of the additions buys nothing, while
two extra products and $2.1\times$ the amplification per level cost about 5\% of the
time per level and $1.7$--$4.4\times$ the error.

\paragraph{What could beat this menu.}
After the campaign we looked for a faster level at the same accuracy without
more hardware, and found the boundary rather than a level
(\path{doc/fastmm_plan.md}, \path{matmul/levels/}, \path{matmul/r346/}).
\begin{itemize}
\item \emph{How much there is to gain.}  At $13824$ the planner's plans spend
  4--8\% of their time in the top level's passes, and their pipelines run
  4--9\% above their GEMM-only time: every pass cost together is 8\% (f64) to
  15\% (f32) of a plan, and no scheme or mixing can gain more at this depth.
  A third level is closed by memory, not by arithmetic: the two-level f64
  plans already move about $112$~GB through a memory system of $130$~GB/s,
  and each further level triples the pass traffic.
\item \emph{No known scheme is better.}  By \eqref{eq:traffic} a level's
  traffic is $3+\kappa(g_1+g_2+g_3)$ words per $n^2$ for a kept fraction $\kappa$: it depends on the scheme only through $\kappa$ and the sum of its grid.  Over the
  $5{,}456$ formats of Sedoglavic's table of best known ranks
  \cite{Sedoglavic} the scheme $\mm{3\times3\times6:40}$ has the least traffic per unit of saving (the traffic divided by $1-\kappa$), $45.9$, at any stream count ($48.0$ for
  $\mm{4\times4\times4:48}$, $51.0$ for $\mm{3\times4\times6:54}$, $54.0$ for
  $\mm{6\times6\times6:153}$).  The schemes of lowest exponent in that table,
  the squares from $16^3$ to $32^3$, are no faster in f32 even at their
  best (their leaves are too small for the GEMM and their passes need
  thousands of streams), and to be no less accurate they would need a
  leaf-error amplification of at most $2.05$--$2.2$ per $4\times4$-equivalent
  saving, below the floor of about $2.24$ of the best class known.  A better level
  takes a rank record: $\mm{3\times3\times6:39}$, $\mm{3\times4\times6:53}$,
  or $\mm{4\times4\times4:47}$ over the rationals.
\item \emph{We did not find one.}  Around the known $\mm{3\times4\times6:54}$
  and $\mm{3\times3\times6:40}$ schemes: flip graphs \cite{KM22} over
  $\mathbb{F}_2$ and $\mathbb{F}_3$, with and without an imposed symmetry
  \cite{MP25}; an exact search for a set of $j$ of a scheme's products whose
  sum $j-1$ products can also compute, which would lower the rank by one,
  for every size $j$ up to $9$ over $\mathbb{F}_2$ and up to $8$ in
  characteristic $0$ (for the rank-$40$ scheme a proof up to $j=5$, and for
  all but $0.07\%$ of the subsets of size up to $8$); and numerical descent on the variety of exact
  decompositions, on which $\mm{3\times3\times6:40}$ is an isolated point and
  $\mm{3\times4\times6:54}$ lies on an $8$-parameter family.  None found a
  scheme of lower rank.  These are searches near known schemes, not lower
  bounds.
\item \emph{A record would not have to be dyadic.}  The accurate schemes
  of the $4\times4$ and $3\times3\times6$ levels
  have coefficients that are powers of two, which scale exactly.  Rescaling
  the products of the $\mm{3\times3\times6:40}$ level by constants such as
  $3$, $\frac13$ and $\frac57$ gives the same scheme with the same growth
  factor and constants that all round: its error against the exact product
  is $0.99$--$1.00$ of the dyadic programs' (geometric mean over five plans,
  two sizes and both precisions; single cells $0.94$--$1.05$) and its pass
  time the same (file~\rf{nondyadic}).  The $\mathbb{Q}(\sqrt3)$ point of the
  $2\times2$ scheme is a second case: irrational constants, and less error
  than the dyadic point by what its amplification predicts
  (Section~\ref{sec:sw}), which is why the library runs it.  A scheme
  pays for its growth factor and its rank, not for its constants.
\item \emph{Operands used in place.}  A factor that is a single block needs
  no pass.  Only a rank-one factor can be one, the factor ranks are
  invariants of the class, and with every such factor free the passes would
  move 8\% ($3\times3\times6$) or 17\% ($4\times4$) less, 1--2\% of a plan,
  at a point that is not the accurate one.
\end{itemize}
What would move f64 on the machine is a rank record or more hardware;
nothing else we measured does.

What is next.  The mixed points of the $3\times6\times3$ level
(Section~\ref{sec:accuracy}) cut the error of the plans that have it above
another level by 9--18\% at no cost in time; they should enter the menu as a
second form of the level.  The error model can now tell the two positions of
a level apart (Section~\ref{sec:errmodel}); what the second form needs is
its own two constants in the calibration.  The same model points at a
cheaper gain: \lv{R336} has the largest rounding of its own, which is what
it costs at the bottom of a plan, and a program for it with less would make
that orientation as good below another level as above.  Fusing the $L$
and $R$ passes into GEMM operand packing, the ``Strassen reloaded'' design,
needs an open GEMM kernel; on the machine the vendor's SME kernels are
closed, so it waits for another machine (x86-64 with BLIS, or the Apple GPU).
The planner's $0.07$~s per new shape in C should be a millisecond.  The error model's largest systematic error is the GEMM's own: beyond the
calibration's size its error grows a little faster than $\sqrt{k}$, which
makes the predicted growth conservative by up to a quarter at $13824$.  Prescaling (Section~\ref{sec:scaled}) should enter the public
interface and the Rust port, skip its factors when the inputs' range is
small, compute them in one read of each input, and perhaps become the
default; a planner could also detect cross structure and fall back to the
GEMM or to a block product.  The
library takes $C=AB$ only: $\alpha$, $\beta$ and transposes are a day's work.

\appendix

\section{Two implementations}\label{sec:impl}

The library is C (\texttt{fmm.h}, \texttt{libfmm.a}: \texttt{fmm\_sgemm},
\texttt{fmm\_dgemm} and their \texttt{\_mkn} forms for an $m\times k$ times
$k\times n$ product with separate leading dimensions, \texttt{fmm\_calibrate},
\texttt{fmm\_plan}, \texttt{fmm\_release}) over Accelerate, with the kernels
generated from the schemes' programs by \texttt{gen\_kernels.py}: every
program in PLinOpt's syntax becomes one fused loop in f32 and in f64, plus an
f64-arithmetic variant for f32 blocks.  Plans are made once per (shape,
precision, budget) and kept; the workspace is kept between calls; a call
outside the calibrated range or without a calibration runs the classical GEMM,
so a call is always correct.  The benchmark driver includes the library, so
both run the same code, and every plan prints its predicted time next to the
measured one.

The Rust port (crate \texttt{fmm}, \texttt{rust/}) is the same library module
for module -- plans and shapes, the calibration in the C file format so that
either library reads the other's cache, the planner with pipelines, budget
and padding, the executor on a rayon pool, the entry points -- with the
fused passes as safe Rust emitted by the same generator (each slice a
separate parameter, cut to the row length first: the compiler then
vectorizes as \texttt{restrict} and the pragma let it in C).  The only unsafe
code is the three FFI declarations for Accelerate behind safe wrappers that
check the slices, the two instructions of the passes' output-line zeroing
(\texttt{DC~ZVA} and the read of its block size) as inline assembly, behind
a safe function that takes the output slice and zeroes only whole lines inside
it, and, with the feature \texttt{gpu}, the four FFI declarations of the C
library's GPU leaf (\texttt{gpu\_leaf.m}, compiled into the crate): wrap a
region for the GPU, unwrap it, ask for a GPU, and run a batch of products.
Their wrappers let the GPU see only memory the crate holds -- a page-aligned
buffer wrapped from its creation to its drop (the workspace, the
calibration's matrices), or the caller's matrices for the duration of one call
of a K level -- and check every product's slices against its extent.  Both
ports carry the error model, the default budget, the planner's tie rules and
the GPU's rules (Section~\ref{sec:planner}).

Cross-checked against the C library on the shared calibration, $32$ cases in
the default builds and $34$ with the GPU (square sizes from $1000$ to $13824$, a leading dimension above
$n$, an error budget, four rectangular shapes, both precisions, and with the
GPU the plan without a budget at $3456$, \lv{K32/G6M}) give the same plans,
the same predicted times to the digit and results bit for bit, in the default
builds and in the builds with the GPU against \texttt{libfmm\_gpu.a}
(file~\rf{crosscheck}).  Bit for bit needed one change to
the C build.  With clang's default contraction of \texttt{a*b+c} into a fused
multiply-add, the one statement of the programs with a factor that is not a
power of two -- \texttt{v59 + (ov1 - (v48 + 2 v46)/5)*5} in the $P$ program of
\lv{R336}, as PLinOpt's optimizer wrote it -- rounded once in C and twice in
Rust, and every plan with an \lv{R336} level differed in the last bits; the C
library is now built with \texttt{-ffp-contract=off}, so both compute the
programs as written.  Of the errors this report quotes, two change in the
third digit (the \lv{R363/R336/G3} plans at $13824$ f64, $6.40$ to
$6.18\cdot10^{-15}$; \lv{R363+4/R336/G6} at $13824$ f32, $3.33$ to
$3.53\cdot10^{-6}$) and the rest not at all
(file~\rf{nocontract}); the tables, all measured since, give the new
values.  (A first crosscheck of the Rust \texttt{DC~ZVA} passes,
file~\rf{rustzva}, found no calibration -- a relative
cache path in a directory the crosscheck ran from -- and compared the classical
GEMM only.)  Speeds match at large sizes
(best of two alternating rounds, file~\rf{emreruns}; f64 $8192$: $0.790$~s in Rust against
$0.806$ in C; f32 $8192$: $0.230$ against $0.227$; f32
$6912\times6912\times3456$: $0.076$ against $0.078$); without the
\texttt{DC~ZVA} zeroing the Rust plans were 1--4\% slower in the earlier run of file~\rf{rustzva}, as
the C plans are.  With the GPU (two alternating rounds, the machine busy with a
system indexing process; file~\rf{rustgpu}) Rust runs
\lv{R363/G6M} at $3456$ in $0.0155$--$0.0160$~s against $0.0151$--$0.0188$ in
C and at $6912$ in $0.087$--$0.099$ against $0.084$--$0.093$, and the
pipelined \lv{R363+4/R336/G6M} at $13824$ 5--10\% slower, $0.611$--$0.628$
against $0.559$--$0.580$.  The Rust passes on small blocks were slower than the C library's until the
way they start was changed.  They handed a pass's $64$ row chunks to a
parallel iterator, which splits them in halves and wakes one sleeping thread
of the pool per split, so that the pool came up over six rounds; a pass now
spawns one task per chunk at once.  Measured as the Rust calibration measures
a pass, against the C library's calibrated entries (three rounds,
file~\rf{rustpass}), the passes took $1.27$, $1.13$ and $1.09$ times the C
library's time in f32 on blocks of side up to $300$, $600$ and $1200$ ($1.12$,
$1.04$ and $1.02$ in f64) and now take $1.08$, $1.00$ and $1.05$ ($1.01$,
$0.96$ and $0.99$).  GCD's \texttt{dispatch\_apply}, which is how the C
library starts a pass, was tried in its place and is slower there ($1.30$,
$1.14$, $1.07$ in f32), because the work before and after a pass is on the
pool.  The difference only shows when Rust calibrates for itself.  Two traps
found on the way: a zeroed \texttt{Vec} gets its pages on first write, so
timed buffers must be written first; and a lone multi-threaded GEMM issued
right after GEMM streams on the rayon pool runs 15--25\% slower and stays so,
which the leaves' timing avoids by running where the library runs a
classical plan.

\section{Commercial value, in software and in hardware}\label{sec:value}

The accuracy gain by itself has little commercial value; it is what makes
the speed gain sellable, and that combination has a real if narrow market.  This appendix says where, and whether the design could be built into a GPU or
a TPU.

\paragraph{Who would use it.}
Large dense products with a large inner dimension, in f32 or f64, whose
accuracy is a concern, occur in scientific and engineering computing: Gram
and covariance matrices over many samples, subspace orthogonalization in
electronic-structure codes, tensor contractions in quantum chemistry,
spectral transforms in weather and climate models, dense boundary-element and
imaging matrices.  They are GEMM-dominated, mostly f64, and their inner
dimensions run to the thousands, which is where the plans pay.  Two large
markets are excluded.  Machine learning runs bf16 or fp8 inputs with f32
accumulation, so its error is the input quantization, not the accumulation;
a more accurate f32 or f64 product is worth nothing there, and that is the
market that drives GEMM hardware.  And the dense factorizations of LAPACK and
the HPL benchmark call GEMM as rank-$k$ updates with $k$ of a few hundred,
where no level applies (Section~\ref{sec:planner}), so the most visible
dense-linear-algebra workloads gain nothing.

\paragraph{What the result is worth.}
Not the last digit of a Gram matrix: the value is speed without the usual
accuracy cost.
Fast matrix multiplication has been known since 1969, yet no vendor GEMM
uses it by default: IBM's ESSL ships Strassen--Winograd only as a separate
routine with a warning about accuracy \cite{ESSL} (its algorithm, in our implementation over Accelerate's GEMM at its fastest depth, is $1.21$--$1.54\times$ slower than the plans with
$1.9$--$9.7\times$ their error; Section~\ref{sec:essl}), and the standard
objection is Higham's \cite{Higham02}, that the Strassen-type algorithms lose
accuracy.  The mixed plans are
$1.4$--$1.9\times$ faster than the vendor's GEMM at $6912$ and $13824$ with
less error than it on well-scaled inputs, in both precisions (and up to
$3.1\times$ in f32 with the GPU as a second engine), under a default that
aims at no less accuracy than the GEMM and measured below the GEMM's error in
every case on those inputs.  That answers half of the objection.  The other
half is the one ESSL's warning names, inputs of divergent scale: there a
fast scheme's error is normwise, not componentwise, and the plans lose up
to nine orders of magnitude in the smallest entries of $C$
(Section~\ref{sec:scaled}).  Exact power-of-two prescaling recovers the
GEMM's accuracy for graded rows, columns and inner dimensions at up to $4\%$
of the time, leaving the plans $1.37$--$1.89\times$ faster, but no diagonal
scaling repairs magnitudes that cross between blocks, and nonnegative
inputs leave the default plans only a quarter below the GEMM's error.  A library could adopt the
plans with prescaling on and the scalings it repairs stated; a vendor
promising the GEMM's componentwise bound for every input could not.  An HPC
centre whose matrices are not of the defeating kind would take about 40\% of
its dense-contraction time back.  The
realistic paths, in decreasing order of likelihood: adoption inside a vendor
or open BLAS as a large-size path (the fused design needs an open kernel, so
BLIS is the natural first target), with reputation rather than revenue as the
return; consulting or a licensed library for codes with large contractions,
a small market whose customers exist and pay; and nothing proprietary, since
the schemes are published \cite{DPS26a,DPS26b} and the engineering is
reproducible from this report.  Two trends limit all of it.  Dense f64 work
is moving to GPUs, where accuracy is already sold in the other direction,
f64 products emulated on low-precision tensor cores by the Ozaki scheme
\cite{Ozaki12}; and on the machine the GPU alone already ties the CPU
plans in f32 at $13824$ (Section~\ref{sec:gpu}), so the gain on the CPU alone
is cleanest in f64, the smallest market of all.  A port to an x86-64 server part with BLIS, measured, is what
commercial traction would need.

\paragraph{In hardware.}
The design could speed up a GPU or a TPU in principle, but the right place
for it is not the multiplier array.  A matrix unit is compute-bound on large
products (cuBLAS reaches 70--80\% of peak), so a fused level of the accurate
$\mm{4\times4\times4:48}$ scheme buys 25\% of the array's cycles, a $\mm{3\times3\times6:40}$ level
26\%, and two levels about 45\%, with passes of O($n^2$) work against the
products' O($n^3$) -- once the blocks are large.  That condition rules the
design out of the tensor core itself: for a $16\times16$ register tile the
level's additions are comparable to its multiplies and the amplification is
paid for nothing.  The level must sit where blocks of hundreds of rows are
resident, which on a GPU is shared memory and L2 and on a TPU the vector unit
and its SRAM beside the systolic array; so it is a kernel and compiler
feature with modest hardware help (wider adders on the operand path, staging
SRAM), not a new multiply-accumulate design.  The software form exists: the
GPU kernels of Huang, Yu and van de Geijn \cite{HYvdG20} fuse a Strassen
level's additions into the operand packing and gain 20--30\% at large sizes,
and Winograd's minimal convolution \cite{LG16}, the same idea for
convolutions, ships in several neural accelerators; vendors know the trick
and have not applied it to GEMM.  Precision decides whether it is worth
building.  For f64 and f32 units the answer is yes: an f64 multiplier is the
most expensive arithmetic block on the chip, so trading a quarter of them per
level for additions is an area and energy lever as well as a time one, and
with the accurate schemes the level is more accurate than the plain product.
For bf16 and fp8 units the answer is no: the operand combinations are not
representable in the input format, so the $L$ and $R$ passes add a rounding
per combination to inputs that already carry a 0.4\% or 6\% quantization
error, and the scheme's amplification then costs more than the 25\% saves;
nobody in that market buys accuracy, and its throughput is better bought with
narrower formats and more multipliers.  Since that market pays for the
silicon, the idea stays unbuilt.  What an HPC-class part would gain is
20--45\% more effective f64 or f32 GEMM throughput per watt at large sizes
with no accuracy loss on well-scaled operands (with prescaling on graded
ones), for a kernel change and some staging logic.  On this
machine the measurement exists in a weaker form: with the passes on the CPU
and the products shared between the CPU's matrix units and the GPU, one or
two accurate levels make the f32 product $1.8\times$ faster than the GPU
alone at $13824$, with half its error (Section~\ref{sec:gpu}).  What this report changes is not the arithmetic
saving, known for decades, but the evidence that it need not cost accuracy
on well-scaled operands and that exact prescaling recovers it on graded
ones, which is the argument a hardware team would need before spending area on it.

\section{Glossary: the machine, its software, and the report's terms}\label{sec:glossary}

The report uses Apple's names for the hardware and the libraries, and a few
of Arm's.  In one place, followed by the report's own terms and abbreviations:

\begin{description}
\item[Apple silicon, M4~Pro.]  Apple's system on a chip for Macs.  The M4~Pro
  of this report has 14 CPU cores in three clusters (two of five performance
  cores, one of four efficiency cores), a 20-core GPU, and 64~GB of memory
  that all of them share (Figure~\ref{fig:machine}).  A cluster is Apple's group of
  cores around one level-2 cache; each performance cluster of the M4 also
  has one matrix unit.
\item[Performance and efficiency cores.]  The two kinds of CPU core.  The
  performance cores run the passes and the GEMM threads here; the efficiency
  cores are slower and share a smaller cache, and the thread counts of
  Section~\ref{sec:machine} are chosen so that timed work stays off them.
\item[SME, Scalable Matrix Extension.]  Arm's instruction set extension for
  matrix arithmetic, implemented in the M4 as a matrix coprocessor with a
  large tile register and an outer-product instruction that multiplies a
  column of $A$ by a row of $B$ and accumulates into the tile.  The M4~Pro has
  one SME unit per performance cluster, shared by that cluster's cores; it
  succeeds Apple's earlier undocumented AMX unit.  Accelerate's GEMM runs on
  it: a GEMM call keeps about two cores busy and reaches $3.3$~TFLOP/s in
  f32, far above what the cores' vector units could give, with the GPU's
  utilization counter at zero.  It accumulates in the order of the
  sequential FMA chain, which is why its result equals the GPU's bit for
  bit.  Its kernels are Apple's and closed, which keeps the ``Strassen
  reloaded'' design of fusing the passes into the GEMM's operand packing off
  the machine.
\item[NEON.]  Arm's 128-bit SIMD vector instruction set, on every core.
  GEMMs on the idle cores with it were tried and not adopted
  (Section~\ref{sec:machine}).
\item[Accelerate.]  Apple's numerical library: BLAS and LAPACK, vDSP and
  more, with the kernels for Apple silicon.  Its BLAS is the vendor GEMM of
  this report, called through the C interface \lv{cblas\_sgemm} and
  \lv{cblas\_dgemm}; the threading inside a call is Accelerate's own, and
  the library's streams are several single-threaded calls side by side.
\item[BLAS, GEMM, sgemm, dgemm.]  The Basic Linear Algebra Subprograms
  interface; GEMM is its general matrix multiply $C \leftarrow \alpha AB +
  \beta C$, sgemm in single precision (f32), dgemm in double (f64).  ``The
  GEMM'' in this report is Accelerate's.
\item[FMA, fused multiply-add.]  The instruction computing $a\cdot b + c$
  with a single rounding; every modern GEMM is built from it.  One output
  accumulated as $c \leftarrow \mathrm{fma}(a_k, b_k, c)$ for $k$ in order is
  the sequential FMA chain that both Accelerate's SME kernels and Apple's GPU
  kernels compute.
\item[f32, f64, f16, bf16.]  IEEE single and double precision (24- and
  53-bit significands), IEEE half precision (11 bits) and bfloat16 (8 bits).
  The CPU paths of this report are f32 and f64; Apple's GPU has f32, f16 and
  bf16 and no f64.
\item[GPU, unified memory.]  The M4~Pro's 20-core graphics processor,
  programmed through Metal.  It reads the same memory as the CPU, so a CPU
  array is handed to it without a copy by wrapping it as a shared Metal
  buffer; the GPU numbers of Section~\ref{sec:gpu} include no transfers.
\item[Metal.]  Apple's GPU programming interface and shading language, the
  counterpart of CUDA or Vulkan: devices, command queues, command buffers,
  compute kernels, buffers.
\item[MPS, Metal Performance Shaders.]  Apple's library of ready GPU kernels
  on top of Metal, among them \lv{MPSMatrixMultiplication}, a GEMM on Metal
  buffers.
\item[MPSGraph.]  Metal Performance Shaders Graph: Apple's framework for
  tensor computation on the GPU, in which a graph of operations (here two
  placeholders and one matrix multiplication) is compiled to Metal kernels
  and run.  It is what Core ML and PyTorch's Apple backend use underneath;
  its GEMM is the faster of Apple's two.
\item[Grand Central Dispatch.]  Apple's thread-pool interface (libdispatch);
  \lv{dispatch\_apply} runs a block over an index range on the pool.
  The library uses both it and POSIX threads; the benchmark's compensated reference product runs on the pool through \lv{dispatch\_apply}.
\item[Apple clang, macOS~26.]  The compiler and operating system of the
  measurements.  The C library is C11 with Accelerate; the GPU tool is
  Objective-C, because the GPU frameworks have no C interface.
\item[TFLOP/s.]  $10^{12}$ floating-point operations per second, counting
  $2mkn$ for a product whatever the algorithm, so a plan's rate is the
  classical count divided by its time.
\end{description}

\begin{description}
\item[GEMM.]  See above: the general matrix multiply of BLAS, here
  Accelerate's, the leaf of every plan and the baseline of every speedup.
\item[Scheme $\mm{g_1\times g_2\times g_3:r}$, rank.]  A bilinear algorithm
  that multiplies a $g_1\times g_2$ by a $g_2\times g_3$ block matrix with $r$
  block products instead of $g_1g_2g_3$; $r$ is its rank.  Its three matrices
  $L$, $R$, $P$ are Section~\ref{sec:levels}'s.
\item[Level, pass, plan, leaf.]  A level is one application of a scheme to
  the current problem, executed as three fused passes ($L$, $R$, $P$;
  Figure~\ref{fig:lrp}) around
  its products; a fused pass is one loop over a block's element positions
  that evaluates a whole linear program in registers, reading each input
  block and writing each output block once (Section~\ref{sec:intro}); a plan is the sequence of levels from the top down to the
  leaf, the GEMM; the notation is in Table~\ref{tab:notation}.
\item[\lv{D48}, \lv{R336}, \lv{R363}, \lv{R633}, \lv{SW}, \lv{T336},
  \lv{T363}, \lv{T633}, \lv{X}.]  The levels of the menu (Table~\ref{tab:levels}):
  the accurate $\mm{4\times4\times4:48}$ scheme (\lv{D} for its authors, Dumas, Pernet and
  Sedoglavic, $48$ for its rank), the accurate $\mm{3\times3\times6:40}$ scheme in
  its three orientations (\lv{R} for rectangular, the digits the grid
  $g_1\times g_2\times g_3$ it cuts the problem into), the $2\times2$ level
  (Strassen's scheme at DPS's accurate point in $\mathbb{Q}(\sqrt3)$, Section~\ref{sec:sw}),
  and the ternary $\mm{3\times3\times6:42}$ scheme of Section~\ref{sec:ternary}
(ternary: coefficients in $\{-1,0,1\}$, from Perminov's catalogue); \lv{X} in
  Table~\ref{tab:ternary} stands for \lv{R} or \lv{T}.
\item[\lv{G}, \lv{G3}, \lv{G4}, \lv{G6}, \lv{B2}, \lv{+p}.]  The leaves and the
  pipeline mark: \lv{G} is one Accelerate GEMM with its own threading,
  \lv{G3}, \lv{G4}, \lv{G6} the products as three, four or six single-threaded
  GEMM streams (the digit is the stream count), \lv{B2} the classical
  $2\times2$ blocked product as a level (eight products, the calibration's
  reference); \lv{+p} after a level runs its passes on $p$ threads overlapped
  with the products below.
\item[\lv{GM}, \lv{G6M}.]  The leaves of the library's GPU build: a level's
  products shared between Accelerate's GEMM (with its own threading, or six
  streams) and the GPU, which take products from the two ends of the list
  until they meet (Section~\ref{sec:machine}).
\item[\lv{K16}, \lv{K32}, \lv{K64}.]  Levels with no scheme: the classical
  product cut into $16$, $32$ or $64$ strips of rows, shared between the two
  engines.  No passes, no saving of products, and the GEMM's own result:
  they measure what the second engine alone buys (Section~\ref{sec:gpu}).
\item[\texttt{DC ZVA}.]  Arm's ``data cache zero by virtual address''
  instruction: it sets a whole cache line ($64$ bytes here) to zero without
  reading it from memory.  The library's passes issue it on the lines they
  are about to overwrite, which saves the read that an ordinary store to a
  line not in the cache would make (Section~\ref{sec:levels}).
\item[Word, traffic, $\mathcal{B}$.]  A word is one matrix element, an f32 or
  an f64 ($w$ bytes, $4$ or $8$); a level's traffic is the number of words its
  three passes move, $W(m,k,n)$ of \eqref{eq:traffic}, quoted for a scheme as
  $W(n,n,n)/n^2$, the words per $n^2$ on a square problem; $\mathcal{B}$ is the
  streaming bandwidth the passes reach, in bytes per second, in
  \eqref{eq:level}.
\item[Streams, pipelines.]  Section~\ref{sec:machine}: several
  single-threaded GEMM calls side by side, and a level's passes overlapped
  with the products of the level below.
\item[Error, max.]  The max-norm relative error against the exact product (Section~\ref{sec:setup}), ``max'' in the tables that also give the componentwise error; the result files print it as \texttt{err64}.  Their \texttt{relerr} is the same quantity against Accelerate's product, which includes the GEMM's own error and is not an error measure.
\item[Kept fraction $\kappa$, saving.]  A level keeps $\kappa=r/(g_1g_2g_3)$ of the classical products and saves $1-\kappa$; ``per $4\times4$-equivalent saving'' rescales a quantity to the saving of one $4\times4$ level, by the power $\ln0.75/\ln\kappa$.
\item[Amplification ratio, growth factor, level factor, error growth.]  The four accuracy quantities of Table~\ref{tab:accnames}.
\item[cw, componentwise error.]  The largest error of an entry of $C$
  divided by $(|A||B|)$ at that entry (Section~\ref{sec:scaled}); the
  classical algorithm keeps it near the unit roundoff for any inputs, a
  fast scheme only on well-scaled ones.
\item[Prescaling; outer, inner, cross.]  Exact power-of-two scaling of the
  rows of $A$ and columns of $B$ (outside) and of the columns of $A$ against
  the rows of $B$ (inside) before a plan, undone on $C$ after it; the input
  classes of Section~\ref{sec:scaled} that it repairs (outer, inner) and
  cannot (cross).
\item[Error growth, \lv{max\_growth}, \lv{auto:}$b$.]  A plan's predicted
  error, by the error model \eqref{eq:errmodel}, divided by the GEMM's; the budget $b$ a call
  sets on it: $0$ the default of $1$, a negative value no budget
  (Section~\ref{sec:planner}).
\item[$\amp$, $\ginf$, $a_X$, $\nu_X$, $\pi_X$, $c$, $\beta$, $N_E$, $u$, $d$.]  The amplification
  measures of a scheme (root-sum-square and worst-case,
  Section~\ref{sec:accuracy}); the error model's constants
  (Section~\ref{sec:errmodel}): a level's amplification ratio and
  participation number, from its scheme, its own rounding, the leaf constant
  and the tail excess, from the calibration; the size at which the model is
  calibrated ($2304$); the unit roundoff; the number of
  fused levels in a plan.
\item[Kronecker composite.]  The scheme obtained as the Kronecker product of
  two schemes (one level of each, fused), screened out in
  Section~\ref{sec:menu}.
\item[DPS.]  Dumas, Pernet and Sedoglavic, the authors of the accurate
  schemes and of PLinOpt \cite{DPS24,DPS25,DPS26a,DPS26b}; DPSST adds
  Stapleton and Tichavsk\'y \cite{DPSST26}.
\item[PLinOpt, SLP, SLPchecker.]  DPS's library of tools for linear
  programs \cite{PLinOpt}: a straight-line program (SLP) is a list of
  assignments computing a linear map; PLinOpt's optimizer shortens one, its
  transpozer transposes one, and SLPchecker verifies one against its matrix,
  which every program in this report passed.
\item[Orbit (class), point, $GL_n$.]  The schemes equivalent to a given one, each of them a point of the orbit, under the
  general linear group $GL_n$ of invertible $n\times n$ matrices acting on
  the three sides \cite{deGroote78} (Section~\ref{sec:intro}); the orbit floors of Section~\ref{sec:accuracy} are minima
  over it, found by L-BFGS (limited-memory Broyden--Fletcher--Goldfarb--Shanno \cite{LN89}), a quasi-Newton minimizer.
\item[RMS, IEEE 754, SIMD.]  Root mean square; the floating-point
  standard (f32, f64, f16); single instruction, multiple data (vector
  instructions).
\item[POSIX threads, FFI, JSON, C11.]  The standard thread interface the C
  library uses; the foreign-function interface through which Rust calls
  Accelerate; the text format of Perminov's catalogue; the 2011 C standard.
\item[BLIS, CUDA, Vulkan, Core ML, PyTorch, MLX.]  Named for comparison only:
  an open BLAS framework (for the x86-64 machine the design of
  Section~\ref{sec:disc} waits for), Nvidia's and Khronos's GPU interfaces,
  Apple's and Meta's machine-learning frameworks, and Apple's array
  framework.
\item[GB/s, TFLOP/s.]  Gigabytes per second of memory traffic; $10^{12}$
  floating-point operations per second as defined above.
\end{description}

\section{Data and code}\label{sec:data}

The result files cited in the text as F1, F2, \dots\ are in \path{matmul/fastmm/results/} of the repository \cite{logic}:
\begin{description}\setlength{\itemsep}{0pt}
\rfitem{camp}{final_campaign_2026-11-22.txt}: the campaign: the calibration, the square problems, the comparison with \texttt{DC~ZVA} off, the GPU, the ternary trial, the library's tests (\path{run_final3.sh});
\rfitem{t42}{t42_exact64_2026-11-23.txt}: the ternary trial's f64 errors against the exact product;
\rfitem{scaled}{scaled_inputs_2026-11-23.txt}: badly scaled inputs;
\rfitem{prescale}{prescale_timing_2026-11-23.txt}: the time prescaling costs;
\rfitem{essl}{essl_gemms_2026-11-23.txt}: the comparison with ESSL's \texttt{\_GEMMS} algorithm;
\rfitem{swdps7}{sw_dps7_2026-11-02.txt}: the $2\times2$ level's programs, dyadic against Winograd's;
\rfitem{r336opt}{r336_opt_vs_direct_2026-10-28.txt}: the $3\times3\times6$ programs, optimized against direct sums;
\rfitem{probes}{pass_probes_2026-11-14.txt}: the probes that led to the \texttt{DC~ZVA} zeroing;
\rfitem{smeleaf}{sme_leaf_2026-11-14.txt}: our own SME kernel as a leaf;
\rfitem{gpushare}{gpu_share_2026-11-14.txt}: a level's products shared with the GPU;
\rfitem{gpugemm}{gpu_gemm_2026-10-28.txt}: the GPU's own GEMMs;
\rfitem{nondyadic}{nondyadic_2026-11-14_summary.txt}: the non-dyadic rescaling test;
\rfitem{crosscheck}{crosscheck_2026-11-23.txt}: C against Rust, with and without the GPU;
\rfitem{rustc}{rust_vs_c_2026-11-23.txt}: the Rust port's times against the C library's;
\rfitem{tie3}{tie3_picks_2026-11-23.txt}: the planner's picks with the tie window at $3\%$, the changed pick measured, the tests and the crosscheck again (\path{run_tie3_picks.sh});
\rfitem{emdata}{errmodel_data_2026-11-23.txt}: the error model's data: the GEMM by inner dimension, every level alone, the $25$ ordered pairs and $17$ three-level plans at $3456$ and $6912$, four instances (\path{run_errmodel_data.sh}); the singles and pairs at $2304$ on six instances in \path{errmodel_data_2304_2026-11-23.txt};
\rfitem{empicks}{errmodel_picks_2026-11-23.txt}: the planner's picks under the error model \eqref{eq:errmodel}, each measured against its competitors, and the model's predictions for the plans of the data (\path{run_errmodel_picks.sh});
\rfitem{emreruns}{essl_gemms_errmodel_2026-11-23.txt}, \path{prescale_timing_errmodel_2026-11-23.txt}, \path{scaled_inputs_errmodel_2026-11-23.txt}, \path{gpu_errmodel_2026-11-23.txt}, \path{rust_vs_c_errmodel_2026-11-23.txt}, \path{crosscheck_errmodel_2026-11-23.txt}: the rows whose default plan the error model changed, again (\path{run_errmodel_reruns.sh});
\rfitem{nocontract}{r336_nocontract_2026-11-16.txt}: the \lv{R336} plans' errors without FMA contraction;
\rfitem{rustzva}{rust_zva_2026-11-15.txt}: the Rust port with the \texttt{DC~ZVA} passes;
\rfitem{rustgpu}{rust_gpu_2026-11-16.txt}: the Rust GPU leaves against C;
\rfitem{fused}{fused_vs_separate_2026-11-23.txt}: fused $2\times2$ levels against one-at-a-time Strassen--Winograd at equal depth (\path{run_fused_vs_separate.sh});
\rfitem{swq3}{sw_q3_2026-11-22.txt}: the $2\times2$ level at the $\mathbb{Q}(\sqrt3)$ point against the dyadic point (\path{run_sw_q3.sh}; the programs by \path{sw_points.py}, the table by \path{summarize_sw_q3.py} in \path{sw_q3_2026-11-22_summary.txt});
\rfitem{campold}{final_campaign_2026-11-15.txt}, \path{final_campaign_2026-11-15b.txt}: the campaign before the $2\times2$ level's move, on the dyadic point (\path{run_final2.sh}, \path{run_final2b.sh}; its calibration in \path{calib_dps7_2026-11-15/});
\rfitem{rustpass}{rust_small_passes_2026-11-22.txt}: the Rust passes on small blocks against the C library's calibrated passes, by pass driver (\path{run_rust_small_passes.sh});
\rfitem{r363pts}{r363_points_2026-11-22.txt}: the $3\times6\times3$ level at other points of its orbit (\path{run_r363_points.sh}; the table by \path{summarize_r363_points.py} in \path{r363_points_2026-11-22_summary.txt});
\rfitem{r363f64}{r363_points_f64pass_2026-11-22.txt}: the same points with the f32 passes computed in f64;
\rfitem{cacheprobe}{probe_cache_2026-11-23.txt}: a plain pass by where its arrays live, the bandwidths of Figure~\ref{fig:machine} (\path{probe_cache.c});
\rfitem{machprobes}{probes_2026-10-24.txt}, \path{neon_2026-10-24.txt}: the GEMM by thread count, our NEON GEMM, and a GEMM stream held on the efficiency cores.
\end{description}

{\sloppy Everything is in the repository \cite{logic}, directory
\path{matmul/fastmm/}: the library (\path{fmm.c}, \path{fmm_impl.h},
\path{fmm.h}), the kernel generator (\path{gen_kernels.py}), the programs
of the $3\times3\times6$ levels (\path{schemes/opt/r*}, PLinOpt-optimized;
\path{schemes/r*}, the direct sums), of the $2\times2$ level
(\path{schemes/sw/q3_*}; the dyadic point's are \path{schemes/opt/dps7_*}, and Winograd's, for comparison, are in
\path{gen_kernels.py}) and of the ternary levels
(\path{schemes/t*}), the Rust crate (\path{rust/}), the benchmark and the
test (\path{bench_fmm.c}, \path{test_fmm.c}), the calibrations
(\path{results/calib5_f32.txt}, \path{calib5_f64.txt}) and the raw
results of every run quoted here.  One campaign on the library as it
stands gives every table of Sections~\ref{sec:intro},
\ref{sec:results} and~\ref{sec:ternary} and the tests:
\path{run_final3.sh} (\path{results/final_campaign_2026-11-22.txt}), the
calibration in both precisions and with the GPU (\path{calib5_f32.txt},
\path{calib5_f64.txt}, \path{calib5gpu_f32.txt}), the square problems, the comparison with the
passes' \texttt{DC~ZVA} switched off, the GPU, the ternary trial, whose f64 rows
were rerun against the exact product by \path{run_t42_exact64.sh}
(\path{results/t42_exact64_2026-11-23.txt}), and the library's tests in both builds.  The
measurements outside the campaign that depend on the $2\times2$ level or on the
planner's choices were repeated on the same library and calibration by
\path{run_switched_reruns.sh} (2026-11-23): the
comparison with ESSL's \texttt{\_GEMMS} algorithm (Section~\ref{sec:essl};
\path{gemms_impl.h}, \path{run_essl_gemms.sh},
\path{results/essl_gemms_2026-11-23.txt}), fused against one-at-a-time
levels (\path{run_fused_vs_separate.sh}), the badly scaled inputs
and prescaling (Section~\ref{sec:scaled}; \path{run_scaled_inputs.sh},
\path{results/scaled_inputs_2026-11-23.txt}, and
\path{run_prescale_timing.sh}, \path{results/prescale_timing_2026-11-23.txt};
the input classes \texttt{FMM\_INPUT}, the componentwise error and the
prescaling switch \texttt{FMM\_PRESCALE} are in \path{bench_impl.h}), the Rust port against C
(\path{run_rust_vs_c.sh}) and the crosscheck of both builds, with and without the GPU
(\path{run_crosscheck.sh}, \path{results/crosscheck_2026-11-23.txt}).  The campaign before the
$2\times2$ level's move to the $\mathbb{Q}(\sqrt3)$ point is
\path{run_final2.sh} and \path{run_final2b.sh}
(\path{results/final_campaign_2026-11-15.txt}, \path{final_campaign_2026-11-15b.txt}),
with the runs of that week on the dyadic library beside it
(\path{scaled_inputs_2026-11-16.txt}, \path{prescale_timing_2026-11-16.txt},
\path{essl_gemms_2026-11-17.txt}, \path{crosscheck_2026-11-16.txt}).  Earlier steps that the text cites: the Rust
port with the passes' \texttt{DC~ZVA} zeroing
(\path{run_rust_zva.sh}, \path{results/rust_zva_2026-11-15.txt}); the Rust
GPU leaves' tests and timing against C (\path{run_rust_gpu.sh},
\path{results/rust_gpu_2026-11-16.txt}); the R336 plans' errors without FMA
contraction (\path{run_r336_nocontract.sh},
\path{results/r336_nocontract_2026-11-16.txt}).  The GPU's own GEMMs are measured by \path{gpu_gemm.m}
(\texttt{make build/gpu\_gemm}; its f16 path in
\path{gpu_gemm_2026-10-28.txt}), the sharing of a level's products by
\path{gpu_share.m} (\path{gpu_share_2026-11-14.txt}), the probes that led to the
\texttt{DC~ZVA} zeroing, \path{probe_strip.c} and \path{probe_nt.c}
(\path{pass_probes_2026-11-14.txt}), and the non-dyadic test by
\path{nondyadic.py} (\path{nondyadic_2026-11-14_summary.txt}); the comparison
of the $2\times2$ level's programs is \path{sw_dps7_2026-11-02.txt}
(Section~\ref{sec:sw}; \path{run_sw_dps7.sh}, the binary with Winograd's
programs by \texttt{make build/bench\_fmm\_winograd}, its calibration
\path{calib5_f32_winograd_2026-10-29.txt},
\path{calib5_f64_winograd_2026-10-29.txt}); the comparison of the
$3\times3\times6$ programs is \path{r336_opt_vs_direct_2026-10-28.txt};
the estimates' failure on uncalibrated shapes is in
\path{rect_api_2026-10-28.txt}; and the campaigns of October 24--26 hold the
streams, pipelines, accuracy and $4\times4$-program measurements of
Sections~\ref{sec:levels} and~\ref{sec:accuracy}.  The menu screen is \path{matmul/levels/}
(\path{levels.py}, \path{levels.csv}, \path{orbit_rect.py},
\path{cost_rect.py}, README).  The accuracy study of the $4\times4$ class
is \path{doc/matmul_4x4_rank48_accurate262.pdf} \cite{S26b}.  The running
log of the track, with every measurement in the order it was made, is
\path{doc/fastmm_plan.md}.\par}


{\small
\begin{thebibliography}{99}
\bibitem{DPS24} J.-G.~Dumas, C.~Pernet, A.~Sedoglavic, \emph{Strassen's
algorithm is not optimally accurate}, Proc.\ ISSAC 2024, 254--263;
arXiv:2402.05630v2 (28~June 2024), whose numbering we cite: the dyadic
point is Eq.~(38) in Remark~22 (Section~6.3), its program Table~4, its
growth factors the column ``Equation~(38)'' of Table~1.
\bibitem{DPS25} J.-G.~Dumas, C.~Pernet, A.~Sedoglavic, \emph{A
non-commutative algorithm for multiplying $4\times4$ matrices using
$48$ non-complex multiplications}, arXiv:2506.13242 (2025).
\bibitem{DPS26a} J.-G.~Dumas, C.~Pernet, A.~Sedoglavic, \emph{A more
accurate rational non-commutative algorithm for multiplying $4\times4$
matrices using $48$ multiplications}, arXiv:2603.18699 (v1 19~March
2026, v3 29~July 2026).
\bibitem{DPS26b} J.-G.~Dumas, C.~Pernet, A.~Sedoglavic, \emph{Towards
automated generation of fast and accurate algorithms for recursive
matrix multiplication}, J.\ Symbolic Comput.\ 134 (2026) 102524;
arXiv:2506.19405.
\bibitem{DPSST26} J.-G.~Dumas, C.~Pernet, A.~Sedoglavic, J.~Stapleton,
P.~Tichavsk\'y, \emph{Fast matrix multiplication via recursive
$\langle4\times4\times4:48\rangle$ algorithms into practice},
arXiv:2609.12027 (v1 10~September 2026, v2 2~October 2026).
\bibitem{PLinOpt} J.-G.~Dumas, B.~Grenet, C.~Pernet, A.~Sedoglavic,
\emph{PLinOpt, a collection of C++ routines handling linear \&
bilinear programs}, \url{https://github.com/jgdumas/plinopt}; data
files \texttt{4x4x4\_48\_accurate\_262} (commit \texttt{b441528},
13~October 2026), \texttt{2x2x2\_7\_DPS-accurate} (the $\mathbb{Q}(\sqrt3)$ point), \texttt{3x3x6\_40\_DPS-accurate}, \texttt{3x6x3\_40\_DPS-accurate},
\texttt{6x3x3\_40\_DPS-accurate}; tools \texttt{optimizer}, \texttt{transpozer},
\texttt{SLPchecker}.
\bibitem{BB15} A.~R.~Benson, G.~Ballard, \emph{A framework for practical
parallel fast matrix multiplication}, Proc.\ PPoPP 2015, ACM, 42--53;
arXiv:1409.2908.
\bibitem{BBDLS16} G.~Ballard, A.~R.~Benson, A.~Druinsky, B.~Lipshitz,
O.~Schwartz, \emph{Improving the numerical stability of fast matrix
multiplication}, SIAM J.\ Matrix Anal.\ Appl.\ 37 (2016) 1382--1418;
arXiv:1507.00687.
\bibitem{BL80} D.~Bini, G.~Lotti, \emph{Stability of fast algorithms for
matrix multiplication}, Numer.\ Math.\ 36 (1980) 63--72.
\bibitem{Brent70} R.~P.~Brent, \emph{Algorithms for matrix
multiplication}, Technical Report STAN-CS-70-157, Stanford University
(1970).
\bibitem{HM19} N.~J.~Higham, T.~Mary, \emph{A new approach to probabilistic
rounding error analysis}, SIAM J.\ Sci.\ Comput.\ 41 (2019), A2815--A2835.
\bibitem{ESSL} IBM, \emph{Engineering and Scientific Subroutine Library
(ESSL): Guide and Reference}, subroutines \texttt{SGEMMS}, \texttt{DGEMMS},
\texttt{CGEMMS}, \texttt{ZGEMMS} (matrix multiplication using Winograd's
variation of Strassen's algorithm).
\bibitem{DHSS94} C.~C.~Douglas, M.~Heroux, G.~Slishman, R.~M.~Smith,
\emph{GEMMW: a portable level 3 BLAS Winograd variant of Strassen's
matrix--matrix multiply algorithm}, J.\ Comput.\ Phys.\ 110 (1994) 1--10.
\bibitem{BDPZ09} B.~Boyer, J.-G.~Dumas, C.~Pernet, W.~Zhou, \emph{Memory
efficient scheduling of Strassen--Winograd's matrix multiplication
algorithm}, Proc.\ ISSAC 2009, 55--62.
\bibitem{deGroote78} H.~F.~de~Groote, \emph{On varieties of optimal algorithms
for the computation of bilinear mappings I.  The isotropy group of a bilinear
mapping}, Theoret.\ Comput.\ Sci.\ 7 (1978) 1--24.
\bibitem{HLJJTT96} S.~Huss-Lederman, E.~M.~Jacobson, J.~R.~Johnson,
A.~Tsao, T.~Turnbull, \emph{Implementation of Strassen's algorithm for
matrix multiplication}, Proc.\ Supercomputing '96, IEEE, 1996.
\bibitem{HRMvdG17} J.~Huang, L.~Rice, D.~A.~Matthews, R.~A.~van de Geijn,
\emph{Generating families of practical fast matrix multiplication
algorithms}, Proc.\ IPDPS 2017; arXiv:1611.01120.
\bibitem{Higham02} N.~J.~Higham, \emph{Accuracy and Stability of Numerical
Algorithms}, 2nd ed., SIAM, 2002 (Chapter 23, fast matrix multiplication).
\bibitem{HSvdG16} J.~Huang, T.~M.~Smith, G.~M.~Henry, R.~A.~van de Geijn,
\emph{Strassen's algorithm reloaded}, Proc.\ SC~'16 (International
Conference for High Performance Computing, Networking, Storage and
Analysis), 2016.
\bibitem{HYvdG20} J.~Huang, C.~D.~Yu, R.~A.~van de Geijn, \emph{Strassen's
algorithm reloaded on GPUs}, ACM Trans.\ Math.\ Softw.\ 46 (2020), 1:1--1:22.
\bibitem{KM22} M.~Kauers, J.~Moosbauer, \emph{Flip graphs for matrix
multiplication}, arXiv:2212.01175 (2022).
\bibitem{KW} M.~Kauers, J.~Wood, \emph{Consequences of the Moosbauer--Poole
algorithms}, preprint (2025).
\bibitem{LN89} D.~C.~Liu, J.~Nocedal, \emph{On the limited memory BFGS method
for large scale optimization}, Math.\ Programming 45 (1989), 503--528.
\bibitem{LG16} A.~Lavin, S.~Gray, \emph{Fast algorithms for convolutional
neural networks}, Proc.\ CVPR 2016, 4013--4021.
\bibitem{MP25} J.~Moosbauer, M.~Poole, \emph{Flip graphs with symmetry and
new matrix multiplication schemes}, preprint (2025).
\bibitem{MSW25} E.~M\r{a}rtensson, P.~Stankovski Wagner, \emph{The number of
the beast: reducing additions in fast matrix multiplication algorithms
for dimensions up to 666}, Proc.\ ACDA 2025; the library
\texttt{fmm\_add\_reduction}, \url{https://github.com/werekorren/fmm_add_reduction}
(commit \texttt{951cd93}, 8~May 2025), whose \texttt{algorithms/} directory
distributes the Kauers--Wood, Moosbauer--Poole--Kemper and AlphaEvolve
schemes used in the screen.
\bibitem{Novikov25} A.~Novikov et al., \emph{AlphaEvolve: a coding agent
for scientific and algorithmic discovery}, arXiv:2506.13131 (2025).
\bibitem{Ozaki12} K.~Ozaki, T.~Ogita, S.~Oishi, S.~M.~Rump, \emph{Error-free
transformations of matrix multiplication by using fast routines of matrix
multiplication and its applications}, Numer.\ Algorithms 59 (2012), 95--118.
\bibitem{Perminov} A.~I.~Perminov, \emph{FastMatrixMultiplication}: a
catalogue of fast matrix multiplication schemes,
\url{https://github.com/dronperminov/FastMatrixMultiplication}, commit
\texttt{3ee3284} (28~October 2026); file
\texttt{schemes/results/addition\_reduced\_ZT/3x3x6\_m42\_cr139\_cn287\_ZT\_reduced.json}.
\bibitem{Perminov25} A.~I.~Perminov, \emph{Parallel heuristic exploration
for additive complexity reduction in fast matrix multiplication},
arXiv:2512.13365 (2025).
\bibitem{Sedoglavic} A.~Sedoglavic, \emph{Yet another catalogue of fast matrix
multiplication algorithms}, \url{https://fmm.univ-lille.fr} (the table of
best known ranks as of November 2026).
\bibitem{Strassen69} V.~Strassen, \emph{Gaussian elimination is not optimal},
Numer.\ Math.\ 13 (1969) 354--356.
\bibitem{Winograd71} S.~Winograd, \emph{On multiplication of $2\times2$
matrices}, Linear Algebra Appl.\ 4 (1971) 381--388.
\bibitem{Accelerate} Apple, \emph{Accelerate}, the vector, BLAS and LAPACK
framework of macOS, \url{https://developer.apple.com/documentation/accelerate}.
\bibitem{AppleM4Pro} Apple, \emph{Apple introduces M4 Pro and M4 Max}, press
release, October 30, 2024,
\url{https://www.apple.com/newsroom/2024/10/apple-introduces-m4-pro-and-m4-max/}.
\bibitem{Smirnov13} A.~V.~Smirnov, \emph{The bilinear complexity and
practical algorithms for matrix multiplication}, Comput.\ Math.\ Math.\
Phys.\ 53 (2013) 1781--1795.
\bibitem{Tichavsky21} P.~Tichavsk\'y, \emph{Characterization of
decomposition of matrix multiplication tensors}, arXiv:2104.05323 (2021).
\bibitem{S26} G.~Sidebottom, \emph{201 operations for rank-48 $4\times4$
matrix multiplication over rings containing $1/2$}, preprint (2026),
\url{https://github.com/gsidebottom/logic/blob/master/doc/matmul_4x4_rank48_201ops.pdf};
PLinOpt data file \texttt{4x4x4\_48\_201}.
\bibitem{S26b} G.~Sidebottom, \emph{The 262-operation
$\langle4\times4\times4:48\rangle$ algorithm is also the more accurate one:
error-bound exponent $9/4$ and about a quarter less error}, draft of
1~November 2026,
\url{https://github.com/gsidebottom/logic/blob/master/doc/matmul_4x4_rank48_accurate262.pdf}.
\bibitem{logic} G.~Sidebottom, \emph{logic}, source repository,
\url{https://github.com/gsidebottom/logic} (2026).
\end{thebibliography}
}
\end{document}